\documentclass[twoside,twocolumn,english,aps,showpacs,prl,superscriptaddress]{revtex4-1}
\usepackage[T1]{fontenc}
\usepackage{color}
\usepackage{bm}
\usepackage{amsmath}
\usepackage{amssymb}
\usepackage{graphicx}
\usepackage{esint}
\usepackage{mathrsfs}
\usepackage{booktabs}
\usepackage{multirow}
\usepackage{makecell}
\usepackage{verbatim}
\usepackage[plainpages = false, pdfpagelabels, 
                 bookmarks,
                 bookmarksopen = true,
                 bookmarksnumbered = true,
                 breaklinks = true,
                 linktocpage,
                 colorlinks = true,
                 linkcolor = blue,
                 urlcolor  = blue,
                 citecolor = blue,
                 anchorcolor = green,
                 hyperindex = true,
                 hyperfigures
                 ]{hyperref} 
\usepackage{dcolumn}   

\makeatletter
\newcommand*{\rom}[1]{\expandafter\@slowromancap\romannumeral #1@}
\makeatother

\usepackage{times}{\Large }

\makeatother

\usepackage{babel}

\begin{document}

\preprint{APS/123-QED}

\title{Measuring statistics-induced entanglement entropy with a Hong-Ou-Mandel interferometer}

\author{Gu Zhang}
\thanks{These authors contributed equally to this work.}
\affiliation{Beijing Academy of Quantum Information Sciences, Beijing 100193, China}
\affiliation{Institute for Quantum Materials and Technologies, Karlsruhe Institute of Technology, 76021 Karlsruhe, Germany}

\author{Changki Hong}
\thanks{These authors contributed equally to this work.}
\affiliation{Braun Center for Submicron Research, Department of Condensed Matter Physics, Weizmann Institute of Science, Rehovot 761001, Israel}

\author{Tomer Alkalay}
\thanks{These authors contributed equally to this work.}
\affiliation{Braun Center for Submicron Research, Department of Condensed Matter Physics, Weizmann Institute of Science, Rehovot 761001, Israel}

\author{Vladimir Umansky}
\affiliation{Braun Center for Submicron Research, Department of Condensed Matter Physics, Weizmann Institute of Science, Rehovot 761001, Israel}

\author{Moty Heiblum}
\email{moty.heiblum@weizmann.ac.il}
\affiliation{Braun Center for Submicron Research, Department of Condensed Matter Physics, Weizmann Institute of Science, Rehovot 761001, Israel}

\author{Igor Gornyi}
\email{igor.gornyi@kit.edu}
\affiliation{Institute for Quantum Materials and Technologies, Karlsruhe Institute of Technology, 76021 Karlsruhe, Germany}

\author{Yuval Gefen}
\email{yuval.gefen@weizmann.ac.il}
\affiliation{Department of Condensed Matter Physics, Weizmann Institute of Science, Rehovot 761001, Israel}

\date{\today}

\begin{abstract}
Despite its ubiquity in quantum computation and quantum information, a universally applicable definition of quantum entanglement remains elusive. The challenge is further accentuated when entanglement is associated with other key themes, e.g., quantum interference and quantum statistics. Here, we introduce two novel motifs that characterize the interplay of entanglement and quantum statistics: an ‘entanglement pointer’ and a ‘statistics-induced entanglement entropy’. The two provide a quantitative description of the statistics-induced entanglement: (i) they are finite only in the presence of quantum entanglement underlined by quantum statistics; (ii) their explicit form depends on the quantum statistics of the particles (e.g., fermions, bosons, anyons). We have experimentally implemented these ideas by employing an electronic Hong-Ou-Mandel interferometer fed by two highly diluted electron beams in an integer quantum Hall platform. Performing measurements of auto-correlation and cross-correlation of current fluctuations of the scattered beams (following ‘collisions’), we quantify the statistics-induced entanglement by experimentally accessing the 
entanglement pointer and the statistics-induced entanglement entropy. Our theoretical and experimental approaches pave the way to study entanglement in various correlated platforms, e.g., those involving anyonic Abelian and non-Abelian states.
\end{abstract}

\maketitle

\textbf{\emph{Introduction.}}
A pillar of quantum mechanics -- quantum entanglement -- prevents us from obtaining a full independent knowledge of subsystem $A$ entangled with another subsystem $B$.  Indeed, the state of subsystem $A$ may be influenced or even determined following a measurement of $B$, even when both are distant apart.
This feature, known as the \textit{non-locality} of quantum entanglement, is at the heart of the fast-developing field of quantum information processing~(see, e.g., Refs.~\cite{HorodeckiRevModPhys09,NielsenChuangBook,WildeBook,GuhneTothReview,StreltsovAdessoPlenioRevModPhys17}).
An apt example is a system comprising two particles with opposite internal magnetic moments (``spin up'’ and ``spin down'’). Imagine we put one particle on Earth (subsystem $A$) and the other on Mars (subsystem $B$). If measurement on $A$ reveals the particle is in the ``up'’ state, this \textit{instantaneously} dictates that the $B$ particle is ``down''.
Following Bell~\cite{Bell64} and CHSH~\cite{CHSH69} inequalities, measurements of the respective spins in different directions may unambiguously demonstrate the quantum nature of the entanglement of $A$ and $B$.

An essential way in which quantum entanglement reveals itself is the entangled subsystem's entropy. The entanglement entropy (EE) of subsystem $A$ can be found when the complete information of $B$ is discarded. This amounts to summing over all possible states of $B$. Formally, the von Neumann EE is defined as $S_\text{ent} = -\text{Tr} (\rho_A \ln \rho_A)$, where $\rho_A = \text{Tr}_B (\rho_{AB})$ is the reduced density matrix of $A$ after tracing over the states of $B$, where \textcolor{red}{$\rho_{AB}$} is the density matrix in the combined Hilbert space $ A \oplus B$. When the subsystems share common entangled pairs of particles, such pairs are effectively counted by the EE.

Another pillar of quantum mechanics is the quantum statistics of indistinguishable particles, whose wavefunction might acquire a non-trivial phase upon exchanging the particles' positions (braiding).
This phase is pertinent in classifying quasiparticles as fermions, bosons, and, most interestingly, anyons.
Being instrumental in realizing platforms for quantum information processing~(see, e.g., Ref.~\cite{SarmaNPJQI15}), it motivated several insightful experiments~\cite{CaminoPRB05,CaminoPRL07,OfekPNAS10,WillettPRL13,NakamuraNatPhys19,BartolomeiScience20,NakamuraNatPhys20} that intended to detect anyonic statistics~\cite{SafiDevilardPRL01,KimPRL05,LawPRB06,CampagnanoPRL12,CampagnanoPRB13,RosenowLevkivskyiHalperinPRL16,LeePRL19,RosenowSternPRL20}.
Among such experimental setups, the Hong-Ou-Mandel (HOM) interferometer~\cite{HongOuMadelPRL87} was employed as one of the simplest platforms to manifest bosonic~\cite{BeugnonNature06}, fermionic (see, e.g., Refs.~\cite{BlanterButtikerPhysRep00,SplettstoesserPRL09}), and anyonic (Laughlin quasiparticles)~\cite{RosenowLevkivskyiHalperinPRL16,BartolomeiScience20,MartinDeltaT20} statistics.
Despite their importance, the interplay of entanglement with quantum statistics 
has hardly been studied, either theoretically or experimentally (see, however, Refs.~\cite{CramerEisertPlenioPRL07,SplettstoesserPRL09}).

In an attempt to address EE in the context of quantum transport, it has been theoretically 
proposed~\cite{KlichLevitovPRL09} to focus on a single quantum point contact (QPC) geometry (with partitioning $\mathcal{T}$ of the incident beam), which allows partial separation of two subsystems (``arms''), $A$ and $B$.
Following partial tracing over states in one subsystem, the EE can, in principle, be obtained indirectly via a weighted summation over even cumulants of particle numbers extracted from the current-noise measurements (see the discussion of noise cumulants in, e.g., Ref.~\cite{LevitovReznikovPRB04}).
However, even measurement of the fourth cumulant is not straightforward~\cite{PortierPRL10} in mesoscopic conductors~\cite{ChristopherRepProgPhys18}.
To our knowledge, no study of EE through measurements of quantum transport has been reported. We note that the EE had been measured in localized atomic systems (see Ref.~\cite{Lewis-SwanNatRevPhys19} for a review).
In addition, the ``impurity entropy'' (not an EE), induced by frustration at quantum criticality was most recently reported in Refs.~\cite{CheolheeSelaPRL22,ChildFolkEntropy22}.

\begin{figure}[ht!]
\includegraphics[width = 0.8 \columnwidth]{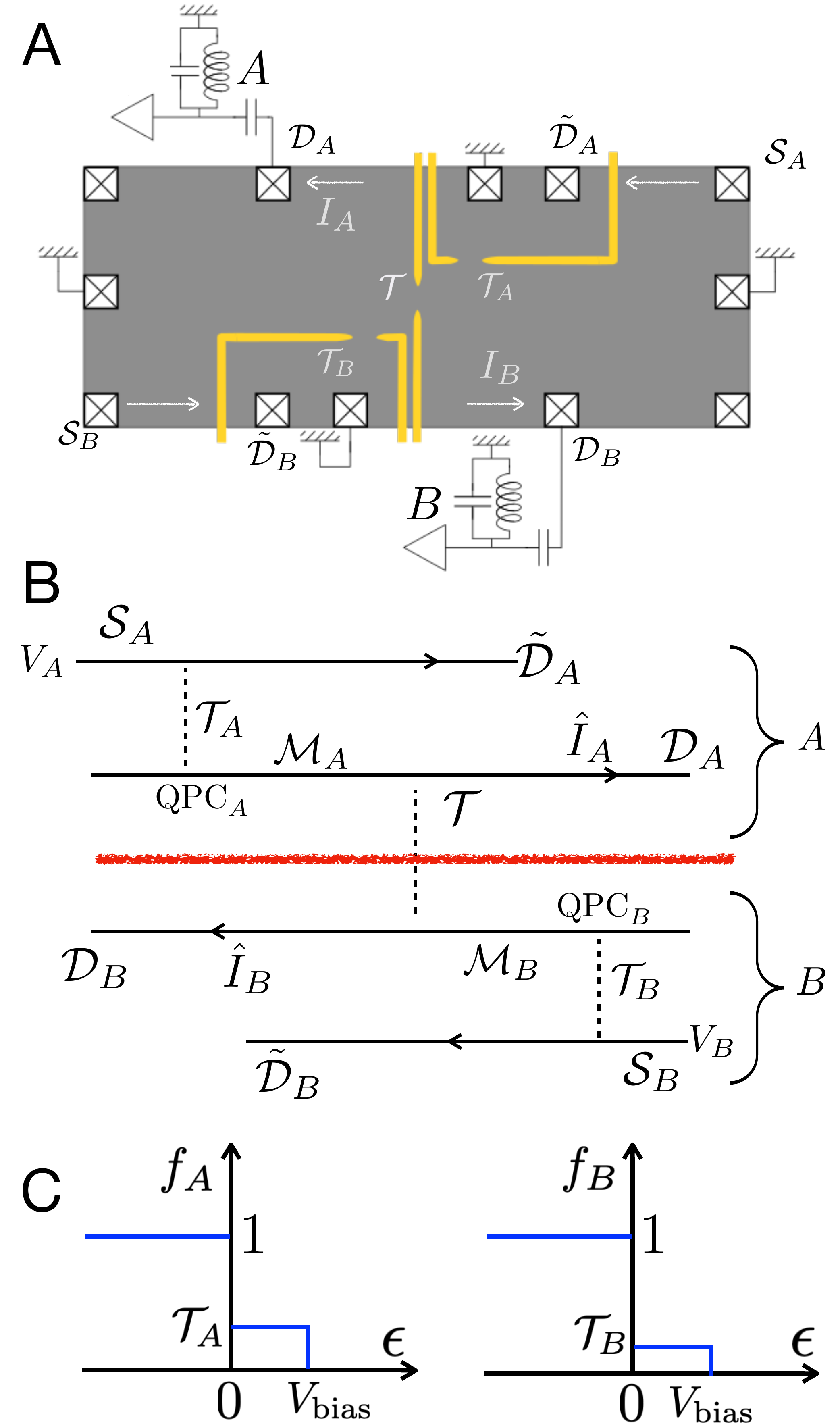}
\caption{\textbf{Schematics of the setup.}
(A) Schematics of the experimental setup.
(B) The corresponding theoretical schematics. The HOM interferometer consists of two sources ($\mathcal{S}_A$ and $\mathcal{S}_B$) and two diluted middle arms ($\mathcal{M}_A$, $\mathcal{M}_B$ ) via transmission probabilities $\mathcal{T}_A$ and $\mathcal{T}_B$ (with two QPC's).
The currents are measured at drains $\mathcal{D}_A$ and $\mathcal{D}_B$.
For later convenience, we call the source arms after the diluters as $\tilde{\mathcal{D}}_A$ and $\tilde{\mathcal{D}}_B$.
For simplicity, the two sources are equally biased: $V_A = V_B = V_\text{bias}$.
The red line separates the two ``entangled'' subsystems, $A$ (the two upper arms with labels ``$A$'') and $B$ (the two lower arms with labels ``$B$''). (C) Distribution functions of the two middle arms ($\mathcal{M}_A$ and $\mathcal{M}_B$) for non-interacting fermions ($f_A$, $f_B$, respectively) at zero temperature (carrying shot noise). 
The double-step distributions are modified when the filling factor is larger than one, with an added
interaction between the two modes on each edge [Eq.~(S34) of SI Section~S2].
}
\label{fig:model}
\end{figure}

\textbf{\emph{Perspective of our analysis.}}
In the present study, we fuse two foundational quantum-mechanical notions: quantum statistics and entanglement, and propose the concept of \textit{statistics-induced entanglement}.
We introduce two functions quantifying \textit{entanglement arising from the quantum statistics} of indistinguishable particles: (i) the ``entanglement pointer'' (EP, $\mathcal{P}_\text{E}$),
and (ii) the statistics-induced entanglement entropy, SEE (denoted as $S_\text{SEE}$). 
Both are derived from correlations of currents fluctuations
in an HOM configuration, and are expressed in Eqs.~\eqref{eq:ep_definition} $\&$ \eqref{eq:dsa}. Importantly, these two functions vanish for distinguishable particles, and are finite when indistinguishable particles emitted from the two sources $\mathcal{S}_A$ and $\mathcal{S}_B$ become entangled following ``collisions''.

Typically entanglement is the outcome of Coulomb interaction between distinct constituents of the system. Here, we focus on the entanglement being solely a manifestation of quantum statistics. If one considers current-current correlators, this contribution to the entanglement may be complemented (or even fully masked) by the effect of Coulomb interactions between the colliding particles.
Below we show, theoretically and experimentally, that with our specially designed function, $S_\text{SEE}$,
the leading contributions of the Coulomb interaction are canceled, with the remaining terms dominated by quantum statistics [cf. Eq.~\eqref{eq:int_pe} and Eq.~(S21) of Supplementary Information (SI) Section~S1].

We note that the acquisition of statistics-induced entanglement is both {\it instantaneous} and {\it  non-local}. It is acquired immediately when two identical particles braid each other, even at a distance. By these features it is universal. By contrast, the Coulomb interaction contribution to the entanglement requires the two particles to directly interact with each other, and depends on the strength and duration of this interaction, hence it is non-universal.
This non-universal influence from interaction becomes dominant in the measured noise (Figs.~\ref{fig:experiment}B and ~\ref{fig:experiment}C), but is negligible in our constructed EP, $\mathcal{P}_\text{E}$.

Turning now to the technicalities of our study, the theoretical derivation of the explicit forms of the EP and SEE (see Methods) employs, respectively, the Keldysh technique~\cite{LevitovReznikovPRB04} and an extended version of the approach of Ref.~\cite{KlichLevitovPRL09} [see Eq.~\eqref{eq:ee}].
The actual measurements were carried out in an HOM configuration~\cite{HongOuMadelPRL87}, fabricated in a two-dimensional electron gas (2DEG) tuned to the integer quantum Hall (IQH) regime. 
Two highly diluted (via weak partitioning in two outer QPCs) edge modes were let to ``collide'' at a center-QPC, and current fluctuations (shot noise) of two scattered diluted beams (Fig.~\ref{fig:model}) were measured.
While the definitions of the EP and SEE are not restricted to a specific range of parameters, expressing SEE in terms of the measured EP is possible only within the limit of highly diluted impinging current beams [Eq.~\eqref{eq:tilde_se}]. As will be shown, the theoretical prediction agrees very well with the experimental data.

\textbf{\emph{The model and the entanglement pointer (EP).}}
Our HOM interferometer consists of four arms, all in the IQH regime (Fig.~\ref{fig:model}).
Two sources $\mathcal{S}_A$ and $\mathcal{S}_B$ are biased equally at $V_A =V_B = V_\text{bias}$, with sources currents weakly scattered by two QPCs, each with dilution $\mathcal{T}_A$ and $\mathcal{T}_B$, respectively. The partitioned beams impinge on a central QPC (from middle arms $\mathcal{M}_A$ and $\mathcal{M}_B$ in Fig.~\ref{fig:model}) with transmission $\mathcal{T}$~\cite{RosenowLevkivskyiHalperinPRL16}.
The two transmitted electron beams are measured at drains $\mathcal{D}_A$ and $\mathcal{D}_B$.

With this setup, we define the first entanglement-quantification function -- EP, $\mathcal{P}_\text{E}$. It is expressed through the cross-correlation of the two current fluctuations, excluding the statistics-irrelevant contribution,
\begin{equation}
\begin{aligned}
&\mathcal{P}_\text{E} (\mathcal{T}_A, \mathcal{T}_B,V_\text{bias}) \equiv \int dt \left[ \langle I_A (t) I_B (0) \rangle_{\text{irr}}\left|_{\mathcal{T}_A,\mathcal{T}_B,V_\text{bias}}
\right.
\right.\\
& \left.  
-  \langle I_A (t) I_B (0) \rangle_{\text{irr}}\left|_{0,\mathcal{T}_B,V_\text{bias}}\right. \!-\! \langle I_A (t) I_B (0) \rangle_{\text{irr}}
\left|_{\mathcal{T}_A,0,V_\text{bias}}\right.\right].
\label{eq:ep_definition}
\end{aligned}
\end{equation}
Here, $I_A$ and $I_B$ refer to the current operators in the corresponding drains $\mathcal{D}_A$ and $\mathcal{D}_B$ (Fig.\,\ref{fig:model}B), and ``irr'' refers to the irreducible correlators (connected correlation function), where the product of the averages is removed. 
Note that the last two terms in Eq.~\eqref{eq:ep_definition} are each evaluated with only one active source (i.e., either $\mathcal{T}_A$ or $\mathcal{T}_B$ is zero), and, thus do not involve the two-particle scattering in the HOM configuration~\cite{BlanterButtikerPhysRep00,VishveshwaraPRL03,CampagnanoPRL12}.
This removal of last two terms has been carried out in Refs.~\cite{ButtikerPRB92,VishveshwaraPRL03}, however without referring to entanglement.
Importantly, Eq.~\eqref{eq:ep_definition} yields zero for distinguishable, non-interacting particles, since the first term is then a superposition of two independent single-source terms.
By contrast, the EP is finite and statistics-dependent for indistinguishable particles.

For instance, for a double-step-like distribution of such particles (e.g., Fig.\,\ref{fig:model}C for fermions), we obtain cross-correlations (CC) of current operators,
\begin{equation}
\begin{aligned}
\text{fermions:}&   \int dt \langle I_A (t) I_B (0) \rangle_{\text{irr}}\Big|_{\mathcal{T}_A,\mathcal{T}_B,V_\text{bias}} = \\
& - \frac{e^3}{h}   \mathcal{T} (1 - \mathcal{T}) \left[ (\mathcal{T}_A - \mathcal{T}_B)^2 + \mathcal{T}_A \mathcal{T}_B P_\text{QPC} \right] V_\text{bias} ,\\
\text{bosons:}&   \int dt \langle I_A (t) I_B (0) \rangle_{\text{irr}}\Big|_{\mathcal{T}_A,\mathcal{T}_B,V_\text{bias}} = \\
& \frac{e^3}{h}   \mathcal{T} (1 - \mathcal{T}) \left[ (\mathcal{T}_A - \mathcal{T}_B)^2 - \mathcal{T}_A \mathcal{T}_B P_\text{QPC} \right] V_\text{bias}.
\end{aligned}
\label{eq:fb_pint}
\end{equation}
Here $P_\text{QPC}$ describes an additional bunching (or anti-bunching) probability induced by Coulomb interactions within the central QPC (cf. SI Section~S2).
Note that for equal diluters, $\mathcal{T}_A= \mathcal{T}_B$, the non-interacting part of the CC vanishes, indicating that the nature of the CC is then solely determined by interactions.
This is however not so for the EP. Indeed,
with Eqs.~\eqref{eq:ep_definition} and \eqref{eq:fb_pint}, we obtain,
\begin{equation}
\begin{aligned}
\text{fermion EP: \quad }& \! \mathcal{P}_\text{E} \!=\! (2 - P_\text{QPC}) \frac{e^3}{h} \mathcal{T} ( 1 - \mathcal{T} ) \mathcal{T}_A \mathcal{T}_B V_\text{bias}, \\
\text{boson EP: \quad }& \! \mathcal{P}_\text{E} \!=\! (-2 - P_\text{QPC} ) \frac{e^3}{h} \mathcal{T} ( 1 - \mathcal{T} ) \mathcal{T}_A \mathcal{T}_B V_\text{bias}.
\label{eq:int_pe}
\end{aligned}
\end{equation}

In the presence of a weak inter-mode interaction among particles within the middle arms $\mathcal{M}_A$ and $\mathcal{M}_B$, $P_\text{QPC}$ is replaced with $P_\text{QPC} + P_\text{frac}$ [see Methods and Eq.~(S51) of SI Section~S2].
The term $P_\text{frac}$ refers to the influence of intra-arm ``charge fractionalization'' that produces ``particle-hole'' dipoles in the two interacting edge modes (Refs.~\cite{LevkiskyiSukhorukovPRB12,WahlMartinPRL14}).
Crucially, the unavoidable Coulomb-interaction contribution to the EP, parameterized by $P_\text{QPC}$ and $P_\text{frac}$ (introduced in Supplementary Eqs.~(S50) and (S51), respectively), appears in terms that are quadratic in the beam dilution ($\mathcal{T}_A$ and $\mathcal{T}_B$) and hence is parametrically smaller than the linear ($\sim \mathcal{T}_A P_A$ and $\sim \mathcal{T}_B P_B$) terms in the noise correlation functions [see Eq.~\eqref{eq:iu_id_methods} in Materials and Methods].
It follows that the EP rids of the undesired effect of Coulomb interactions, hence truly reflecting the state’s statistical nature.

\begin{figure}[h!]
\includegraphics[width = \columnwidth]{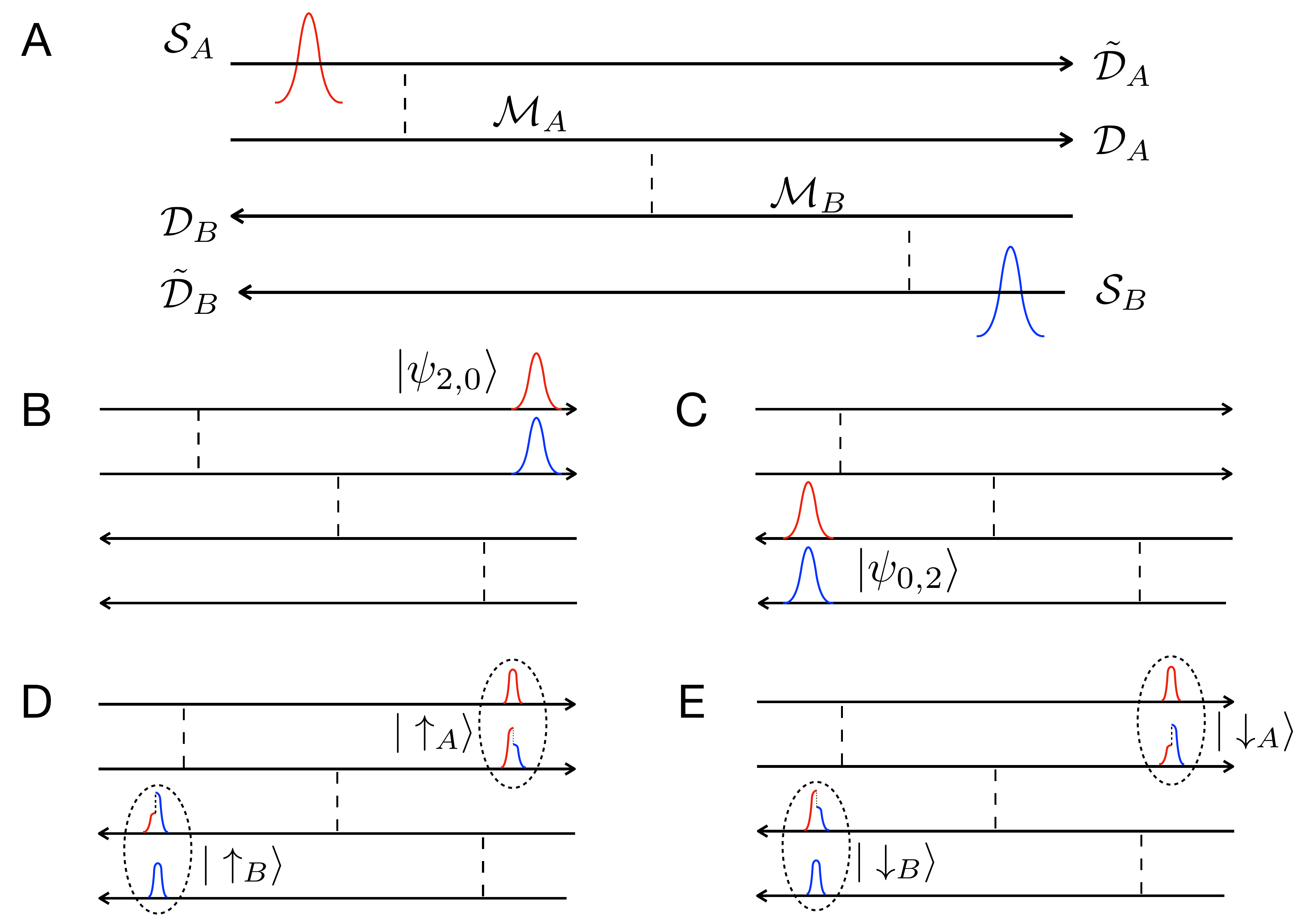}
\caption{\textbf{Two components $|\tilde{\psi}\rangle$ and $|\psi\rangle$, of two-particle wavefunctions, shown in the schematics of Fig.~\ref{fig:model}B}. (A) Pre-collision configurations. Two particles (red and blue pulses) are injected from $\mathcal{S}_A$ and $\mathcal{S}_B$, respectively. Hereafter, post-scattering quasi-particles comprise contributions from both incident particles (blue and red). (B) and (C): Constituents of  the state $|\tilde{\psi}\rangle$ $= |\psi_{2,0}\rangle + |\psi_{0,2}\rangle$. (D) and (E): Constituents of the ``Bell pair'' state $|\psi\rangle$ $= {\alpha{|\!\uparrow_A\rangle}{|\!\uparrow_B\rangle}+\beta{|\!\downarrow_A\rangle}{|\!\downarrow_B\rangle}}$, Eq.~\eqref{eq:bell_state}. In comparison to $|\tilde{\psi}\rangle$ configurations, particle states of $|\psi\rangle$ configurations are entangled, both within (indicated by dashed ellipses) and between (indicated by the red-blue mixed pulses) subsystems.}
\label{fig:psi_quantum_classic}
\end{figure}

\textbf{\emph{Entanglement entropy from statistics.}}
The second entanglement quantifier is the ``statistics-induced entanglement entropy'' (SEE),
which is defined in a similar spirit to the EP (by removing the statistics-irrelevant single-source contributions to the EE),
\begin{equation}
S_\text{SEE} (\mathcal{T}_A,\mathcal{T}_B)\! \equiv \!- \!\left[S_\text{ent} (\mathcal{T}_A,\mathcal{T}_B)\! -\! S_\text{ent} (\mathcal{T}_A,0)\! -\! S_\text{ent} (0,\mathcal{T}_B)\right]. \label{eq:dsa}
\end{equation}
To illustrate the relation between the SEE and Bell-pair entanglement, we consider the case of two incoming fermions (Fig.~\ref{fig:psi_quantum_classic}A).
The pure two-particle state at the output of our device is represented as (see Fig.~\ref{fig:psi_quantum_classic})
\begin{equation}
    |\Psi\rangle = |\tilde{\psi}\rangle + |\psi \rangle.
\end{equation}

Here, $|\tilde{\psi}\rangle \equiv |\psi_{2,0}\rangle+|\psi_{0,2}\rangle$ denotes a state where both particles end up in either subsystem $A$ (2,0) or $B$ (0,2) (Figs.~\ref{fig:psi_quantum_classic}B and \ref{fig:psi_quantum_classic}C, respectively).
In either case, obeying Pauli’s blockade, two electrons must occupy the two arms of the same subsystem: for example, $\tilde{\mathcal{D}}_A$ and $\mathcal{D}_A$ for the state $(2,0)$, which can be written as $(1,1,0,0)$ in the basis of drain arms $\tilde{\mathcal{D}}_A,\mathcal{D}_A,\mathcal{D}_B,\tilde{\mathcal{D}}_B$.
Any coupling between the arms within one subsystem cannot change the $1,1$ arrangement for subsystem $A$ in
$|\psi_{2,0}\rangle$.
This implies that no quantum manipulations on subsystem $A$, leading to Bell's inequalities (Refs.~\cite{Bell64,CHSH69}) are possible with $|\tilde{\psi}\rangle$ \textbf{alone}, i.e., without coupling the present setup to extra channels.
The same holds for subsystem $B$. 
Nevertheless, $|\tilde{\psi}\rangle$ is a non-product state with nonlocal~\cite{GisinPLA91,WisemanJonesDohertyPRL07} entanglement:
if the two particles are detected in subsystem $A$, this automatically implies that no particles are to be detected in subsystem $B$.
In principle, Bell's inequalities can be tested with $|\tilde{\psi}\rangle$ using modified devices akin to those proposed for a similar bosonic state (a NOON state) in, e.g.,  Refs.~\cite{JonesWisemanPRA11,WisemanFurusawaNC15} and Refs.~\cite{TanWallsCollettPRL91,HessmoPRL04}, after the introduction of external states.

By contrast,
\begin{equation}
\begin{aligned}
|\psi\rangle = \alpha | \uparrow_A \rangle | \uparrow_B \rangle + \beta | \downarrow_A \rangle | \downarrow_B \rangle \equiv |\psi_{1,1}\rangle,
\end{aligned}
\label{eq:bell_state}
\end{equation}
represents an effective Bell pair (with amplitudes $\alpha$ and $\beta$), where one particle leaves the device through subsystem $A$ and the other through $B$ (hence $|\psi_{1,1}\rangle$, as opposed to $|\psi_{2,0}\rangle$ and $|\psi_{0,2}\rangle$, see Figs.~\ref{fig:psi_quantum_classic}D and ~\ref{fig:psi_quantum_classic}E).
Here, ${|\!\uparrow_A\rangle}, 
{|\!\downarrow_A\rangle}$ are certain (mutually orthogonal) linear combinations of the states $|\tilde{\mathcal{D}}_A\rangle$ and $|\mathcal{D}_A\rangle$; ${|\!\uparrow_B\rangle},{|\!\downarrow_B\rangle}$ are defined similarly for subsystem $B$ [see Eq.~(S67) of SI Section~S3].
The quantum superposition in Eq.~\eqref{eq:bell_state} allows for rotating the ``pseudospin'' (i.e., rotation between orthonormal bases ${|\!\uparrow_A\rangle}$ and ${|\!\downarrow_A\rangle}$) of subsystem $A$, hence the measurement of pseudospins in a ``transverse direction'' is possible, as required by Bell's inequalities.
It is also worth noting that the states ${|\!\uparrow_A\rangle}$, ${|\!\downarrow_A\rangle}$ are nonlocal: each of them is constructed out of states in the arms associated with $\tilde{\mathcal{D}}_A$ and $\mathcal{D}_A$, which are spatially separated by the middle arm $\mathcal{M}_A$ of length about $2\mu$m in the real setup, see Fig.~\ref{fig:experiment}A.
The detailed forms of $|\tilde{\psi}\rangle$ and $|\psi\rangle$ are manifestations of quantum statistics, which underlines the statistics-induced entanglement, captured by the function $S_\text{SEE}$.

Technically, the building blocks of SEE, $S_\text{ent}$ (Eq.~\eqref{eq:dsa}) can be obtained~\cite{KlichLevitovPRL09,KlichLevitovAIP09,SongLehurPRB11} by calculating the ``generating function'' $\chi (\lambda) \equiv \sum_q \exp(i\lambda q) P_q$ of the full counting statistics (FCS)~\cite{LevitovReznikovPRB04} (see Methods). Here $P_q$ refers to the probability to transport charge $q$ between the subsystems $A$ and $B$ over the measurement time. In a steady state, $S_\text{SEE}$ is proportional to the dwell time $\tau$ (see SI Section~S5), which corresponds to the shortest travel time from the central QPC to an external drain.
This should be replaced by the coherence time, $\tau_\varphi$, if the latter is shorter than the dwell time. The EE grows linearly with the coherent  arm’s length (beyond this coherence length the particles become dephased, hence  disentangled). In the following analysis we neglect externally-induced dephasing along the arms, as the dephasing length in this type of IQH devices is known to be longer than the arm’s length.

While the function $S_\text{SEE}$ is
in principle measurable, it becomes readily accessible in the strongly diluted limit $\mathcal{T}_A, \mathcal{T}_B \ll 1$.
In this limit, $S_\text{SEE}$ is approximately equal to a function $\tilde{S}_\text{SEE}$,
which is proportional to the EP of free fermions and bosons (see Methods): 
\begin{equation}
\begin{aligned}
S_\text{SEE} \xrightarrow{\mathcal{T}_A,\mathcal{T}_B \ll 1} - \frac{1}{2e^2} \mathcal{P}_\text{E} \ln (\mathcal{T}_A\mathcal{T}_B \mathcal{T}^2) \tau \equiv \tilde{S}_\text{SEE}.
\end{aligned}
\label{eq:tilde_se}
\end{equation}
Note the statistics-sensitive factor, $\mathcal{P}_\text{E}$. Since the non-interacting (purely statistical) contribution to $\mathcal{P}_\text{E}$ (cf. Eq.~\eqref{eq:int_pe}) for bosons is opposite in sign to its fermionic counterpart, so is the corresponding $\tilde{S}_\text{SEE}$.

\begin{figure}[h!]
\includegraphics[width= \columnwidth]{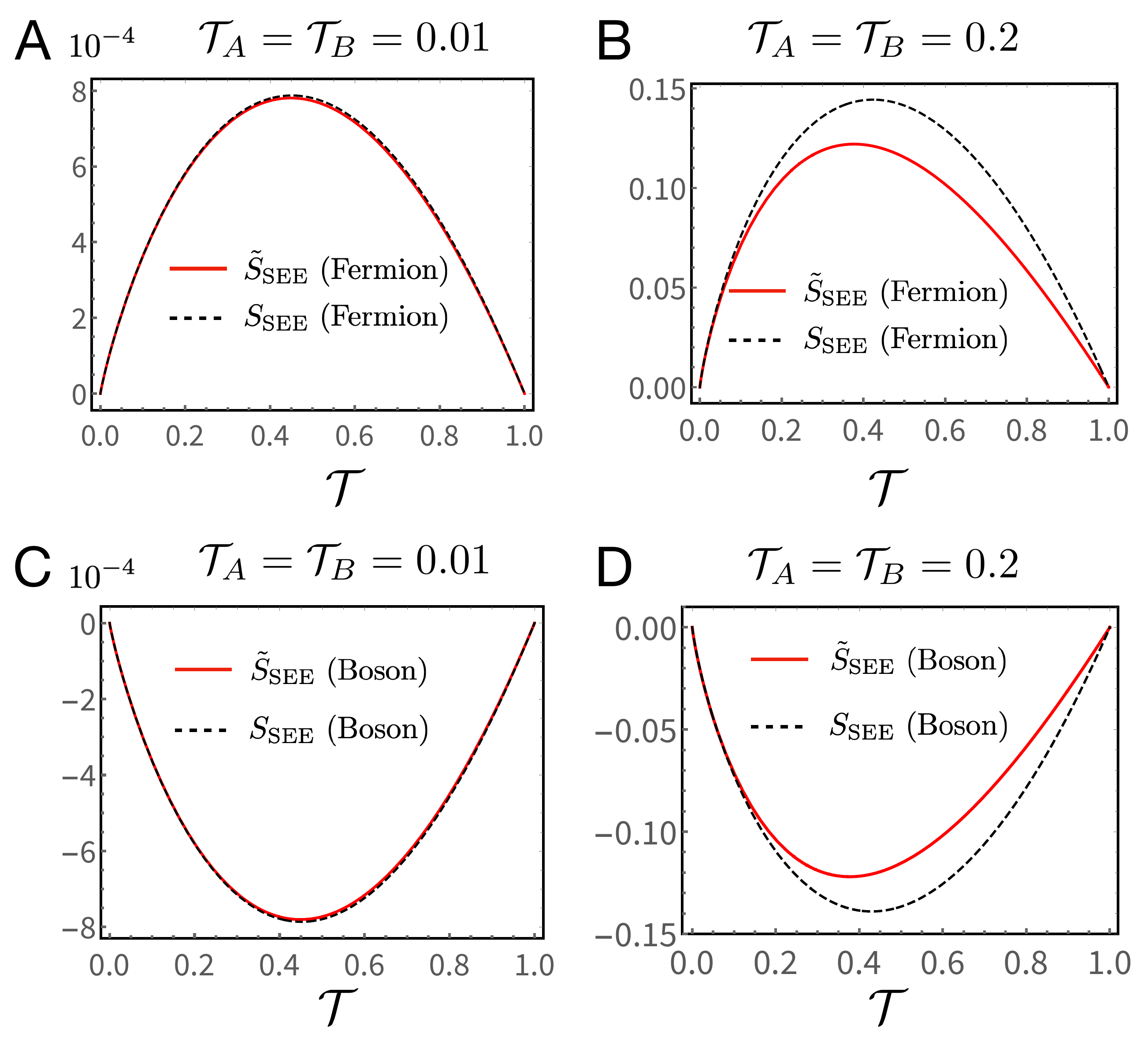}
\caption{\textbf{Comparison between theoretical values of $ S_\text{SEE}$ and $\tilde{S}_\text{SEE}$.} Free fermions [panels (A) and (B)] and free bosons [panels (C) and (D)].
These functions perfectly overlap for the entire range of $\mathcal{T}$ when $\mathcal{T}_A = \mathcal{T}_B = 0.01$ in (A) and (C). When $\mathcal{T}_A = \mathcal{T}_B = 0.2$ [(B) and (D)], a finite but small difference begins to show up between them. The bias used $V_\text{bias} = 20.7$µV. We take the dwell time $\tau_\text{dwell} = 0.01$ns (see SI Section~S5 for the evaluation of $\tau_\text{dwell}$).}
\label{fig:comparisons}
\end{figure}

Importantly, addressing SEE (compared to $S_\text{ent}$) has two obvious upsides. First, extracting $\mathcal{P}_\text{E}$ via current cross-correlation measurements, this quantity is easy to obtain in the strongly diluted particle beam limit.
Second, it allows us to rid of most of the undesired effects of Coulomb interactions, and clearly single out quantum statistics contributions. To validate Eq.~\eqref{eq:tilde_se} we compare $S_\text{SEE}$ [calculated according to Eq.~(S21) of SI Section~S1] and $\tilde{S}_\text{SEE}$ for free fermions (Figs.~\ref{fig:comparisons}A $\&$ \ref{fig:comparisons}B) and free bosons (Figs.~\ref{fig:comparisons}C $\&$ \ref{fig:comparisons}D). We next compare out theoretical predictions with experiment.

\textbf{\emph{Experimental results.}}
The experimental structure was fabricated in uniformly doped GaAs/AlGaAs heterostructure, with an electron density of 9.2 ${\times}$ 10$^{10}$ cm$^{-2}$ and 4.2 K dark mobility 3.9 ${\times}$ 10$^{6}$ cm$^{2}$V$^{-1}$s$^{-1}$.
The 2DEG is located $125$nm below the surface. Measurements were conducted at an electron temperature $\sim14$mK.
The structure is shown in Fig.~\ref{fig:model}A (schematically) and in Fig.~\ref{fig:experiment}A (electron micrograph).
Two QPCs are used to dilute the two electron beams, which ``collided'' at the central QPC located 2µm away.
Two amplifiers, each with an LC circuit tuned to 730 KHz (with bandwidth 44 KHz) measuring the charge fluctuations, are placed at a large distance (around 100µm) from the 2D Hall bar.
The outer-most edge mode of filling factor $\nu = 3$ of the IQH was diluted by the two external QPCs.

Cross-correlation of the current fluctuations of the reflected diluted beams from the central QPC ($\mathcal{T} = 0.53$), with $\mathcal{T}_A = \mathcal{T}_B= 0.2$, is plotted in Fig.~\ref{fig:experiment}B. The corresponding single source CC, with $\mathcal{T}_A = 0.2\  \&\  \mathcal{T}_B=0$, is plotted in Fig.~\ref{fig:experiment}C. Though the data is rather scattered, the agreement with the theoretically expected CC is reasonable.
For both cases, the measured data displays a clear deviation from the non-interacting curve: an
evidence of strong interaction influence. Importantly, for the equal-source situation ($\mathcal{T}_A = \mathcal{T}_B = 0.2$, Fig.~\ref{fig:experiment}B), the CC is entirely produced by interactions within a single source [see SI Eq.~(S48)], indicating the inadequacy of CC to quantify entanglement.

The measured data, with the applied source voltage larger than the electron's temperature ($eV > k_B T$) was used to calculate the EP (Fig.~\ref{fig:experiment}D) and the SEE (Fig.~\ref{fig:experiment}E), and then compared with the expected EP and SEE. The measurement results conformed with the theoretical prediction of the EP. More data for $\nu = 3$ and $\nu = 1$ situations is provided in SI Sections~S8 and S9, respectively.

\begin{widetext}

\begin{figure*}[h!]
\vspace*{-0.5cm}
\includegraphics[width= 0.95\columnwidth]{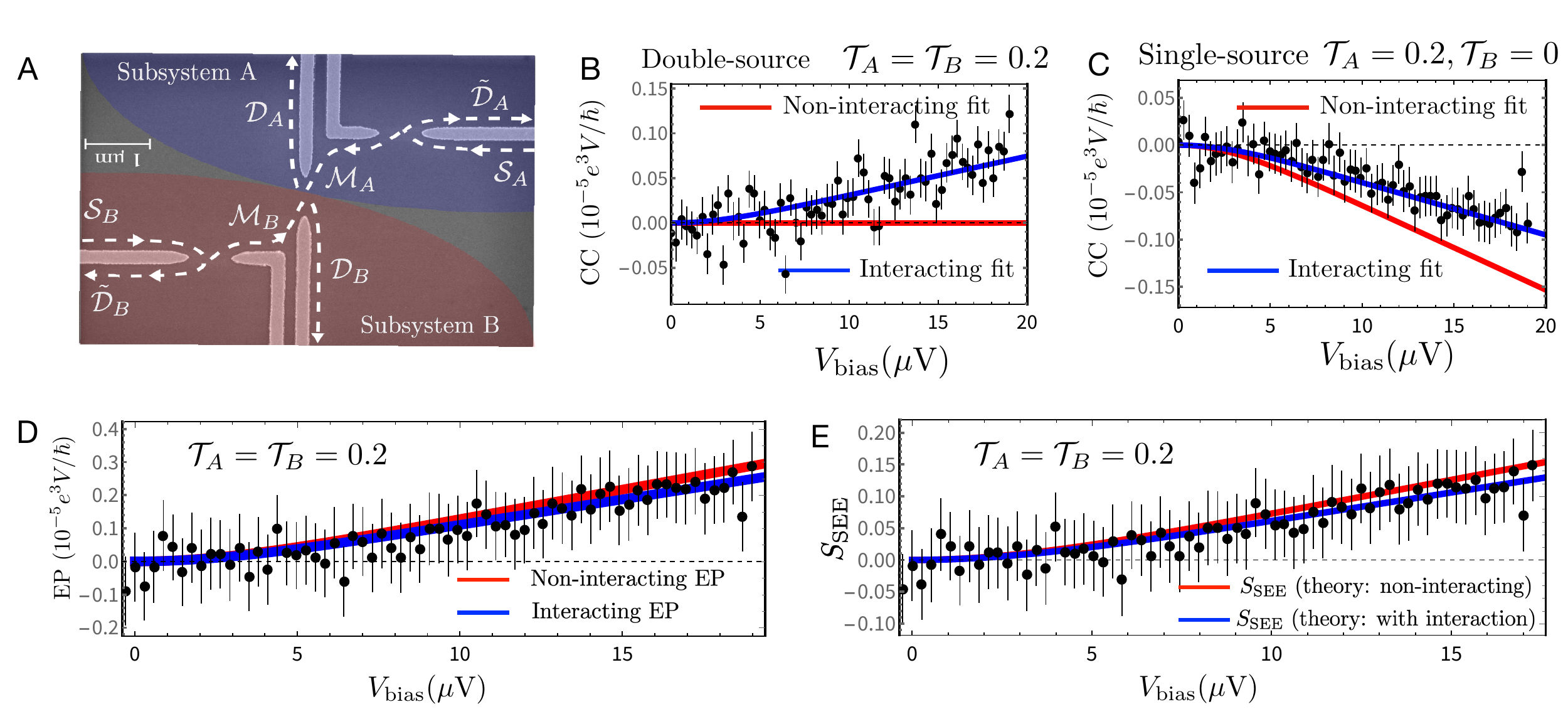}
\vspace*{-0.5cm}
\caption{\textbf{Experimental setup and experiment-theory comparisons.}
(A) SEM (scanning electron microscope) micrograph of the central part of the fabricated sample. Subsystems $A$ and $B$ (cf.  Fig.~\ref{fig:model}B) are highlighted by shaded blue and shaded red areas, respectively. Transport directions of edge states in the arms associated with the sources $\mathcal{S}_{A,B}$
and drains $\mathcal{D}_{A,B}$,  $\tilde{\mathcal{D}}_{A,B}$, 
as well as in the diluted ``middle arms'' $\mathcal{M}_{A,B}$, are indicated by dashed black arrows.
(B), (C) Double-source and single-source cross-correlations (CCs), respectively [see Eq. (S48) of the SI for expressions that include interaction contributions]. In both cases, theoretical curves with interaction taken into account (blue curves) agree better with the experimental data (black dots). (D), (E) Measured data for the EP and SEE. Panel (D) compares the measured EP (black dots) 
with theoretical curves: including interaction contributions  (blue) and non-interacting particles  (red). Although the
interaction strength is the same as in (B) and (C), the difference between the theoretically calculated values of the EP with and without interaction  contribution is  much smaller than  for the cross-current correlations, panels B and C. This demonstrates that the expression for the EP subtracts the interaction contribution to leading order. 
Panel (E): Comparison between the experimentally measured data (black dots) and
the theoretical dependence of $S_\text{SEE}$ on the source current (here $\tau_\text{dwell} = 0.01$ns
as in Fig.~4).
Experimental data points are obtained in two steps (see SI Section S1D for details): (i) we evaluate $\tilde{S}_\text{SEE}$ (using Eq.~\eqref{eq:tilde_se})
with the measured EP from panel (D), and, (ii) relying on the fact that at the experimental value $\mathcal{T} = 0.53$ the ratio $S_\text{SEE}/\tilde{S}_\text{SEE} \approx 1.22$ in Fig.~4B,
we use this ratio to scale the measured $\tilde{S}_\text{SEE}$ to reconstruct $S_\text{SEE}$.
Both the interacting (blue) and non-interacting (red) theoretical curves for  $S_\text{SEE}$ agree remarkably well with the experimental data.
}
\label{fig:experiment}
\end{figure*}

\end{widetext}

As discussed above, the current fluctuations are also influenced by two sources of Coulomb interactions: (i) inter-mode interaction at the same edge, and (ii) interaction within the central QPC in the process of the two-particle scattering (Figs.~\ref{fig:experiment}B $\&$ \ref{fig:experiment}C).
However, the influence of these interactions on the EP and SEE is negligible in our setup, which is in great contrast to, e.g., Refs.~\cite{SaragaPRL04,SaragaPRB05}, where entanglement is purely interaction-induced; see also Refs.~\cite{PeresBook,SchliemannPRA01}, as Figs.~\ref{fig:experiment}D $\&$ \ref{fig:experiment}E demonstrate (see also SI Section~S2).
Thus, the measured current noise indeed yields the information on statistics-induced entanglement.

\textbf{\emph{Summary and outlook.}}
Entanglement and exchange statistics are two cornerstones of the quantum realm. Swapping quantum particles affects the many-body wavefunction by introducing a statistical phase, even if the particles do not interact directly. We have shown that quantum statistics induces genuine entanglement of indistinguishable particles, and developed theoretical and experimental tools to unambiguously quantify this effect.  Our success in ridding of the contribution of the local Coulomb interaction, facilitates a manifestation of the foundational property of quantum mechanics – nonlocality.  It also presents the prospects of generalizing our protocol to a broad range of correlated systems, including those hosting anyons (Abelian and non-Abelian, see e.g., SI Section S7), and exotic composite particles (e.g. ``neutralons'' at the edge of topological insulators~\cite{KaneFisherPolchinskPRL94,WangMeirGefenPRL13}). Our protocol may also be generalized to include setups based on more complex edge structures, and platforms where the quasi-particles involved are spinful. Another intriguing direction is to explore the interplay of statistic-induced entanglement and quantum interference~(e.g., similar structures considered in SI Section S8).
\\

\textbf{\large{{Materials and Methods:}}}

In Materials and Methods, we provide (i) the sample fabrication; (ii) the experimental setups; (iii) an outline of the SEE derivation; (iv) an physical interpretation on the EP-SEE connection, and (v) a physical picture on the effect of interaction along the arms: charge fractionalization.\\

\textbf{\emph{Sample Fabrication.}}
The GaAs/AlGaAs heterostructure we use consists of a 2DEG layer (with 125 nm depth) and a donor layer (with 92 nm depth). We perform the MESA layer (with thickness 100 nm) etching with a wet solution H$_3$PO$_4$:H$_2$O$_2$:H$_2$O = 1:1:50. Ge/Ni/Au materials (with a stacked total thickness 450 nm) serve as a rapid thermal annealed standard Ohmic contacts. A high-$\kappa$ oxide layer (HfO$_2$, thickness 30 nm) is fabricated with an atomic layer deposition machine. Ti/Au quantum point contact (QPC) gates with 200 nm width and an 800 nm gap, are deposited on the oxide layer. Before the deposition of the last 300 nm Ti/Au metal contact pad, the sample is etched for reactive-ion etching with Ar and BCl$_3$ gas window for the oxide layer for the contact.

\textbf{\emph{Experimental setups.}}
Base temperature for the measurement was below 10 mK by CMN temperature sensor and Rutinum Oxide sensor. Low frequency 13 Hz used for the transmission and gate voltage versus conductance measurement by lock-in amplifier (NF corporation LI 5655). For the noise measurement, a voltage source with 1 G${\Omega}$ resistor series connected for the DC currents to the source ohmic contacts. Each amplifier line used around $\sim$730 KHz resonance frequency LC resonant circuit with blocking capacitor and amplified by ATF-34143 HEMT based home-made voltage preamplifier at 4K plate. At room temperature, the noise amplified room temperature voltage amplifier (NF corporation SA-220F5) and measured the noise with digital multi-meters (HP 34401A) after a home-made analogue cross-correlator.

\textbf{\emph{Derivation of SEE.}}
The system EE $S_\text{ent}$ (and the ensued SEE) can be obtained through its connection with full counting statistics (FCS)~\cite{KlichLevitovPRL09,KlichLevitovAIP09,SongLehurPRB11}, i.e.,
\begin{equation}
S_\text{ent} = \frac{1}{\pi} \int_{-\infty}^{\infty} du \frac{u}{\cosh^2 u} \text{Im} \left[ \ln \chi(\pi - \eta - 2 i u) \right],
\label{eq:ee}
\end{equation}
where $\chi$ refers to the generating function that fully describes the tunneling between subsystems, and $\eta$ is a positive infinitesimal.

Of the non-interacting situation, the zero-temperature generating function equals~\cite{LevitovReznikovPRB04,GogolinKomnikPRB06},
\begin{equation}
\begin{aligned}
\ln \chi(\lambda_\text{FCS}) & = e V_\text{bias}\tau \ln \left\{ 1 + \mathcal{T} \left[ \mathcal{T}_A (1 - \mathcal{T}_B) \left( e^{i \lambda_\text{FCS}} - 1 \right)\right] \right. \\
& \left. +\mathcal{T} \left[ \mathcal{T}_B (1 - \mathcal{T}_A) \left( e^{-i \lambda_\text{FCS}} - 1 \right) \right] \right\}/h,
\label{eq:chi_qpc}
\end{aligned}
\end{equation}
where $\tau$ is the dwell time and $\lambda_\text{FCS}$ is an auxiliary field (the ``measuring field'') introduced in FCS.
To the second order of dilutions, it approximately becomes
\begin{equation}
\begin{aligned}
& \frac{S_\text{ent} (\mathcal{T}_A ,\mathcal{T}_B)}{\tau} \! \approx \! \frac{eV_\text{bias}}{h} \left\{ \mathcal{T}
 [ \mathcal{T}_A\! +\! \mathcal{T}_B \! - \! \mathcal{T}_A \ln (\mathcal{T}_A \mathcal{T}) \!  \right.\\
 &  -\!   \mathcal{T}_B \ln (\mathcal{T}_B \mathcal{T} ) ] - \frac{1}{2} \mathcal{T}^2(\mathcal{T}_A^2 + \mathcal{T}_B^2)\\
 & + \left.  \mathcal{T} (1 - \mathcal{T}) \mathcal{T}_A \mathcal{T}_B \ln (\mathcal{T}^2 \mathcal{T}_A \mathcal{T}_B) \!+\! O(\mathcal{T}_{A,B}^3) \right\}.
\end{aligned}
\label{eq:diff}
\end{equation}
After the removal of the single-source contributions, Eq.~\eqref{eq:diff} reduces to Eq.~\eqref{eq:tilde_se}, i.e.,
\begin{equation}
\begin{aligned}
S_\text{SEE} & = -[  S_\text{ent} (\mathcal{T}_A, \mathcal{T}_B) - S_\text{ent} (\mathcal{T}_A, 0) - S_\text{ent}(0 , \mathcal{T}_B)]\\
& \approx - \frac{eV_\text{bias}}{h} \mathcal{T} (1 - \mathcal{T}) \mathcal{T}_A \mathcal{T}_B \ln (\mathcal{T}^2 \mathcal{T}_A \mathcal{T}_B).
\end{aligned}
\label{eq:complete_see_expression}
\end{equation}
The non-perturbative SEE expressions can be found from Eqs.~(S21) of SI Section~S1 and (S60) of SI Section~S2.

\textbf{\emph{Understanding the relation between EP and SEE.}}
To understand the EP-SEE relation Eq.~\eqref{eq:tilde_se},
we start with a single-particle situation where one particle from source $\mathcal{S}_A$ contributes to the entanglement
\begin{equation}
\begin{aligned}
    S_{\text{1e}} & = \mathcal{T}_A \mathcal{T} \ln (\mathcal{T}_A \mathcal{T}) + (1 - \mathcal{T}_A) \ln (1 - \mathcal{T}_A \mathcal{T}) \\
    & + \mathcal{T}_A (1 - \mathcal{T}) \ln (1 - \mathcal{T}_A \mathcal{T})
\end{aligned}
     \nonumber
\end{equation}
that
has three contributions.
The second one $(1 - \mathcal{T}_A) \ln (1 - \mathcal{T}_A \mathcal{T})$ comes from the case when the particle stays in the upper source $\mathcal{S}_A \to \tilde{\mathcal{D}}_A$.
This term has no contribution to SEE as the involved particle has no chance to join a two-particle scattering.
Of the remaining two, the first one $\mathcal{T}_A \mathcal{T} \ln (\mathcal{T}_A \mathcal{T})$ dominates in the strongly diluted regime.
It can be considered as the product of the conditioned EE $\ln (\mathcal{T}_A \mathcal{T})$ and its corresponding probability $\mathcal{T}_A \mathcal{T}$.

Now we move to two-particle scatterings to see the role of statistics.
Following the single-particle analysis, a two-particle scattering event produces the leading conditioned EE $[\ln (\mathcal{T} \mathcal{T}_A) + \ln (\mathcal{T} \mathcal{T}_B) ]$ when two particles enter the same drain (i.e., bunching) after the scattering.
Consequently, for a two-particle scattering event of free particles, the difference in the EE emerges between indistinguishable (fermions and bosons) and distinguishable cases, 
\begin{equation}
\begin{aligned}
\text{fermion:}\quad &(P_\text{fermion}^b - P_\text{dis}^b )  \left[\ln (\mathcal{T} \mathcal{T}_A) + \ln (\mathcal{T} \mathcal{T}_B) \right]/2 = \\
& \quad -\mathcal{T}(1 - \mathcal{T})\left[\ln (\mathcal{T} \mathcal{T}_A) + \ln (\mathcal{T} \mathcal{T}_B) \right],\\
\text{boson:}\quad &(P_\text{boson}^b - P_\text{dis}^b )  \left[\ln (\mathcal{T} \mathcal{T}_A) + \ln (\mathcal{T} \mathcal{T}_B) \right]/2 = \\
& \quad \mathcal{T}(1 - \mathcal{T})\left[\ln (\mathcal{T} \mathcal{T}_A) + \ln (\mathcal{T} \mathcal{T}_B) \right],
 \end{aligned}
\label{eq:reasoning}
\end{equation}
where $P^{b}_{\text{fermion}}$, $P^{b}_{\text{boson}}$ and $P^{b}_\text{dis}$ refer to the bunching probabilities of fermions, bosons and distinguishable particles, respectively.
In more realistic considerations, the two-particle scattering rate, i.e., $\mathcal{T}_A \mathcal{T}_B$ should be included as the prefactor of Eq.~\eqref{eq:reasoning}, leading to the statistics-induced entropy of $2e$-scattering processes $ -\mathcal{P}_\text{E} \left[\ln (\mathcal{T} \mathcal{T}_A) + \ln (\mathcal{T} \mathcal{T}_B) \right]/2$
for fermions and bosons alike. It equals the auxiliary function $\tilde{S}_\text{SEE}$ [Eq.~\eqref{eq:tilde_se}] after the integral over energy and time.

Notice that arguments above on the EP-SEE connection rely on the fact that both quantities can describe the tunneling between two subsystems. As a consequence, this EP-SEE connection remains valid if EP is instead defined after replacing $I_A$ and $I_B$ of Eq.~\eqref{eq:ep_definition} by the total current of two subsystems $A$ and $B$. This flexibility in the definition of EP enhances the potential range of applicability of our theory.

\textbf{\emph{Effect of interaction along the arms: charge fractionalization.}}
In the main text, we mention that interaction along the arms influences correlation functions, EP and SEE via the introduction of charge fractionalization~\cite{LevkiskyiSukhorukovPRB12,WahlMartinPRL14}.
To understand this phenomenon, we consider two chiral fermionic channels 1 and 2. Both channels (with corresponding fields $\phi_{1}$ and $\phi_2$) are described by free 1D Hamiltonians
\begin{equation}
H_1 = \frac{v_F}{4\pi\hbar} \int dx (\partial_x \phi_1)^2,\quad H_2 = \frac{v_F}{4\pi\hbar} \int dx (\partial_x \phi_2)^2.
\nonumber
\end{equation}
Fermions in channels also Coulomb-interact, leading to \begin{equation}
    H_{12} = \frac{v}{2\pi\hbar} \int dx (\partial_x \phi_1) (\partial_x \phi_2).
    \nonumber
\end{equation}

The total Hamiltonian can be diagonalized via the rotation $\phi_{\pm} \equiv (\phi_1 \pm \phi_2)/\sqrt{2}$, after which two modes $\phi_\pm$ travel at different velocities $v_F \pm v$.
Consequently, after entering one middle arm, an electron gradually splits into two (spatially-separated) wave packets.
With fractionalization taken into consideration, the cross current-current correlation becomes (see SI Section~S2)
\begin{equation}
\begin{aligned}
   & \int\! dt  \langle I_A (t) I_B (0) \rangle_\text{irr} \!=\! -\! \frac{e^3}{h}   \mathcal{T} (1 - \mathcal{T}) \left[ (\mathcal{T}_A - \mathcal{T}_B)^2 \right.\\
   &\quad \left. + \mathcal{T}_A P_A + \mathcal{T}_B P_B + \mathcal{T}_A \mathcal{T}_B ( P_\text{QPC} + P_\text{frac} ) \right]\! V_\text{bias} ,
\end{aligned}
\label{eq:iu_id_methods}
\end{equation}
where
\begin{equation}
\begin{aligned}
P_A & = -  \left[ (1 - \mathcal{T}_A) \left( 1 - \frac{1}{2 - 2 \mathcal{T}}\! \right) +  \frac{l^2}{\lambda^2} \right]  \frac{v^2}{v_F^2} , \\
P_B &  = -  \left[ (1 - \mathcal{T}_B) \left( 1 - \frac{1}{2 - 2 \mathcal{T}} \right) +  \frac{l^2}{\lambda^2} \right]  \frac{v^2}{v_F^2},
\end{aligned}
\label{eq:pu_pd_methods}
\end{equation}
refer to the modification of correlation function due to the particle fractionalization in each arm, and 
\begin{equation}
P_\text{frac} = 2 \frac{v^2}{v_F^2} \frac{l^2}{\lambda^2},
\label{eq:pfrac_methods}
\end{equation}
only contributes when both sources are on.
In these expressions, $l$ and $\lambda$ refer to the distance from the diluter to the central QPC, and the half-width of the diluted fermionic wave packet, respectively.
Following equations above, the extent of fractionalization in our system depends on the interaction amplitude $v$ and the distance $l$ from the diluter to the central QPC (around 2$\mu$m in our setup). Based on experimental data (cf. SI Section~S2), we expect the fractionalization to be minimal before the packets arrive at the central QPC.

\begin{acknowledgments} 
\textbf{{\em Acknowledgments.}} We acknowledge Dong E. Liu, Gabriele Campagnano, Christian Glattli, and Janine Splettstoesser for useful comments, and Yunchul Chung and Hyungkook Choi for the discussion of the experiments. IG and YG acknowledge the support from the DFG grant No. MI658/10-2 and German-Israeli Foundation (GIF) grant No. I-1505-303.10/2019. YG acknowledges support from the Helmholtz International Fellow Award, the DFG
Grant RO 2247/11-1, CRC 183 (project C01), the US-Israel Binational Science Foundation, and the Minerva Foundation.
MH acknowledges the continuous support of the Sub-Micron Center staff and the support of the European Research Council under the European Union’s Horizon 2020 research and innovation programme (grant agreement number 833078).

\textbf{{\em Author contribution.}} GZ, IG, and YG conducted all theoretical calculations and analysis. 
CH and TA fabricated the structures, performed measurements, and analyzed the data. VU designed and grew the heterostructures by molecular-beam epitaxy. MH supervised the experiments. All the authors participated in writing the manuscript.
\end{acknowledgments}

\textbf{\em{{Competing interests.}}}
The authors declare no competing interests.

\newpage

\widetext
\clearpage

\renewcommand{\bibnumfmt}[1]{[S#1]}
\renewcommand{\citenumfont}[1]{S#1}
\global\long\def\theequation{S\arabic{equation}}
\global\long\def\thefigure{S\arabic{figure}}
\setcounter{equation}{0}
\setcounter{figure}{0}

\begin{center}
\textbf{\large Supplementary Information for ``Measuring statistics-induced entanglement entropy with a Hong-Ou-Mandel interferometer''
	\vspace{5pt}}\\
\vspace{15pt}
Gu Zhang, Changki Hong, Tomer Alkalay, Vladimir Umansky, Moty Heiblum, Igor Gornyi, and Yuval Gefen\\
(Dated: December 25, 2023)
\end{center}
\vspace{10pt}

In this Supplementary Information, we provide details on: (i) Deriving entanglement entropy from full counting statistics, (ii) Influence of Coulomb interactions on statistics quantification functions, (iii) Statistics-induced entanglement in terms of Bell pairs, (iv) The dwell time and the measuring time, (v) Additional experimental data on $\nu = 3$ and $\nu = 1$ measurements, (vi) Possible SEE application: quantifying entanglement of a mixed state, and (vii) EE and SEE with a finite scattering area.

\section*{S1. Deriving entanglement entropy from full counting statistics}
\subsection*{S1A. Generating function of full counting statistics}
\label{sec:fcs_generating}

In this section, we express the FCS-generating function of a free-particle HOM interferometer in terms of the spectrum function $\mu(z)$ of the correlation matrix $M$ after scatterings [see its definition after \eqref{eq:chi_final}].
Before scatterings, the system state is described by the diagonal zero-temperature correlation matrix $n_0$ with elements $\langle \epsilon,\alpha | n_0 | \epsilon', \alpha'\rangle = \Theta(V_\alpha - \epsilon) \delta_{\alpha,\alpha'} \delta(\epsilon - \epsilon')$ labeled by the energy $\epsilon$ and arm $\alpha$ indices, with $\Theta(\epsilon)$ the step function. Here $V_\alpha = V_\text{bias}$ for two sources $\alpha = \mathcal{S}_A,\mathcal{S}_B$ and is zero for two grounded middle arms $\alpha = \mathcal{M}_A,\mathcal{M}_B$.

After scatterings, following Ref.\,\cite{SKlichLevitovPRL09}, we write down the general expression of the generating function
\begin{equation}
\chi (\lambda_\text{FCS})= \text{det} (1 - n + n U^{\dagger}_\mathcal{T} e^{i\lambda_\text{FCS} P_A} U_\mathcal{T} e^{-i\lambda_\text{FCS} P_A}),
\label{eq:chi_begin}
\end{equation}
where $U_\mathcal{T}$ refers to the scattering matrix of the central QPC, $\lambda_\text{FCS}$ is the measuring field in FCS, $P_A$ is the projection operator onto the subsystem $A$, and $n = U_0 n_0 U^{\dagger}_0$ is the original correlation matrix $n_0$, after the transformation
\begin{equation}
\begin{aligned}
U_0 = 
\begin{pmatrix}
r_{A} & t_{A} &0 & 0 \\
-t^*_A & r_A^* & 0 & 0  \\
0 & 0 & r_{B}& t_{B}\\
0 & 0 & -t_B^* & r_B^*
\end{pmatrix}
\end{aligned}
\label{eq:u0}
\end{equation}
that represents scatterings at two diluters.
It acts on a four-dimensional spinor with indexes respectively referring to the upper source $\mathcal{S}_A$, two middle arms $\mathcal{M}_A$, $\mathcal{M}_B$, and the lower source $\mathcal{S}_B$.
In Eq.~\eqref{eq:u0}, the matrix elements $|t_A|^2 = \mathcal{T}_A$ and $|t_B|^2 = \mathcal{T}_B$ for transmission and $|r_A|^2 = 1-\mathcal{T}_A$ and $|r_B|^2 = 1-\mathcal{T}_B$ for reflections.
To proceed, we rewrite Eq.~\eqref{eq:chi_begin} as
\begin{equation}
\begin{aligned}
\chi (\lambda_\text{FCS}) & = \text{det} \left(U_0U^{\dagger}_0 - U_0 n_0 U^{\dagger}_0 + U_0 n_0 U^{\dagger}_0 e^{-i\lambda_\text{FCS} P_A} U_0 U_0^{\dagger} U^{\dagger} e^{i\lambda_\text{FCS} P_A} U (U_0 U_0^{\dagger}) \right)\\
& = \text{det}\left[ 1 -  n_0  +  n_0 \left( U^{\dagger}_0 e^{-i\lambda_\text{FCS} P_A} U_0\right) \left( U_c^{\dagger} e^{i\lambda_\text{FCS} P_A} U_c\right) \right] \\
& = \text{det} \left[ 1 - n_0 + n_0 e^{-i\lambda_\text{FCS} P_A} \left( U_c^{\dagger} e^{i\lambda_\text{FCS} P_A} U_c\right) \right],
\end{aligned}
\label{eq:chi_sim1}
\end{equation}
where $U_c = U_\mathcal{T} U_0$.
In the first line of Eq.~\eqref{eq:chi_sim1}, we simply inserted identities $U_0  U^{\dagger}_0  =1$.
In the second line, we have used the fact that $\text{det} (m_1 m_2 m_3) = \text{det} (m_1) \text{det} (m_2) \text{det} (m_3)$ for square matrices, and that $\text{det}(U_0) = 1$.
Finally, in the last line, we have used the fact that $P_A$ commutes with $U_0$.
The derivation that follows is quite similar to that of Ref.\,\cite{SKlichLevitovPRL09}:
\begin{equation}
\begin{aligned}
\chi (\lambda_\text{FCS}) & = \text{det} \left\{e^{i\lambda_\text{FCS} P_A n_0}  \left[1 - n_0 + n_0 e^{-i\lambda_\text{FCS} P_A} \left( U_c^{\dagger} e^{i\lambda_\text{FCS} P_A} U_c\right) \right] e^{-i\lambda_\text{FCS} P_A n_0} \right\}\\
& =  \text{det} \left[ \left(1 - n_U + n_U e^{i\lambda_\text{FCS} P_A} \right) U_c e^{-i\lambda_\text{FCS} P_A n_0} U_c^{\dagger} \right]  \\
& =  \text{det} \left[ \left( 1 + P_A n_U (e^{i\lambda_\text{FCS}} - 1)  P_A \right) U_c e^{-i\lambda_\text{FCS} P_A n_0} U_c^{\dagger} \right]\\
& = \text{det} \left[ \left( 1 + M (e^{i\lambda_\text{FCS}} - 1)  \right) U_c e^{-i\lambda_\text{FCS} P_A n_0} U_c^{\dagger} \right],
\end{aligned}
\label{eq:chi_final}
\end{equation}
where $n_U = U_c n_0 U_c^{\dagger}$ and $M = P_A n_U P_A$. After the substitution $z\equiv (1 - e^{i\lambda_\text{FCS}})$, we arrive at the expression
\begin{equation}
\chi(z) = \text{det} \left[ (z - M) \frac{1}{z}  e^{-i\lambda_\text{FCS}(z) U_cP_A n_0 U_c^{\dagger}}  \right].
\label{eq:chi_final2}
\end{equation}
Notice that if we add a small imaginary part to $z$, the spectral function of $M$ can be found as
\begin{equation}
\mu(z) = \frac{1}{\pi} \text{Im} \partial_z \left[ \ln \chi(z-i\eta)\right] + \mathcal{C}_1 \delta(z) + \mathcal{C}_2 \delta(z - 1),
\label{seq:muz}
\end{equation}
where $\eta >0$ is an infinitesimal, and $\mathcal{C}_1$, $\mathcal{C}_2$ are two constant numbers.
The term $\mathcal{C}_1 \delta(z)$ originates from the $1/z$ term of Eq.~\eqref{eq:chi_final2}, and that of $\mathcal{C}_2$ comes from the exponential phase factor.

Equation~\eqref{seq:muz} almost coincides with the expression for the generating function in Refs.~\cite{SKlichLevitovPRL09,SKlichLevitovAIP09} obtained for a single-QPC model, except for the values of $\mathcal{C}_1$ and $\mathcal{C}_2$. However, these two constant numbers will not enter the integral over the spectral function, and are thus irrelevant to the EE. We also emphasize that Eq.~\eqref{seq:muz} is valid for any finite number of co-propagating channels bridged by finite numbers of independent scattering QPCs, given the measuring time is much longer than the traveling time between any two QPCs.

\subsection*{S1B. The EE evaluated from the spectral function of the correlation matrix}
\label{app:ee}

In S1A, we have shown the connection between the FCS generating function $\chi(z)$ and the spectrum function $\mu(z)$ of the correlation function matrix $M$.
In this section, we relate $\mu(z)$ to the EE, through which the EE-FCS connection can be established.

We consider a general case where a non-interacting subsystem is described by the reduced density matrix $\rho_\text{sub}$
\begin{equation}
\rho_\text{sub} = \frac{1}{Z} \exp \left( -\tilde{H}_{ij} a_i^{\dagger} a_j  \right),
\label{eq:rhos}
\end{equation}
with the matrix elements $\tilde{H}_{ij}$.
The operators $a_i^{\dagger}$ and $a_j^{\dagger}$ act on states that belong to the subsystem we consider.
In Eq.~\eqref{eq:rhos}, the factor $Z$
\begin{equation}
Z = \text{det} \left( 1 + e^{-\tilde{H}} \right)
\end{equation}
is $\tilde{H}$-dependent, and is introduced due to the normalization requirement $\text{Tr}(\rho_\text{sub}) = 1$.

Following Eq.~\eqref{eq:rhos}, we define an $m$ matrix whose element
\begin{equation}
m_{ij} = \text{Tr} (\rho_\text{sub} a_i^{\dagger} a_j) = \left[ (e^{\tilde{H}} + 1)^{-1} \right]_{ij}
\label{eq:mh_relation}
\end{equation}
is the correlation function of particles in states $i$ and $j$.
Now, the entanglement entropy equals
\begin{equation}
\begin{aligned}
S_\text{ent} = -\text{Tr} (\rho_\text{sub} \ln \rho_\text{sub}) & \!= \! - \text{Tr} \left[ \rho_\text{sub} \left( -\sum_{ij} \tilde{H}_{ij} a^{\dagger}_i a_j \right) \right] \! + \! \text{Tr} \left[ \rho_\text{sub} \ln Z \right]\\
& =  \sum_{ij} \tilde{H}_{ij} m_{ij } + \ln \left[ \text{det} \left( 1 + e^{-\tilde{H}} \right) \right] \\
& =  \text{Tr} ( \tilde{H} m ) +\text{Tr} \left[ \ln  \left( 1 + e^{-\tilde{H}} \right) \right].
\end{aligned}
\label{eq:ee_h_form}
\end{equation}
Finally, we rewrite Eq.~\eqref{eq:ee_h_form} in terms of $m$ defined in Eq.~\eqref{eq:mh_relation}
\begin{equation}
\tilde{H} = \ln (1 - m) - \ln (m), 
\end{equation}
to arrive at the expression
\begin{equation}
\begin{aligned}
S_\text{ent} & = -\text{Tr} (\rho_\text{sub} \ln \rho_\text{sub}) = \text{Tr} \left[ m \ln(1-m)  - m \ln(m)  \right] - \text{Tr} \left[ \ln (1-m) \right]\\
& = -\text{Tr} \left[ (1-m) \ln (1-m) + m\ln (m)  \right]
\label{seq:ee_general}
\end{aligned}
\end{equation}
that bridges the EE with the matrix $m$.
Notice that Eq.~\eqref{seq:ee_general} is the general expression for the EE in non-interacting subsystems with the density matrix Eq.~\eqref{eq:rhos} and the corresponding $m$ matrix defined in Eq.~\eqref{eq:mh_relation}.

In our case, after turning on three QPCs, the $m$ matrix becomes $M$ for the subsystem $A$ (discussed in S1A).
We then arrive at the following expression for the EE:
\begin{equation}
S_\text{ent}  = -\text{Tr} \left[ (1-M) \ln (1-M) + M\ln M  \right].
\label{eq:ee_our_case}
\end{equation}
Equations~\eqref{eq:chi_final2} and \eqref{eq:ee_our_case} build the connection between FCS and EE in an HOM interferometer.

\subsection*{S1C. Calculating SEE from the generating function of FCS}
\label{app:ee_and_fcs}

A combination of Eqs.~\eqref{eq:chi_final2} and \eqref{eq:ee_our_case} successfully bridges the system EE and the generating function of FCS. These two equations are extensions of that of Refs.\,\cite{SKlichLevitovPRL09,SKlichLevitovAIP09}: now the EE and the related FCS generating function are defined for subsystems $A$ and $B$, rather than for two single channels.
After scattering at two diluters, distribution functions become double-step-like, as shown in Fig.\,1C of the main text, where the heights of middle steps respectively equal the transmission probabilities $\mathcal{T}_A(\epsilon)$ and $\mathcal{T}_B(\epsilon)$ at two diluters. Here we assume that both sources $\mathcal{S}_A$ and $\mathcal{S}_B$ are biased at the same voltage $V_\text{bias}$.
Actually, when two biases are unequal, contributions from particles with energies between two biases will be removed following the definition of the entanglement pointer and the SEE.

Since Eqs.~\eqref{eq:chi_final2} and \eqref{eq:ee_our_case} share the form with that of Refs.\,\cite{SKlichLevitovPRL09,SKlichLevitovAIP09,SSongLehurPRB11}, the EE can be calculated via the integral form
\begin{equation}
S_\text{ent} = \frac{1}{\pi} \int_{-\infty}^{\infty} du \frac{u}{\cosh^2 u} \text{Im} \left[ \ln \chi(\pi - \eta - 2 i u) \right],
\label{eq:see_integral_form}
\end{equation}
where $\eta$ is a positive infinitesimal introduced to avoid possible divergence.
The generating function of FCS of our system can be calculated following Ref.\,\cite{SGogolinKomnikPRB06}
\begin{equation}
\begin{aligned}
\ln \chi(\lambda_\text{FCS}) &= \frac{\tau}{h} \int_0^{e V_\text{bias}} d\epsilon \ln \left\{ 1 + \mathcal{T} \left[ \mathcal{T}_A (1 - \mathcal{T}_B) \left( e^{i \lambda_\text{FCS}} - 1 \right)  + \mathcal{T}_B (1 - \mathcal{T}_A) \left( e^{-i \lambda_\text{FCS}} - 1 \right) \right] \right\},
\end{aligned}
\label{eq:sgenerating_function_expression}
\end{equation}
where $h$ is the Planck constant and $\tau$ refers to the measuring time. In Eq.~\eqref{eq:sgenerating_function_expression}, we have used the fact that the non-equilibrium distributions $\delta f_A (\epsilon) = \mathcal{T}_A$ and $\delta f_B (\epsilon)= \mathcal{T}_B$ for energies $0 < \epsilon < e V_\text{bias}$, and equals zero otherwise.

With Eqs.~\eqref{eq:see_integral_form} and \eqref{eq:sgenerating_function_expression}, the EE of our system becomes
\begin{equation}
\int_{u_0}^{\infty} du \frac{u}{\cosh^2u} = \ln\left[  2 \cosh (u_0) \right] - u_0 \tanh(u_0)
\end{equation}
where $u_0$ is the solution of 
\begin{equation}
\alpha_{12}(u_0) \equiv 1 - \mathcal{T} \mathcal{T}_A (1 - \mathcal{T}_B) (1 + e^{2u_0} e^{-i\delta}) - \mathcal{T} \mathcal{T}_B (1 - n_A) (1 + e^{-2u_0} e^{i\delta}) = 0.
\label{eq:uc}
\end{equation}
Equation above has two solutions
\begin{equation}
\begin{aligned}
& u_1 (\mathcal{T}_A, \mathcal{T}_B) = \frac{1}{2} \ln\left[ \frac{1-a_1-a_2 + \sqrt{(a_1+a_2 - 1)^2 - 4 a_1 a_2}}{2a_1} \right] \\
& u_2 (\mathcal{T}_A, \mathcal{T}_B) = \frac{1}{2} \ln\left[\frac{1-a_1-a_2 + \sqrt{(a_1+a_2 - 1)^2 - 4 a_1 a_2}}{2a_2} \right],
\end{aligned}
\end{equation}
where $a_1 = \mathcal{T} \mathcal{T}_A (1 - \mathcal{T}_B)$ and $a_2 = \mathcal{T} \mathcal{T}_B (1 - \mathcal{T}_A)$.
With this notation, the entire EE (no interaction) comprises statistical contributions and contributions from single-source beams. It is given by
\begin{equation}
    \begin{aligned}
    S_\text{ent} (\mathcal{T}_A,\mathcal{T}_B) = \frac{\tau}{h} \int d\epsilon \sum_{i = 1,2} \left\{ \ln\left[  2 \cosh u_i (\mathcal{T}_A,\mathcal{T}_B) \right] - u_i (\mathcal{T}_A,\mathcal{T}_B) \tanh[u_i (\mathcal{T}_A,\mathcal{T}_B) ] \right\}.
    \end{aligned}
    \label{eq:exact_solotion_two_source}
\end{equation}
Expanding Eq.~\eqref{eq:exact_solotion_two_source} to the leading order of the dilutions $\mathcal{T}_A$ and $\mathcal{T}_B$, the zero-temperature $S_\text{ent}$ of a non-interacting system approximately equals
\begin{equation}
\begin{aligned}
&S_\text{ent} (\mathcal{T}_A ,\mathcal{T}_B,\epsilon) h /(e \tau V_\text{bias})  \approx \mathcal{T}
 \left[ \mathcal{T}_A + \mathcal{T}_B - \mathcal{T}_A \ln (\mathcal{T}_A \mathcal{T}) - \mathcal{T}_B \ln (\mathcal{T}_B \mathcal{T}) \right] -\frac{1}{2} \mathcal{T}^2(\mathcal{T}_A^2 + \mathcal{T}_B^2) \\
 &  + \mathcal{T} (1 - \mathcal{T}) \mathcal{T}_A \mathcal{T}_B \ln (\mathcal{T}^2 \mathcal{T}_A \mathcal{T}_B) + O(\mathcal{T}_{A,B}^3).
\end{aligned}
\label{eq:sdiff_ee}
\end{equation}
Notice that the terms of the first line contain only single-particle contributions. These terms are removed once we calculate the statistics-induced EE.
Performing the integration over energy, we arrive at Eq.~(10) in Methods.

Inserting Eq.~\eqref{eq:exact_solotion_two_source} into Eq.~(4) of the main text, we obtain an expression for the SEE,
\begin{equation}
\begin{aligned}
   & S_\text{SEE}  = - \frac{\tau}{h} \int d\epsilon  \left\{ \ln [ 2 \cosh u_1 (\mathcal{T}_A ,\mathcal{T}_B ) ] - u_1 (\mathcal{T}_A ,\mathcal{T}_B ) \tanh u_1 (\mathcal{T}_A ,\mathcal{T}_B ) \right.\\
    & + \left.  \ln [ 2 \cosh u_2 (\mathcal{T}_A ,\mathcal{T}_B ) ] - u_2 (\mathcal{T}_A ,\mathcal{T}_B ) \tanh u_2 (\mathcal{T}_A ,\mathcal{T}_B )  \right.\\
    & + \left.  \mathcal{T}_A \mathcal{T} \ln (\mathcal{T}_A \mathcal{T}) + (1 - \mathcal{T}_A \mathcal{T}) \ln (1 - \mathcal{T}_A \mathcal{T}) + \mathcal{T}_B \mathcal{T} \ln (\mathcal{T}_B \mathcal{T}) + (1 - \mathcal{T}_B \mathcal{T}) \ln (1 - \mathcal{T}_B \mathcal{T}) \right\}.
\end{aligned}
\label{eq:see_expression}
\end{equation}

\begin{figure}[h!]
\vspace*{-0.25cm}
\includegraphics[width= 0.9\columnwidth]{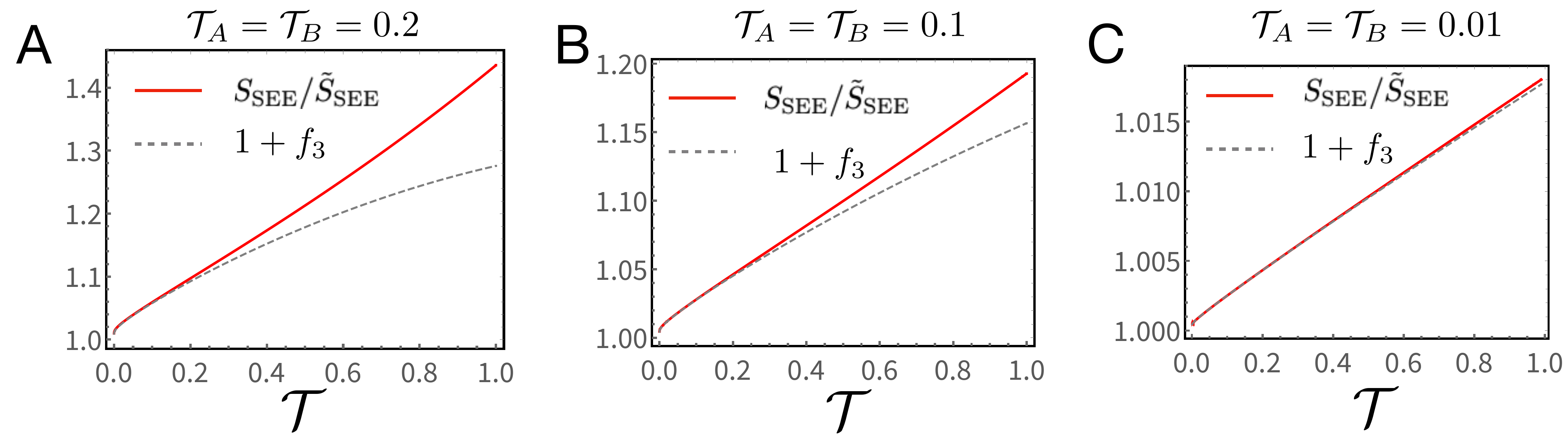}
\vspace*{-0.5cm}
\caption{The ratio between $\tilde{S}_\text{SEE}$ and $S_\text{SEE}$ for different dilutions: A. $\mathcal{T}_A = \mathcal{T}_B = 0.2$, B. $\mathcal{T}_A = \mathcal{T}_B = 0.1$ and C. $\mathcal{T}_A = \mathcal{T}_B = 0.01$. With a stronger dilution, the expansion to the third order (i.e., $1 + f_3$) agrees better with the exact result of $\tilde{S}_\text{SEE}/S_\text{SEE}$.}
\label{fig:3rd_order_see}
\end{figure}

We can further expand the SEE in powers of the diluters transmission coefficients, leading to
\begin{equation}
    S_\text{SEE} \approx - \frac{eV_\text{bias}}{h} \mathcal{T} (1 - \mathcal{T}) \mathcal{T}_A \mathcal{T}_B \ln (\mathcal{T}^2 \mathcal{T}_A \mathcal{T}_B) \left[ 1 + f_3 (\mathcal{T},\mathcal{T}_A,\mathcal{T}_B) \right],
    \label{eq:3rd_order}
\end{equation}
where the first part, $- eV_\text{bias} \mathcal{T} (1 - \mathcal{T}) \mathcal{T}_A \mathcal{T}_B \ln (\mathcal{T}^2 \mathcal{T}_A \mathcal{T}_B)/h$, is the leading-order SEE [i.e., Eq.~(11) of the main text] for strong dilution. This term is proportional to the product of the diluters transmission coefficients. 
The factor
\begin{equation}
    f_3 (\mathcal{T},\mathcal{T}_A,\mathcal{T}_B) \equiv \frac{\mathcal{T}_A + \mathcal{T}_B}{2} \frac{-1 + 3 \mathcal{T} + 2 \mathcal{T} \ln (\mathcal{T}^2 \mathcal{T}_A \mathcal{T}_B)}{\ln (\mathcal{T}^2 \mathcal{T}_A \mathcal{T}_B)}
    \label{eq:f3}
\end{equation}
refers to corrections from the higher-order terms in the diluters transmission coefficients.
With this correction included, we plot the ratio of $\tilde{S}_\text{SEE}$ and $S_\text{SEE}$ in Fig.~\ref{fig:3rd_order_see} (red solid line).
This ratio calculated for $\mathcal{T} = 0.53$ and $\mathcal{T}_A =\mathcal{T}_B = 0.21$ is the factor we use to rescale the SEE data in Fig.~4E of the main text.
For any given values of $\mathcal{T}$, $\mathcal{T}_A$, $\mathcal{T}_B$, this ratio $\tilde{S}_\text{SEE}/S_\text{SEE}$ is well determined theoretically, and thus can always be used for the rescaling of SEE data.
We also compare the exact ratio $\tilde{S}_\text{SEE}/S_\text{SEE}$ with the leading-order approximation $1 + f_3 (\mathcal{T},\mathcal{T}_A,\mathcal{T}_B)$ (the dashed line) defined in Eq.~\eqref{eq:f3}.
In agreement with our anticipation, the approximated value $1 + f_3 (\mathcal{T},\mathcal{T}_A,\mathcal{T}_B)$ agrees well with the ratio $\tilde{S}_\text{SEE}/S_\text{SEE}$, when both $\mathcal{T}_A, \mathcal{T}_B$ are not too large (the difference between the solid and dashed curves is about $5\%$ for the experimental values of the parameters); note that this approximation becomes even better with decreasing $\mathcal{T}$.

With interaction involved, the major structure of the result Eq.~\eqref{eq:see_expression} remains valid, after replacing $\mathcal{T}_A$ and $\mathcal{T}_B$ by the effective distribution functions $\tilde{f}_A$ and $\tilde{f}_B$ [see Eq.~\eqref{eq:effective_distributions} of Sec.~S2 for details].

\subsection*{S1D. Rescaling the experimental data of SEE}

In Fig.~5E of the main text, we compare the theoretical SEE with experimental data points.
Since transmission coefficients of the diluters $\mathcal{T}_A = \mathcal{T}_B = 0.2$ are not small enough, $\tilde{S}_\text{SEE}$ obtained from the experimental data on the EP cannot precisely represent the real SEE, $S_\text{SEE}$ (see Fig.~4B of the main text to see the difference between $\tilde{S}_\text{SEE}$ and $S_\text{SEE}$).
To compare the experimental SEE with the theoretical curves, we thus rescale $\tilde{S}_\text{SEE}$ of the data, following the $S_\text{SEE} - \tilde{S}_\text{SEE}$ difference shown in Fig.~4B.

More specifically, experimental data is obtained with transmission amplitudes $\mathcal{T}_A = \mathcal{T}_B = 0.2$ and $\mathcal{T} = 0.53$. Following Fig.~4B of the main text, and Fig.~\ref{fig:ratio_change_dilution}, with these transmission parameters, $\mathcal{S}_\text{SEE}/\tilde{\mathcal{S}}_\text{SEE} \approx 1.22$.
We thus rescale $\tilde{S}_\text{SEE}$ of the experimental data by the ratio above, to obtain the experimental values of $S_\text{SEE}$ in Fig.~5E of the main text.

\begin{figure}[h!]
\vspace*{-0.25cm}
\includegraphics[width= 0.4\columnwidth]{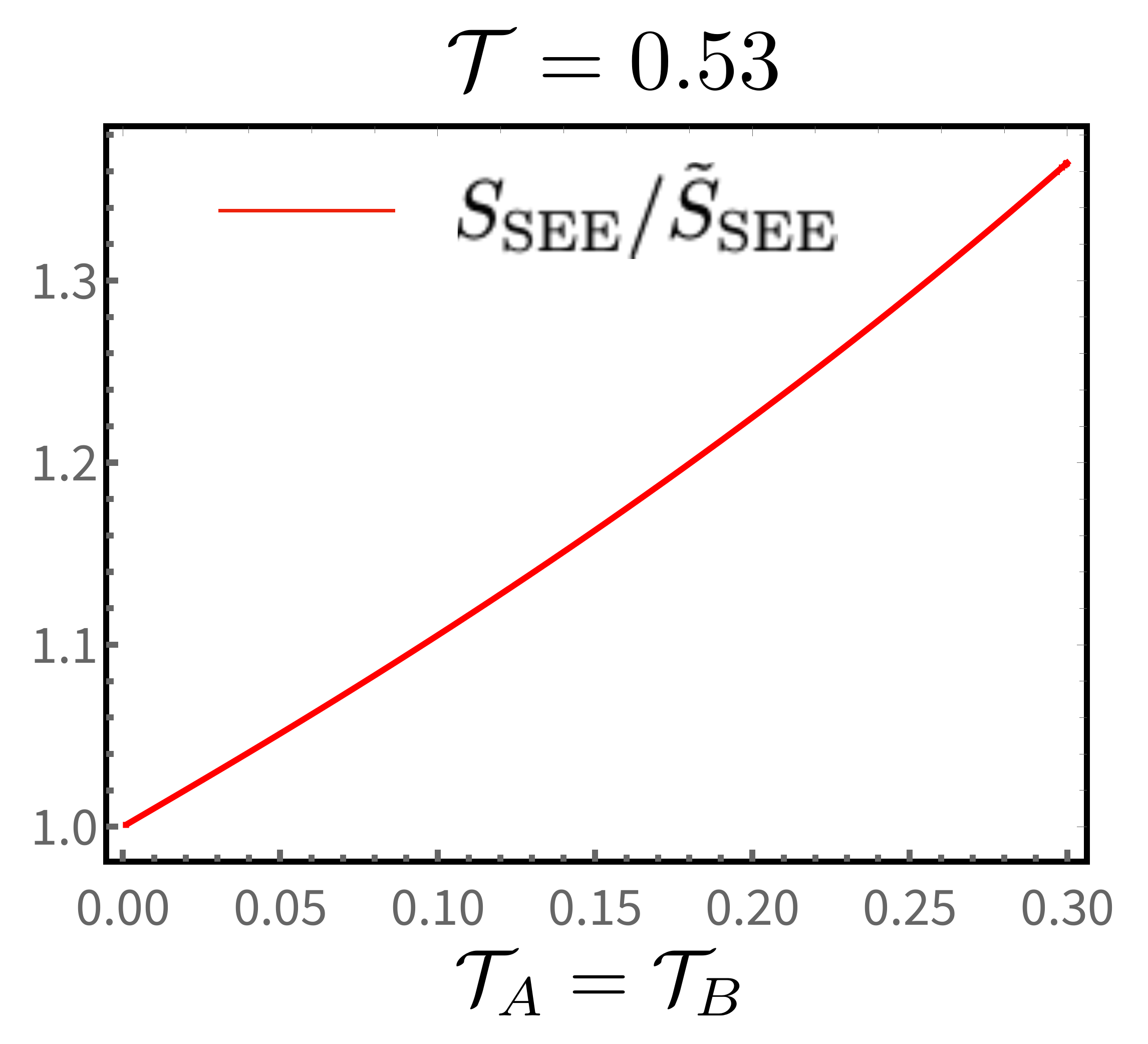}
\vspace*{-0.5cm}
\caption{The ratio of $S_\text{SEE}$ and $\tilde{S}_\text{SEE}$ as a function of $\mathcal{T}_A = \mathcal{T}_B$, when $\mathcal{T} = 0.53$.}
\label{fig:ratio_change_dilution}
\end{figure}

\section*{S2. Influence of Coulomb interactions}

\subsection*{S2A. Influence of Coulomb interaction along the transport on the entanglement pointer}
\label{sec:ep_interacting}

\begin{figure}[h!]
\includegraphics[width= 0.7  \columnwidth]{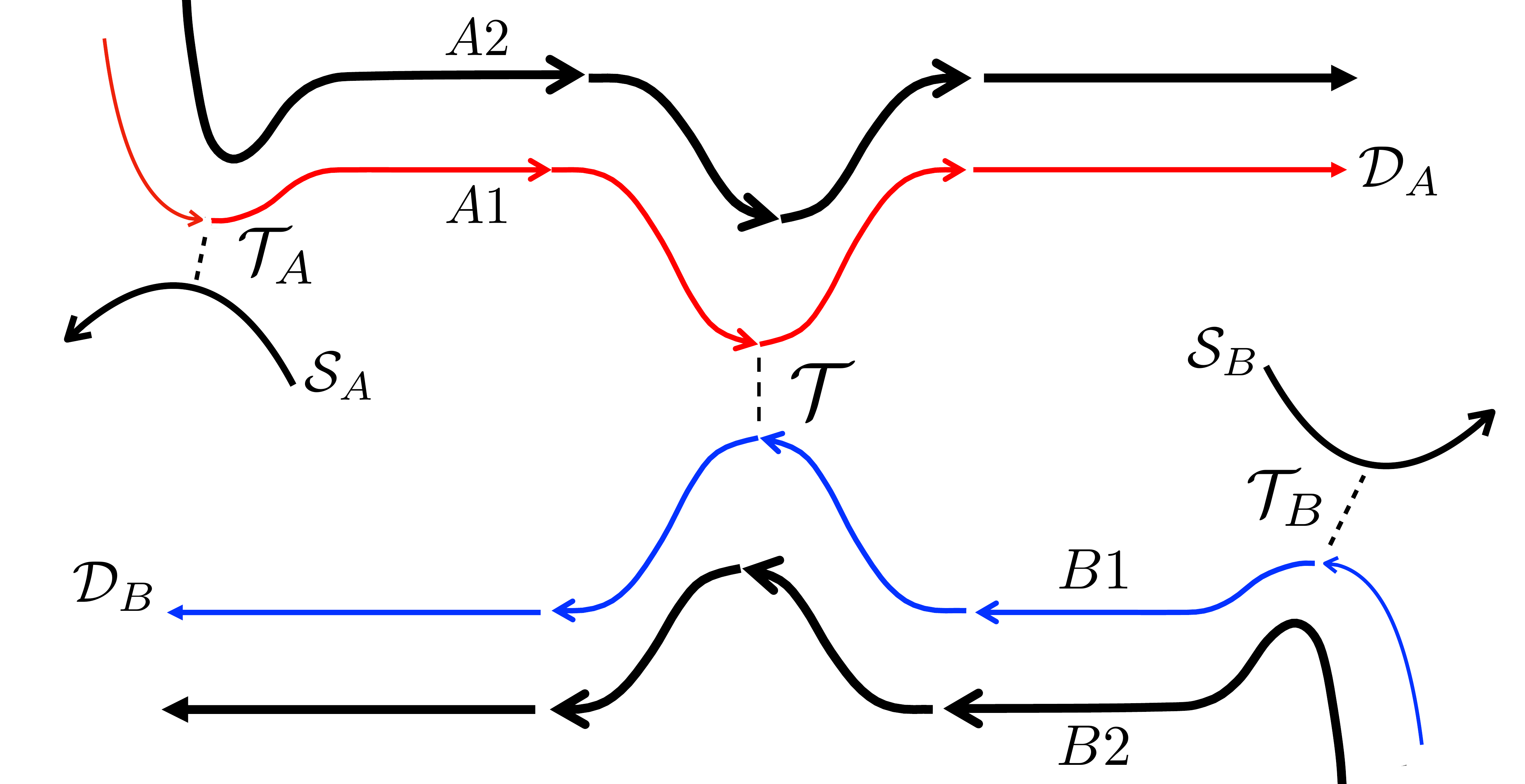}
\caption{The IQH sample that hosts two chiral channels at each edge. Currents in outer channels is diluted at QPCs with transmission probabilities $\mathcal{T}_A$ and $\mathcal{T}_B$. At the central QPC with the transmission $\mathcal{T}$, only the outer edges $A1$ and $B1$ communicate. The central QPC is located at the position $x = l$, with the distance $l$ to two diluters. Only current of two outer channels are measured at drains $\mathcal{D}_A$ and $\mathcal{D}_B$.}
\label{fig:experimental_structure}
\end{figure}

In real experiments, a sample with filling factor one might suffer from non-linearity that increases the complexity of measurement and theoretical analysis (cf. Sec.~S5B). The samples with higher filling factors are more suitable in this respect. However, in such samples the inter-channel interactions are present, which should be taken into consideration.
In this section, we consider a system with the filling factor 2, shown in Fig.\,\ref{fig:experimental_structure}, where each edge hosts two chiral channels.
Only the outer channels (i.e., channels $A1$ and $B1$) receive signals that are diluted at two diluters located at $x = 0$, with the corresponding transmission probabilities $\mathcal{T}_A$ and $\mathcal{T}_B$, respectively.
Only fermions in channels $A1$ and $B1$ tunnel through the central QPC (located at $x = l$) with the transmission probability $\mathcal{T}$, through which a non-vanishing entanglement between two subsystems (the upper subsystem $A1$, $A2$, and the lower subsystem $B1$, $B2$) is established.

Fermions in the inner two channels (i.e., $A2$ and $B2$), in contrast, pass through two diluters with full transmission, and cannot directly witness the presence of the central QPC. They influence our results via the Coulomb interaction with fermions in the two outer channels.
In this section, we consider only the interaction along the middle arms of the device. The influence of interaction in the central QPC will be included in Sec.~S2C.

The above assumptions lead to the following system Hamiltonian
\begin{equation}
\begin{aligned}
& H_0 = \sum_{\alpha = A,B,j =1,2 } \frac{v_F}{4\pi} \int dx \left( \partial_x \phi_{\alpha j} \right)^2 ,\quad H_\text{int} = \frac{v}{2\pi} \sum_{\alpha = A,B}\int dx \left( \partial_x \phi_{\alpha 1}\right) \left( \partial_x \phi_{\alpha 2}\right),\\
& H_{\text{T}} = t \left[ \psi^{\dagger}_{A1}(l) \psi_{B1}(l) + \psi^{\dagger}_{B1}(l) \psi_{A1}(l) \right], \\
& H_\text{source} = t_A \left[ \psi^{\dagger}_{A1}(0) \psi_{sA}(0) + \psi^{\dagger}_{sA}(0) \psi_{A1}(0) \right] + t_B \left[ \psi^{\dagger}_{B1}(0) \psi_{sB}(0) + \psi^{\dagger}_{sB}(0) \psi_{B1}(0) \right] ,
\label{eq:orignial_ham}
\end{aligned}
\end{equation}
where $v_F$ is the Fermi velocity of free fermions, $v$ is the interaction amplitude, and the phase $\phi_\alpha$ is introduced through bosonization
\begin{equation}
    \psi_\alpha = \frac{F_\alpha}{\sqrt{2\pi a}} e^{i\phi_\alpha}
\end{equation}
of the fermion in channel $\alpha$.
Here $F_\alpha$ is the Klein factor, and $a$ is the short-distance cutoff.
Operators $\psi_{sA}$ and $\psi_{sB}$ refer to fermions in sources $\mathcal{S}_A$ and $\mathcal{S}_B$, respectively.

Following Ref.\,\cite{SWahlMartinPRL14}, with inter-channel interactions present, $H_0 + H_\text{int}$ becomes diagonal
\begin{equation}
\begin{aligned}
H & = \frac{v_+}{4\pi} \int dx \left[ \left( \partial_x \phi_{\mathbf{A}+} \right)^2 +\left( \partial_x \phi_{\mathbf{S}+} \right)^2 \right] +  \frac{v_-}{4\pi} \int dx \left[ \left( \partial_x \phi_{\mathbf{A}-} \right)^2 +\left( \partial_x \phi_{\mathbf{S}-} \right)^2 \right],
\end{aligned}
\label{eq:rotation}
\end{equation}
in terms of fast $(+)$ and slow $(-)$ fields
\begin{equation}
\begin{aligned}
& \phi_{\mathbf{A} \pm} = \pm \frac{1}{2} \left[ \left( \phi_{A1} - \phi_{B1} \right) \pm  \left( \phi_{A2} - \phi_{B2} \right) \right], \\
& \phi_{\mathbf{S} \pm} = \pm \frac{1}{2} \left[ \left( \phi_{A1} + \phi_{B1} \right) \pm  \left( \phi_{A2} + \phi_{B2} \right) \right],
\end{aligned}
\label{eq:rotation_2}
\end{equation}
that travel with different velocities $v_\pm = v_F \pm v$.
As a consequence, when an electron enters e.g., channel $A1$, it fractionalizes into two modes: a charge mode that travels at $v_+$, and a neutral (or dipole) mode that travels at $v_-$.
By fractionalization, we refer to the spatial separation of these two modes after their propagation along the arms (see Methods and the Animation).
In the $S$ and $A$ modes, fields in the upper ($A1$ or $A2$) two arms have the same and different signs, respectively, as those of the corresponding lower ($B1$ or $B2$) ones.
With fields defined in Eq.~\eqref{eq:rotation_2}, the relation $\phi_{A1} - \phi_{B1} = \phi_{\mathbf{A}+} - \phi_{\mathbf{A}-}$ holds, and the tunneling operator transforms as
\begin{equation}
    \psi^\dagger_{B1} \psi_{A1} \to \psi^\dagger_{\mathbf{A}-} \psi_{\mathbf{A}+}.
    \label{eq:tunneling_new_fields}
\end{equation}
Here, 
$$\psi_{\mathbf{A}\pm}\sim \exp(i\phi_{\mathbf{A}\pm})/\sqrt{2\pi a}$$ refers to the operator after refermionization.
Importantly, as $\psi_{\mathbf{A}\pm}$ are effectively non-interacting, their scattering at the central QPC can be studied with the scattering matrix formalism, with the scattering matrix
\begin{equation}
\begin{pmatrix}
\psi_{\mathbf{A}+} (l^+) \\
\psi_{\mathbf{A}-} (l^+)
\end{pmatrix}=
\begin{pmatrix}
r & -i t\\
-i t & r
\end{pmatrix}
\begin{pmatrix}
\psi_{\mathbf{A}+} (l^-) \\
\psi_{\mathbf{A}-}  (l^-)
\end{pmatrix}
\label{eq:scattering_matrix}
\end{equation}
that relates the fields right before ($l^- \equiv l - \eta$) and after ($l^+ \equiv l + \eta$) the scattering at the central QPC located at $x = l$ ($\eta$ is again a positive infinitesimal). In Eq.~\eqref{eq:scattering_matrix}, we take $t = \sqrt{\mathcal{T}}$ and $r = \sqrt{1 - \mathcal{T}} = \sqrt{\mathcal{R}}$ as real numbers.
In this scattering matrix, $t$ refers to the tunneling amplitude between states labeled $\mathbf{A}_+$ and $\mathbf{A}_-$.
Following discussions in Ref.~\cite{SWahlMartinPRL14}, it also equals the tunneling amplitude between channels $A1$ and $B1$.

To calculate the current correlation function, we write down the current operators
\begin{equation}
\begin{aligned}
 I_{A1}(l^+) &= i [H, Q_{A1}(x>l^+)] =  \frac{e }{2\pi} \left[ v_F \partial_x\phi_{A1}(l^+) + v \partial_x\phi_{A2}(l^+) \right],\\
 I_{B1} (l^+) &= i [H, Q_{B1}(x>l^+)] = \frac{e }{2\pi} \left[ v_F \partial_x\phi_{B1}(l^+)+ v \partial_x\phi_{B2}(l^+)  \right],
\end{aligned}
\label{eq:original_current}
\end{equation}
after the central QPC, in terms of the fields before the rotation.
Here $Q_{A1} (l^+)$ and $Q_{B1} (l^+)$ are the operators for the particle numbers corresponding to charges after the central QPC, in channels $A1$ and $B1$, respectively.
Similar treatment of interaction and equation-of-motion method were used in Refs.~\cite{SChamonWenPRB94,SKaneFisher95,SSafi1999}.

To proceed, we write Eq.~\eqref{eq:original_current} in terms of the non-interacting fields~\eqref{eq:rotation_2}, through which one can easily relate the operators before and after the central QPC with the scattering matrix Eq.~\eqref{eq:scattering_matrix}.
After that, we transform back to the original basis, through which current operators after the central QPC become
\begin{equation}
\begin{aligned}
I_{A1}(l^+) & = \frac{e}{2\pi} \left\{ v_F \mathcal{R} \partial_x \phi_{A1} (l^-) \!+ v_F \mathcal{T}  \partial_x \phi_{B1} (l^-) + v \partial_x \phi_{A2} (l^-) \right. \\
& \left.  + i\frac{v_F}{a} \sqrt{\mathcal{T} \mathcal{R}}  \exp[-i \phi_{A1}(l^-) + i\phi_{B1} (l^-) ] - i\frac{v_F}{a} \sqrt{\mathcal{T}\mathcal{R}} \exp[i \phi_{A1}(l^-) - i\phi_{B1} (l^-) ]  \right\},\\
I_{B1}(l^+) & = \frac{e}{2\pi} \left\{ v_F \mathcal{T} \partial_x \phi_{A1} (l^-) \!+ v_F \mathcal{R}  \partial_x \phi_{B1} (l^-) + v \partial_x \phi_{B2} (l^-) \right. \\
& \left.  - i\frac{v_F}{a} \sqrt{\mathcal{T} \mathcal{R}}  \exp[-i \phi_{A1}(l^-) + i\phi_{B1} (l^-) ] + i\frac{v_F}{a} \sqrt{\mathcal{T}\mathcal{R}} \exp[i \phi_{A1}(l^-) - i\phi_{B1} (l^-) ]  \right\},
\end{aligned}
\label{eq:current_op_original}
\end{equation}
in terms of bosonic operators before the central QPC.
The first lines of both $I_{A1}$ and $I_{B1}$ consist of density operators that can be expressed in terms of current operators in front of the central QPC.
The second lines in expressions (\ref{eq:current_op_original}), on the other hand, display the quantum feature of transport at the central QPC: these operators produce finite contribution to the noise but zero contribution to the current averages.
Using Eqs.~(\ref{eq:current_op_original}), we arrive at the time-dependent correlation function in the presence of interaction and obtain the difference of correlation functions entering the EP:
\begin{equation}
\begin{aligned}
\langle I_{A1}(l^+,t) I_{B1} (l^+,0) \rangle\Big|_{\mathcal{T}_A,\mathcal{T}_B} 
&- \langle I_{A1}(l^+,t) I_{B1} (l^+,0) \rangle\Big|_{\mathcal{T}_A,0} - \langle I_{A1}(l^+,t) I_{B1} (l^+,0) \rangle\Big|_{0,\mathcal{T}_B} \\
& = \frac{e^2}{4\pi^2} \Bigg\{  \textcolor{red}{ v_F^2 \mathcal{T} \mathcal{R} \langle  \partial_x \phi_{A1} (l^-, t) \partial_x \phi_{A1} (l^-, 0) \rangle } + \textcolor{blue}{v_F^2 \mathcal{T} \mathcal{R} \langle \partial_x \phi_{B1} (l^-, t) \partial_x \phi_{B1} (l^-, 0) \rangle}  \\
& +  \textcolor{red}{vv_F \mathcal{T} \langle \partial_x \phi_{A2} (l^-,t) \partial_x \phi_{A1} (l^-,0) \rangle} + \textcolor{blue}{vv_F \mathcal{T}\langle \partial_x \phi_{B1} (l^-,t) \partial_x \phi_{B2} (l^-,0)\rangle} \\
& \left.  -   \frac{v_F^2}{a^2} \mathcal{T}\mathcal{R} \left[ \langle e^{-i\phi_{A1} (l^-, t) }  e^{i\phi_{A1} (l^-, 0) } \rangle\Big|_{\mathcal{T}_A,\mathcal{T}_B} \langle e^{i\phi_{B1} (l^-, t) } e^{-i\phi_{B1} (l^-, 0) } \rangle\Big|_{\mathcal{T}_A,\mathcal{T}_B} \right. \right. \\
& \left. + \langle e^{i\phi_{A1} (l^-, t) }  e^{-i\phi_{A1} (l^-, 0) } \rangle\Big|_{\mathcal{T}_A,\mathcal{T}_B} \langle e^{-i\phi_{B1} (l^-, t) } e^{i\phi_{B1} (l^-, 0) } \rangle\Big|_{\mathcal{T}_A,\mathcal{T}_B} \right] \Bigg\}\\
&-\frac{e^2}{4\pi^2} \Bigg\{  \textcolor{red}{ v_F^2 \mathcal{T} \mathcal{R} \langle \partial_x \phi_{A1} (l^-, t) \partial_x \phi_{A1} (l^-, 0) \rangle + vv_F\mathcal{T} \langle \partial_x \phi_{A2} (l^-,t) \partial_x \phi_{A1} (l^-,0) \rangle}  \\
& \left.  -   \frac{v_F^2}{a^2} \mathcal{T}\mathcal{R} \left[ \langle e^{-i\phi_{A1} (l^-, t) }  e^{i\phi_{A1} (l^-, 0) } \rangle\Big|_{\mathcal{T}_A,0} \langle e^{i\phi_{B1} (l^-, t) } e^{-i\phi_{B1} (l^-, 0) } \rangle\Big|_{\mathcal{T}_A,0} \right. \right. \\
&  \left. + \langle e^{i\phi_{A1} (l^-, t) }  e^{-i\phi_{A1} (l^-, 0) } \rangle\Big|_{\mathcal{T}_A,0} \langle e^{-i\phi_{B1} (l^-, t) } e^{i\phi_{B1} (l^-, 0) } \rangle\Big|_{\mathcal{T}_A,0} \right] \Bigg\}\\
&-\frac{e^2}{4\pi^2} \Bigg\{ \textcolor{blue}{ v_F^2 \mathcal{T} \mathcal{R} \langle \partial_x \phi_{B1} (l^-, t) \partial_x \phi_{B1} (l^-, 0) \rangle + vv_F\mathcal{T} \langle \partial_x \phi_{B2} (l^-,t) \partial_x \phi_{B1} (l^-,0) \rangle}  
\\
& \left.  -   \frac{v_F^2}{a^2} \mathcal{T}\mathcal{R} \left[ \langle e^{-i\phi_{A1} (l^-, t) }  e^{i\phi_{A1} (l^-, 0) } \rangle\Big|_{0,\mathcal{T}_B} \langle e^{i\phi_{B1} (l^-, t) } e^{-i\phi_{B1} (l^-, 0) } \rangle\Big|_{0,\mathcal{T}_B} \right. \right. \\
&  \left. + \langle e^{i\phi_{A1} (l^-, t) }  e^{-i\phi_{A1} (l^-, 0) } \rangle\Big|_{0,\mathcal{T}_B} \langle e^{-i\phi_{B1} (l^-, t) } e^{i\phi_{B1} (l^-, 0) } \rangle\Big|_{0,\mathcal{T}_B} \right] \Bigg\}.\\.
\end{aligned}
\label{eq:current_correlation_original}
\end{equation}
As a reminder, Eq.~\eqref{eq:current_correlation_original} leads to our defined EP after the integral over time.
The terms in red and blue of Eq.~\eqref{eq:current_correlation_original} are correlations between operators from the same source: from $S_A$ and $S_B$, respectively.
They cancel out following the definition of the EP [Eq.~(1) of the main text]. The other terms become
\begin{equation}
\begin{aligned}
 - \frac{e^2}{h} \mathcal{T} (1 - \mathcal{T}) \int & d\epsilon\  \left\{ f_{A1} (l^-,\mathcal{T}_A,\epsilon)  \left[ 1 - f_{B1} (l^-,\mathcal{T}_B,\epsilon)\right]  + f_{B1} (l^-,\mathcal{T}_B,\epsilon) \left[ 1 - f_{A1} (l^-,\mathcal{T}_A,\epsilon) \right]\right. \\
&\left. - f_{A1} (l^-,\mathcal{T}_A,\epsilon)  \left[ 1 - f_{B1} (l^-,0,\epsilon) \right] - \left[ 1 - f_{B1} (l^-,0,\epsilon) f_{A1} (l^-,\mathcal{T}_A,\epsilon) \right] \right. \\
&\left. - f_{A1} (l^-,0,\epsilon)  \left[ 1- f_{B1} (l^-,\mathcal{T}_B,\epsilon) \right] - f_{B1} (l^-,\mathcal{T}_B,\epsilon) \left[ 1- f_{A1} (l^-,0,\epsilon)\right] \right\}
\end{aligned}
\label{eq:vertex_contribution}
\end{equation}
after the integration over time,
where
\begin{equation}
f_{\alpha} (l^-,\mathcal{T}_\alpha,\epsilon) = v_F\int dt e^{-i\epsilon t} \langle \psi^{\dagger}_{\alpha} (l^-,t) \psi_{\alpha} (l^-,0) \rangle\Big|_{\mathcal{T}_\alpha}
\label{eq:interacting_distribution}
\end{equation}
refers to the distribution evaluated in channel $\alpha$, with the corresponding transmission probability $\mathcal{T}_\alpha$.

A reorganization of \eqref{eq:vertex_contribution} gives us the EP of the interacting situation
\begin{equation}
    \mathcal{P}_\text{E} = 2 \frac{e^2}{h} \mathcal{T}(1 - \mathcal{T}) \int d\epsilon\ \delta f_{A1} (l^-,\mathcal{T}_A,\epsilon) \delta f_{B1} (l^-,\mathcal{T}_B,\epsilon),
    \label{eq:sep_interacting}
\end{equation}
which depends on the non-equilibrium parts of particle distributions, 
$$\delta f_{A1} (l^-,\mathcal{T}_A,\epsilon)\!=\!f_{A1} (l^-,\mathcal{T}_A,\epsilon) - f_{A1} (l^-,0,\epsilon), \qquad
\delta f_{B1} (l^-,\mathcal{T}_B,\epsilon)\!=\!f_{B1} (l^-,\mathcal{T}_B,\epsilon) - f_{B1} (l^-,0,\epsilon)$$
in front of the central QPC.
As has been stated in the main text, Eq.~\eqref{eq:sep_interacting} shares the basic structure of the version without interaction along the arms, i.e., Eq.~(3) of the main text, after replacing the double-step distribution functions with $\delta f_\alpha$.

Following this fact, we see two important pieces of information on the effect (on the EP) of interaction along the arms. As the first message, the influence of interaction on the EP becomes negligible once the distance between the fast and slow wave packets from the same fermion is much smaller than the distance between the diluters and the central QPC: in this situation, the central QPC cannot feel the fractionalization of fermionic packets.
Luckily, this is indeed the situation of our setup, with which we measure EP and the SEE [see Supplementary sections~S2B and S2C, e.g., Fig.~\ref{fig:wave_packs} and discussions around Eq.~\eqref{eq:iu_id_with_interactions} for details].
More importantly, Eq.~\eqref{eq:sep_interacting} tells us that even for the interacting situation, EP is a quantity that is determined by the two-particle scattering rate [proportional to $ \int d\epsilon \delta f_{A1}(\epsilon) \delta f_{B1}(\epsilon)$], through which the statistical information is manifested.

\subsection*{S2B. Correlation functions with interaction present}
\label{sec:correlation_interaction}

In this section, we investigate the influence of interaction on correlation functions.
We focus on the weak-interaction limit $|v| \ll v_F$.
Following Eq.~\eqref{eq:current_op_original}, when expressing current operators after the central QPC in terms of that before the QPC, the operators consist of two categories of contributions, i.e., the current operators [e.g., $v_F \partial_x \phi_{A1} (l^-)$], and the vertex operators [e.g., $\psi^{\dagger}_{A1} (l^-) \psi_{B1} (l^-) $].
For later convenience, we define current operators of these two categories, following
\begin{equation}
\begin{aligned}
I_{A1}^\text{class}(l^+) & = \frac{e}{2\pi} \left[ v_F \mathcal{R} \partial_x \phi_{A1} (l^-) \!+ v_F \mathcal{T}  \partial_x \phi_{B1} (l^-) + v \partial_x \phi_{A2} (l^-) \right], \\
I_{A1}^\text{quant} (l^+) & =   i\frac{e v_F}{2\pi a} \sqrt{\mathcal{T}\mathcal{R}} \left\{   \exp[-i \phi_{A1}(l^-) + i\phi_{B1} (l^-) ] -  \exp[i \phi_{A1}(l^-) - i\phi_{B1} (l^-) ]  \right\},\\
I_{B1}^\text{class} (l^+) & = \frac{e}{2\pi} \left[ v_F \mathcal{T} \partial_x \phi_{A1} (l^-) \!+ v_F \mathcal{R}  \partial_x \phi_{B1} (l^-) + v \partial_x \phi_{B2} (l^-) \right], \\
I_{B1}^\text{quant} (l^+) & =   -i\frac{e v_F}{2\pi a} \sqrt{\mathcal{T}\mathcal{R}} \left\{   \exp[-i \phi_{A1}(l^-) + i\phi_{B1} (l^-) ] -  \exp[i \phi_{A1}(l^-) - i\phi_{B1} (l^-) ]  \right\},
\end{aligned}
\label{eq:classical_quantum_current}
\end{equation}
through which the previous current operators can be written as $I_{A1} = I_{A1}^\text{class} + I_{A1}^\text{quant}$ and $I_{B1} = I_{B1}^\text{class} + I_{B1}^\text{quant}$.
In Eq.~\eqref{eq:classical_quantum_current}, we call current operators defined with density operators as ``classical'' and those with vertex operators as ``quantum'', for two reasons.
Firstly, the ``quantum'' terms do not contribute to the current averages after the QPC: indeed, they influence only the current correlation.
More importantly, following Ref.~\cite{SIdrisovLevkivskyiX22}, the ``classical'' part reflects the noise produced at two diluters before the central QPC. This part of noise, after being produced at two diluters, simply propagates along the arms, before the current has arrived at the central QPC.
The ``quantum'' part instead encodes the noise generated by the scattering at the central QPC.
As another advantage of Eq.~\eqref{eq:classical_quantum_current}, one ``quantum'' and one ``classical'' operators have a vanishing correlation. One can thus calculate correlation functions, e.g., $\langle I_{A1} I_{B1}\rangle$, by addressing the classical and quantum contributions individually.

To calculate the correlation of classical current operators, we further express the density operators in terms of the current operators before the QPC, leading to
\begin{equation}
\begin{aligned}
    \partial_x \phi_{A1} (l^-) & = \frac{2\pi}{e} \frac{v_F I_{A1} (l^-) - v I_{A2} (l^-)}{v_F^2 - v^2}, \quad \partial_x \phi_{A2} (l^-) = -\frac{2\pi}{e} \frac{ v I_{A1} (l^-) - v_F I_{A2} (l^-)}{v_F^2 - v^2},\\
    \partial_x \phi_{B1} (l^-) & = \frac{2\pi}{e} \frac{v_F I_{B1} (l^-) - v I_{B2} (l^-)}{v_F^2 - v^2}, \quad \partial_x \phi_{B2} (l^-) = -\frac{2\pi}{e} \frac{ v I_{B1} (l^-) - v_F I_{B2} (l^-)}{v_F^2 - v^2}.
\end{aligned}
\label{eq:density_current_pre-qpc}
\end{equation}
With expressions in Eq.~\eqref{eq:density_current_pre-qpc}, the classical part of the current becomes
\begin{equation}
\begin{aligned}
I_{B1}^\text{class} (l^+) \equiv & \frac{e}{2\pi} \left\{ v_F \mathcal{R} \partial_x \phi_{A1} (l^-) \!+ v_F \mathcal{T}  \partial_x \phi_{B1} (l^-) + v \partial_x \phi_{A2} (l^-)  \right\}\\
= & \frac{1}{v_+ v_-} \left[ (v_F^2 \mathcal{R} - v^2) I_{A1} (l^-) + v_F v \mathcal{T} I_{A2} (l^-) + v_F^2 \mathcal{T} I_{B1} (l^-) - v_F v \mathcal{T} I_{B2} (l^-) \right] ,\\
I_{B1}^\text{class} (l^+) \equiv & \frac{e}{2\pi} \left\{ v_F \mathcal{T} \partial_x \phi_{A1} (l^-) \!+ v_F \mathcal{R}  \partial_x \phi_{B1} (l^-) + v \partial_x \phi_{B2} (l^-) \right\}\\
= & \frac{1}{v_+ v_-} \left[ (v_F^2 \mathcal{R} - v^2) I_{B1} (l^-) + v_F v \mathcal{T} I_{B2} (l^-) + v_F^2 \mathcal{T} I_{A1} (l^-) - v_F v \mathcal{T} I_{A2} (l^-) \right].
\end{aligned}
\label{eq:currents_density}
\end{equation}
As a consequence of charge fractionalization, post-QPC current operators in $A1$ and $B1$ become dependent on current operators in all four arms, although direct tunneling from $A2$ and $B2$ to $A1$ and $B1$ are not allowed.
However, in our system, currents from the sources only enter the middle two arms $A1$ and $B1$, leading to $\langle I_{A2} \rangle = \langle I_{B2} \rangle = 0$. As a result, most contributions of $\langle I_{A1}^\text{class}(l^+, t) I_{B1}^\text{class} (l^+, 0) \rangle_\text{irr} $ vanish, except for that from auto-correlations $\langle I_{A1}(l^-,t) I_{A1}(l^-,0) \rangle_\text{irr}$ and $\langle I_{B1}(l^-,t) I_{B1}(l^-,0) \rangle_\text{irr}$. Indeed, now the correlation function between the classical operators becomes
\begin{equation}
\begin{aligned}
   & \langle I_{A1}^\text{class} (l^+,t) I_{B1}^\text{class} (l^+,t) \rangle_\text{irr} \! = \! \frac{( v_F^2 \mathcal{R}\! -\! v^2 ) v_F^2 \mathcal{T} }{(v_F^2 - v^2)^2} [\langle I_{A1} (l^-,t) I_{A1} (l^-,0) \rangle_\text{irr} \! + \! \langle I_{B1} (l^-,t) I_{B1} (l^-,0) \rangle_\text{irr}]\\
    &\approx \mathcal{T} \mathcal{R} \left[ 1 + \frac{v^2}{v_F^2} \left( 2 - \frac{1}{\mathcal{R}} \right) \right] [\langle I_{A1} (l^-,t) I_{A1} (l^-,0) \rangle_\text{irr} + \langle I_{B1} (l^-,t) I_{B1} (l^-,0) \rangle_\text{irr}],
\end{aligned}
\label{eq:density_current_correlations}
\end{equation}
where in the second line we have expanded the correlation to the leading order of interaction $v^2/v_F^2$.
After the integration over time, the zero-frequency noise becomes
\begin{equation}
\begin{aligned}
    &\int dt \langle I_{A1} (l^-,t) I_{A1} (l^-,0) \rangle_\text{irr} = \frac{e^3 V_\text{bias}}{h} \mathcal{T}_A (1 - \mathcal{T}_A),\\
    &\int dt \langle I_{B1} (l^-,t) I_{B1} (l^-,0) \rangle_\text{irr} = \frac{e^3 V_\text{bias}}{h} \mathcal{T}_B (1 - \mathcal{T}_B).
\end{aligned}
\end{equation}
which does not depend on the amplitude of interaction $v$.
With this knowledge, Eq.~\eqref{eq:density_current_correlations} indicates that intra-edge interaction between neighboring channels generates a leading correction
\begin{equation}
   \frac{e^3 V_\text{bias}}{h} \mathcal{T}( 1- \mathcal{T} ) \frac{v^2}{v_F^2} \left( 2 - \frac{1}{1-\mathcal{T}} \right) \left[ \mathcal{T}_A (1 - \mathcal{T}_A) + \mathcal{T}_B (1 - \mathcal{T}_B) \right] ,
   \label{eq:extra_correlation_classic}
\end{equation}
to the current correlation $\langle I_{A1}^\text{class}(l^+, t) I_{B1}^\text{class} (l^+, 0) \rangle_\text{irr} $.
The correction given by Eq.~\eqref{eq:extra_correlation_classic} is positive if the transmission at the central QPC $\mathcal{T} < 1/2$, and becomes negative if $\mathcal{T} > 1/2$.
Notice that this interaction-induced correction to the cross-correlation function is linear in the dilution $\mathcal{T}_A$ or $\mathcal{T}_B$, and remains finite even with only one source on.

To understand how charge fractionalization introduces this extra correlation function, we remind ourselves that, following the effective free Hamiltonian \eqref{eq:rotation}, two ``plus'' modes $\mathbf{A}_+$ and $\mathbf{S}_+$ are traveling with the speed $v_+$, which is larger than that ($v_-$) of two ``minus'' modes $\mathbf{A}_-$ and $\mathbf{S}_-$.
In this basis, the currents carried by symmetric modes $\mathbf{S}_+$ and $\mathbf{S}_-$ are not affected by the central QPC [see Eq.~\eqref{eq:tunneling_new_fields}].
By contrast, the wave packets of the other two modes $\mathbf{A}_+$ and $\mathbf{A}_-$ tunnel to each other at the central QPC.
Most interestingly, when an $\mathbf{A}_\pm$ wave packet ``transmits'' into an $\mathbf{A}_\mp$ packet after scattering at the central QPC, the speed of the wave packet suddenly changes from $v_\pm$ to $v_\mp$.
After including contributions of both asymmetric modes, the current transmitted (between $A1$ and $B1$) at the central QPC becomes larger than that in the non-interacting case.
Indeed, with only the upper source on ($\mathcal{T}_B = 0$), classical current operators become
\begin{equation}
\begin{aligned}
    I_{A1}^\text{class} (l^+) \sim (1 - \mathcal{T}_A - \mathcal{T}_v ) I_{A1} (l^-), \\
    I_{B1}^\text{class} (l^+) \sim (\mathcal{T}_A + \mathcal{T}_v ) I_{A1} (l^-),
\end{aligned}
\label{eq:explain_classic}
\end{equation}
where we have neglected $I_{A2}$ that does not contribute to the correlation function.
In Eq.~\eqref{eq:explain_classic}, the ``dressed transmission'' 
$$\mathcal{T}_v \equiv v^2 \mathcal{T}/(v_F^2 - v^2) > 0$$ 
is positive, indicating an extra amount of transmitted current.
With these operators, the correlation becomes
\begin{equation}
    \langle I_{A1}^\text{class} (l^+) I_{B1}^\text{class} (l^+) \rangle = \left[ \mathcal{T R} - (\mathcal{T} - \mathcal{R}) \mathcal{T}_v - \mathcal{T}_v^2 \right] \langle I_{A1} (l^-) I_{B1} (l^-) \rangle.
\end{equation}
The sign of the leading order correction ($\propto \mathcal{T}_v$) 
depends on the transmission amplitude $\mathcal{T}$, in agreement with Eq.~\eqref{eq:extra_correlation_classic}.

Although an extra amount of transmitted noise induces an extra ``classical'' noise, one cannot incorporate the interaction effect by a simple modification of the transmission $\mathcal{T}$.
Indeed, as will be shown shortly, modifications on the correlation of the quantum part cannot be described by a modified $\mathcal{T}$.

Now we evaluate the correlation of the quantum part, i.e., vertex operators, by inspecting the correlation function $\langle \psi^\dagger_{A1} (l,t) \psi_{A1}(l,0)\rangle$ to the leading order of interaction $v$. Following Ref.~\cite{SLevkiskyiSukhorukovPRB12}, one can calculate the correlation function by writing the fields
\begin{equation}
    \begin{aligned}
    & \phi_{A1}(l, t) = -\frac{\pi}{e} [ Q_{A1} (t - l/v_+) + Q_{A2} (t - l/v_+) + Q_{A1} (t - l/v_-) - Q_{A2} (t - l/v_-) ],\\
    & \phi_{A2}(l, t) = -\frac{\pi}{e} [ Q_{A1} (t - l/v_+) + Q_{A2} (t - l/v_+) - Q_{A1} (t - l/v_-) + Q_{A2} (t - l/v_-) ],
    \end{aligned}
\end{equation}
in terms of the operators $Q_{A1} (t)$ and $Q_{A2} (t)$,
\begin{equation}
    Q_{\alpha}(t) = \int_{-\infty}^t dt' j_{\alpha} (t'),
\end{equation}
which refer to the number of charges that have entered the interfering arms at time $t$. Here, $j_{\alpha}(t)$ is the current operator at time $t$ right after the diluter in arm $\alpha$.
In terms of the charge operators, the correlation of bosonized operators becomes
\begin{equation}
\begin{aligned}
    \langle e^{-i\phi_{A1} (t) } e^{i\phi_{A1} (0)} \rangle = & \Big\langle e^{i\frac{\pi}{e} \left[ Q_{A1} (t - l/v_+) + Q_{A1} (t - l/v_-) \right]} e^{-i\frac{\pi}{e} \left[ Q_{A1} ( - l/v_+) + Q_{A1} ( - l/v_-) \right]} \Big\rangle \\
    \approx & \langle e^{2i\frac{\pi}{e} Q_{A1} (t - l/v_+)} e^{-2i\frac{\pi}{e} Q_{A1} (- l/v_+)} \rangle \left[ 1 + \frac{4\pi^2l^2 v^2}{e^2 v_F^4} \langle j_{A1}(t) j_{A1}(0) \rangle_\text{irr} \right]\\
    \approx & \langle e^{2i\frac{\pi}{e} Q_{A1} (t - l/v_+)} e^{-2i\frac{\pi}{e} Q_{A1} (- l/v_+)} \rangle \left( 1 - \frac{l^2 v^2}{v_F^4 t^2}  \right).
\end{aligned}
\end{equation}
From this expression, we see that interaction introduces a (negative) correction, i.e.,
\begin{equation}
\langle \psi^\dagger_{A1} (t) \psi_{A1}(0)\rangle \to \langle \psi^\dagger_{A1} (t) \psi_{A1}(0)\rangle \left( 1 - \frac{l^2 v^2}{v_F^4 t^2} \right),
\label{eq:interact_vertex}
\end{equation}
to the time-dependent correlation of vertex operators.
This perturbative expansion is valid when $vl/v_F \ll v_F t \sim \lambda$, where $\lambda \sim e v_F/I$ describes the wave-packet width that depends on the amplitude of nonequilibrium current $I$. The factor $vl/v_F$ is indeed the distance between the centers of two wave packets (charge and neutral modes) fractionalized from a single fermion.

\begin{figure}[h!]
\includegraphics[width=0.8  \columnwidth]{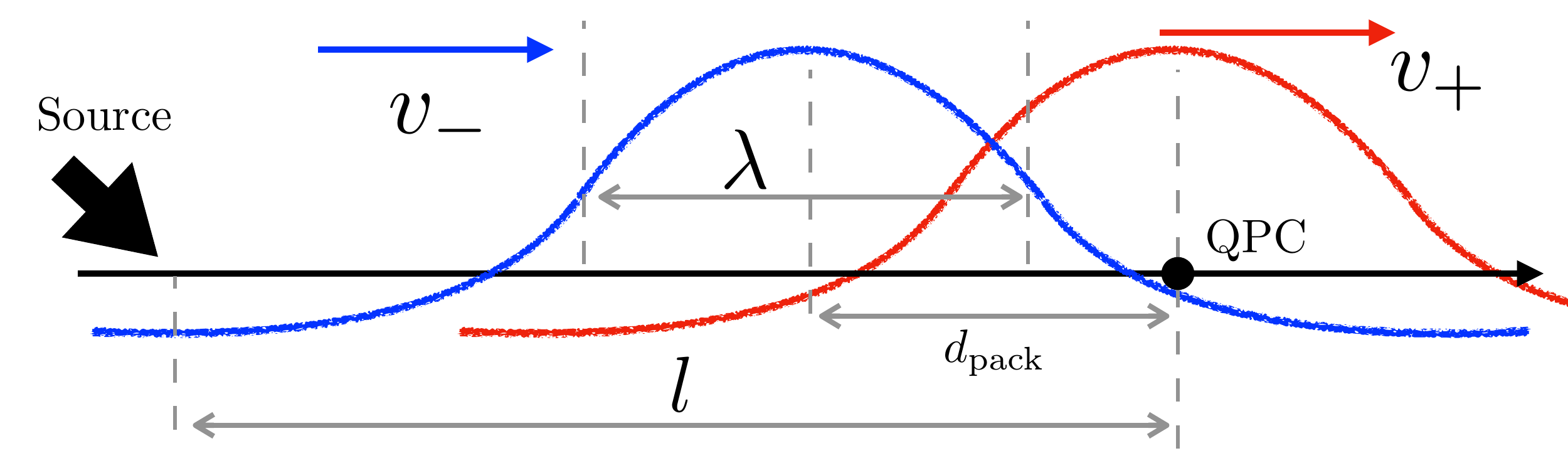}
\caption{Length scales of an interacting system. If $d_\text{pack} \ll \lambda$, two wave packets (red and blue) splitted from the same fermion becomes indistinguishable. Of this case, the interaction influences on EP and SEE are weak.}
\label{fig:wave_packs}
\end{figure}

For a more intuitive understanding of the validity of the perturbative expansion, we look at Fig.\,\ref{fig:wave_packs} for an interacting quantum Hall edge with filling factor 2. Here three length scales have been highlighted.
Under the influence of interaction, a fermion that enters $A1$ fractionalizes into a charge and a dipole packets that travel at different velocities $v_\pm$, as illustrated by the two wave-packets (of width $\lambda$) in Fig.\,\ref{fig:wave_packs}.
When the faster mode (charge mode) has arrived at the QPC, the centers of these two packets are separated by the distance $d_\text{pack} \approx vl/v_F$.
The importance of interaction is then determined by the relation between the distance between the packets $d_\text{pack}$ and their width $\lambda$. Specifically, when $\lambda \gg d_\text{pack}$, the fermion can not sense its fractionalization, so that the interaction influence is negligible. On the contrary, two modes are well-separated in the opposite limit $\lambda \ll d_\text{pack}$, and the effect of interaction becomes significant.
Actually, in Eq.~\eqref{eq:interact_vertex}, the factor that quantifies the significance of interaction, i.e., $vl/(v_F \lambda) = d_\text{pack}/\lambda$ becomes important only when $d_\text{pack} \gg \lambda$, in agreement with our qualitative analysis.

To display the effect of interaction, we provide an experiment-theory comparison of cross-correlation functions in Figs.\,5C and D of the main text, for the diluter transmission coefficients equal $\mathcal{T}_A = \mathcal{T}_B = 0.2$.
Interaction generates extra cross-correlation that equals to
\begin{equation}
\begin{aligned}
    &  \int dt \langle I_A (t) I_B (0) \rangle_\text{irr} \Big|_v - \int dt \langle I_A (t) I_B (0) \rangle_\text{irr} \Big|_{v= 0} \\
    &= \frac{e^2}{h} \mathcal{T} (1 - \mathcal{T}) \int d\epsilon \, \left( 2 - \frac{1}{1 - \mathcal{T}} \right) \frac{v^2}{v_F^2} \\
&\times  \Big\{ \delta f_{A1} (l^-,\mathcal{T}_A,\epsilon) [1 - 2 f_{A1} (l^-,0,\epsilon) - \delta f_{A1}(l^-,\mathcal{T}_A,\epsilon)]  + \delta f_{B1} (l^-,\mathcal{T}_B,\epsilon) [1 - 2 f_{B1} (l^-,0,\epsilon) - \delta f_{B1}(l^-,\mathcal{T}_B,\epsilon)] \Big\} \\
    & + \frac{e^2}{h} \mathcal{T} (1 - \mathcal{T})  \int d\epsilon \, \frac{v^2}{v_F^2} \frac{l^2}{\lambda^2}
    \\
    &\times \Big\{   \delta f_{A1} (l^-,\mathcal{T}_A,\epsilon) [1 - 2 f_{A1} (l^-,0,\epsilon)] + \delta f_{B1} (l^-,\mathcal{T}_B,\epsilon) [1 - 2 f_{B1} (l^-,0,\epsilon)]  - 2 \delta f_{A1} (l^-,\mathcal{T}_A,\epsilon) \delta f_{B1} (l^-,\mathcal{T}_B,\epsilon)\Big\} .
\end{aligned}
\label{eq:interaction_corrections_noise}
\end{equation}
These modifications are incorporated in Sec.~S2C, through the introduction of parameters $P_A$, $P_B$, and $P_\text{frac}$.
Importantly, the linear-in-$\delta f$ terms on the right-hand side of Eq.~\eqref{eq:interaction_corrections_noise} perfectly cancel when calculating the EP.
The last term, which is quadratic in $\delta f$, induces a finite correction to the EP. It is, however, comparatively smaller than the linear-in-$\delta f$ terms in the strongly diluted limit, as (i) it is proportional to the product of transmission probabilities at both diluters, i.e., $\mathcal{T}_A \mathcal{T}_B$, and (ii) it has a smaller prefactor $d_\text{pack}^2/\lambda^2$.
The SEE also receives a minor influence from interactions, as it is proportional to the EP in the strongly diluted limit.

In cross-correlation functions, the linear terms $\mathcal{T}_A P_A$ and $\mathcal{T}_B P_B$ remain, and the interaction-induced corrections are rather manifest, as can be seen from Figs.\,5C and D of the main text.
Indeed, in the presence of interaction, correlation functions display apparent deviations from the non-interacting ones: the former ones fit significantly better (than the non-interacting ones) the experimental data.
We have taken $v/v_F = 0.4$ and $l/\lambda = 0.5$ in this fitting.
The interaction-induced modification of the correlation functions is in strong contrast to the minor changes on the EP displayed in Fig.\,5E of the main text.
This comparatively much weaker sensitivity of the entanglement pointer to interactions originates from the cancellation of leading terms in Eq.~\eqref{eq:interaction_corrections_noise}.

For later convenience, we can also write down the correction to the auto-correlation as
\begin{equation}
\begin{aligned}
    & \int dt \langle I_A (t) I_A (0) \rangle_\text{irr} \Big|_v - \int dt \langle I_A (t) I_A (0) \rangle_\text{irr} \Big|_{v= 0}   \\
     &= \frac{e^2}{h} \int d\epsilon\, 
      \left(2 -\frac{1}{1 - \mathcal{T}} \right) \frac{v^2}{v_F^2} 
      \\
      &\times
     \Big\{ \mathcal{T}^2 \delta f_{A1} (l^-,\epsilon) [1 - 2 f_{A1} (l^-,0,\epsilon) - \delta f_{A1}(l^-,\epsilon)] + ( 1 - \mathcal{T} )^2 \delta f_{B1} (l^-,\epsilon) [1 - 2 f_{B1} (l^-,0,\epsilon) - \delta f_{B1}(l^-,\epsilon)] \Big\}\\
    & - \frac{e^2}{h} \mathcal{T} (1 - \mathcal{T}) \int d\epsilon\,  \frac{v^2}{v_F^2} \frac{l^2}{\lambda^2}
    \\
    &\times 
    \Big\{  \delta f_{A1} (l^-,\epsilon) [1 - 2 f_{A1} (l^-,0,\epsilon)] + \delta f_{B1} (l^-,\epsilon) [1 - 2 f_{B1} (l^-,0,\epsilon)] - 2 \delta f_{A1} (l^-,\epsilon) \delta f_{B1} (l^-,\epsilon)\Big\},
\end{aligned}
\label{eq:interaction_auto_corrections_noise}
\end{equation}
which is of the same order as the interaction-induced correction to the cross-correlation.
However, the correction Eq.~\eqref{eq:interaction_auto_corrections_noise} is experimentally less visible in comparison to that of the cross-correlation, since the non-interacting autocorrelation is normally much larger than the interaction-induced correction Eq.~\eqref{eq:interaction_auto_corrections_noise} in strongly diluted systems.

\subsection*{S2C. Parameterization of the interaction: Correlation functions, entanglement pointer and SEE with two sources of interaction}
\label{sec:ep_see_interaction}

In experiments, all quantities, including correlation functions, EP, and SEE are influenced by two sources of interactions: the interaction along the arms and that at the central QPC.
The former source of interaction, as has been shown in previous sections [more specifically in Eqs.~\eqref{eq:interaction_corrections_noise} and \eqref{eq:interaction_auto_corrections_noise}], can be obtained from the interaction-induced modifications of either the cross-correlation and autocorrelation functions.
These modifications can be described by parameters $P_A$, $P_B$, and $P_\text{frac}$ (introduced below) that have explicit connections to microscopic parameters under either a strong (see, e.g., Ref.~\cite{SIdrisovLevkivskyiX22}) or weak (Sec.~S2B) fractionalization.
By contrast, the influence of interaction near the QPC is hard to evaluate from microscopic parameters, as the interaction depends on non-universal details of the sample, including e.g., the size and shape of the QPC, how strongly the QPC has been depleted, and also on whether the tunneling across the QPC is adiabatic or not (see e.g., Ref.~\cite{SRyuX22} as a recent example). 
Therefore, we investigate the influence of interaction at the central QPC by introducing a phenomenological parameter $P_\text{QPC}$: when $P_\text{QPC}$ is positive (or negative), it indicates an extra bunching (or anti-bunching) probability of two-particle scatterings induced by the interaction at the central QPC.
Fortunately, as will be shown in this section, interaction parameters of both types can be obtained from the theory-experiment comparison.
Most importantly, we show that the leading interaction-induced effects on cross-correlations cancel out when calculating the EP and the corresponding SEE.
Consequently, the EP and SEE are less sensitive to interactions and thus better (than the correlation functions) quantify the statistics-related entanglement of the system.

We begin by considering the effect of the phenomenologically introduced parameter $P_\text{QPC}$. Briefly, as $P_\text{QPC}$ is generated by two-particle scattering, it is only finite when both sources are on.
For instance, for a fermionic system where particles only interact at the central QPC, the zero-temperature cross-correlation becomes
\begin{equation}
   \int dt \langle I_A (t) I_B (0) \rangle_\text{irr} \Big|_{v = 0}= - \frac{e^3}{h}   \mathcal{T} (1 - \mathcal{T}) \left[ (\mathcal{T}_A - \mathcal{T}_B)^2 + \mathcal{T}_A \mathcal{T}_B P_\text{QPC} \right] V_\text{bias} ,
   \label{eq:fermion_pint}
\end{equation}
where the second term is the correction from finite $P_\text{QPC}$. The prefactor of $P_\text{QPC}$ [i.e., $ \mathcal{T} (1 - \mathcal{T}) \mathcal{T}_A \mathcal{T}_B $] is the probability that two distinguishable particles bunch after a two-particle scattering event.
In Eq.~\eqref{eq:fermion_pint}, only a positive $P_\text{QPC}$, i.e., an extra interaction-induced bunching is allowed, as free fermions perfectly anti-bunch during two-particle scatterings.
A positive value of $P_\text{QPC}$ has indeed been shown numerically in, e.g., Ref.~\cite{SBellentaniPRB19}, where fermions Coulomb-interact near the central QPC.
This extra bunching can be understood as a spatial separation (of two incoming fermions) generated by the repulsive Coulomb interaction near the central QPC. By contrast, for bosons, interaction near the QPC would possibly induce an extra anti-bunching. It is then another non-trivial issue to build a confirmative connection between the interaction Hamiltonian and the value of $P_\text{QPC}$ for different types of particles.

\begin{figure}[ht!]
\includegraphics[width =  \columnwidth]{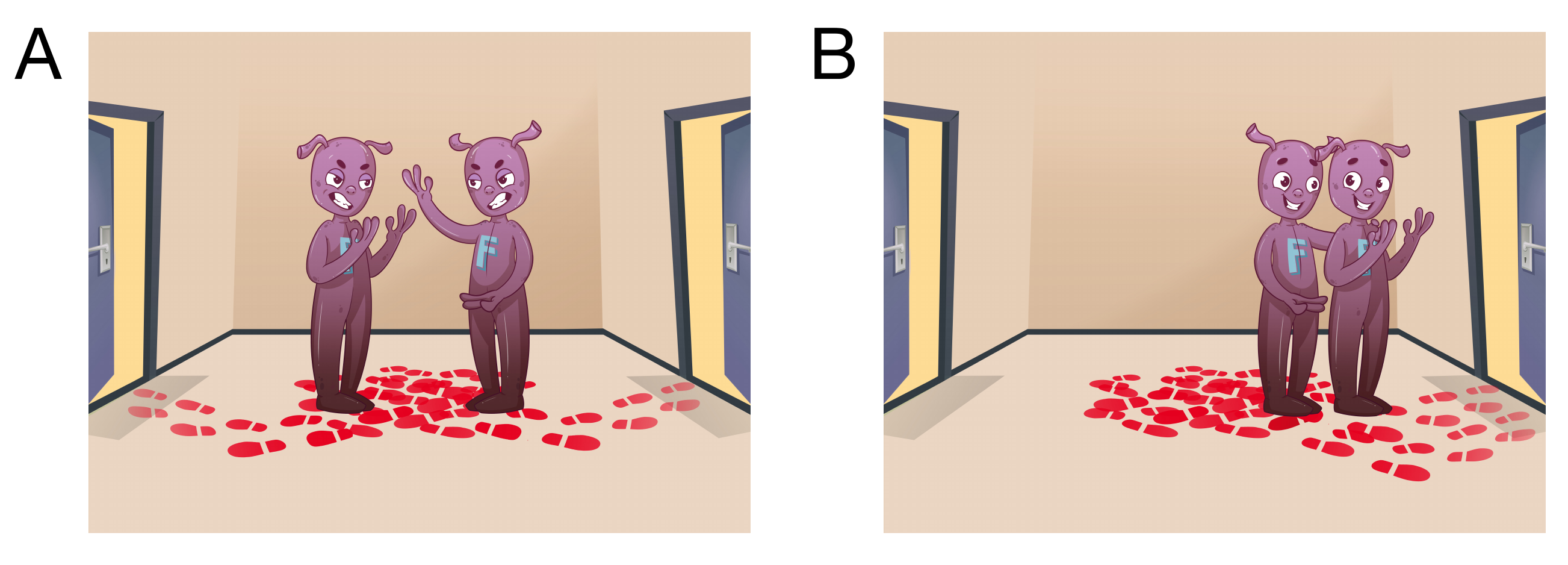}
\caption{\textbf{Competing effects in entanglement: interaction-induced entanglement blurs statistics-induced entanglement.}
(A) When two aliens from the star ``F'' meet, they instinctively hate each other, and opt to leave the room through different doors (alluding to statistics-induced anti-bunching). (B) After communicating with each other (alluding to Coulomb interaction), the two ``F''s may change their mutual sentiment, and decide to bunch together, leaving through the same door.}
\label{fig:statistics_cartoon}
\end{figure}

The following caricature (Fig.~\ref{fig:statistics_cartoon}) illustrates two competing effects, particle statistics and interaction, in two-particle scatterings.
Consider two alien civilizations, one from star ``F'' (where aliens hate each other) and the other from star ``B'' (where aliens like each other). 
Two aliens from the same star enter a meeting room from two opposite doors. Upon entering the room, they immediately recognize the other alien.
The ``F''s would like to leave through opposite doors (Fig.~\ref{fig:statistics_cartoon}A),  (the ``B''s would naturally prefer leaving the room together through the same door). This would be a representation of instantaneously acquired ``statistical entanglement''.
However, if the two aliens spend some time in the meeting room and interact with each other (alluding to Coulomb interaction), their instinctive preference (statistical entanglement) may be modified by interaction-acquired entanglement, affecting the probability of leaving ``bunched'' together [Fig.~\ref{fig:statistics_cartoon}B, which is explicitly quantified by $P_\text{QPC}$ of Eq.~\eqref{eq:fermion_pint}] (for the ``B''s--- anti-bunched).
Similarly, the measured current-current correlations may manifest contributions (to the entanglement) from both statistics and Coulomb interactions.
Following the experimental data, entanglement pointer of our model contains mainly contributions from statistics.

Although the value of $P_\text{QPC}$ is generically non-universal and hard to evaluate theoretically, it can be extracted from the theory-experiment comparison, after including interactions both along the arms and at the central QPC.
Similarly to $P_\text{QPC}$ introduced for the interaction at the QPC, we can introduce extra factors to describe the influence of interactions along the arms. Indeed, following Eq.~\eqref{eq:interaction_corrections_noise}, after including interactions along the arms, the zero-temperature cross-correlation function becomes (included in Methods)
\begin{equation}
   \int\! dt \langle I_A (t) I_B (0) \rangle_\text{irr}\! = - \frac{e^3}{h}   \mathcal{T} (1 \!-\! \mathcal{T}) \left[ (\mathcal{T}_A \!-\! \mathcal{T}_B)^2 \!+\! \mathcal{T}_A P_A \!+\! \mathcal{T}_B P_B + \mathcal{T}_A \mathcal{T}_B ( P_\text{QPC} + P_\text{frac} ) \right] V_\text{bias} ,
   \label{eq:iu_id_with_interactions}
\end{equation}
where
\begin{equation}
\begin{aligned}
P_A & = -  \left[ (1 - \mathcal{T}_A) \left( 1 - \frac{1}{2 - 2 \mathcal{T}} \right) +  \frac{l^2}{\lambda^2} \right]  \frac{v^2}{v_F^2} , \\
P_B & = -  \left[ (1 - \mathcal{T}_B) \left( 1 - \frac{1}{2 - 2 \mathcal{T}} \right) +  \frac{l^2}{\lambda^2} \right]  \frac{v^2}{v_F^2},
\end{aligned}
\label{eq:pu_pd}
\end{equation}
quantify the modification of the correlation function due to the particle fractionalization in each arm [the linear-in-$\delta f$ terms in Eq.~\eqref{eq:interaction_corrections_noise}].
These two terms are finite as long as the corresponding source is on.
By contrast, the other part $\mathcal{T}_A \mathcal{T}_B P_\text{frac}$, with
\begin{equation}
P_\text{frac} = 2 \frac{v^2}{v_F^2} \frac{l^2}{\lambda^2},
\label{eq:pfrac}
\end{equation}
is only finite when both sources are on [the last term in  Eq.~\eqref{eq:interaction_corrections_noise}].
We further define 
\begin{equation}
    P_\text{int} \equiv P_\text{frac} + P_\text{QPC}
\end{equation} 
as the parameter that reflects double-source contributions to the correlation function.
With these parameters and the definition [Eq.~(1) of the main text], the EP of fermions now reads as
\begin{equation}
    \mathcal{P}_\text{E} = (2 - P_\text{int}) \frac{e^3}{h} \mathcal{T} ( 1 - \mathcal{T} ) \mathcal{T}_A \mathcal{T}_B V_\text{bias}.
    \label{eq:ep_with_pint}
\end{equation}
In our case, as follows from the theory-experiment comparison, $P_A = P_B \approx -0.057$ and $P_\text{int} \approx 0.08$ with the values of $v/v_F$ and $l/\lambda$ evaluated in Sec.~S2 [see lines below Eq.~\eqref{eq:interaction_corrections_noise}].
In the strongly diluted limit, the single-source contributions $\mathcal{T}_A P_A$ and $\mathcal{T}_B P_B$ are generically much larger than the two-source contribution $\mathcal{T}_A \mathcal{T}_B P_\text{int}$, as $\mathcal{T}_A P_A$ and $\mathcal{T}_B P_B$ are linear functions of the diluter transmission coefficients, while $\mathcal{T}_A \mathcal{T}_B P_\text{int}$ is quadratic in diluters transmission.
Meanwhile, the values of $P_A$ and $P_B$ are generically larger than $P_\text{frac}$ from fractionalization, when $l\ll \lambda$ for strong enough dilution.
Indeed, contributions from $\mathcal{T}_A P_A$ and $\mathcal{T}_B P_B$ are the leading terms that introduce the manifest modification of correlation functions in Figs.\,5C and 5D of the main text.
However, despite their leading importance for correlation functions, $\mathcal{T}_A P_A$ and $\mathcal{T}_B P_B$ perfectly cancel out when calculating the entanglement pointer and the ensued SEE, as has been discussed after Eq.~\eqref{eq:interaction_corrections_noise}.
This cancellation, importantly, indicates that the entanglement pointer introduced in our work has much stronger resistance against interactions: the leading interaction-induced corrections to correlation functions cancel out in the EP, and a strong signal of statistics now becomes manifest.

The fact that $P_\text{QPC}$ and $P_\text{frac}$ only appear in two-source correlations helps the extraction of interaction parameters.
Indeed, with the above knowledge, in experiments, we can first evaluate the values of $P_A$ and $P_B$ from single-source correlations, and equivalently obtain the value $v$ along two interfering arms, together with the value of $P_\text{frac}$.
Afterwards, one moves to the two-source correlation to obtain $P_\text{QPC}$ that represents the effect of interaction at the central QPC.
As shown in Figs.\,5C and 5D of the main text, the interaction in the arm (i.e., $v$) suffices to produce remarkable theory-experiment matching, indicating a rather small value of $P_\text{QPC}\approx 0.02$ for our sample. 

It is worth noticing that the interaction amplitudes $v$ and $P_\text{int}$ can also be obtained from the autocorrelation function. For instance, the autocorrelation of current operators in the upper arm, with interaction included, becomes [see Eq.~\eqref{eq:interaction_auto_corrections_noise}]
\begin{equation}
\begin{aligned}
\int dt \langle I_A(t) I_A(0) \rangle_\text{irr} & = \frac{e^3}{h} V_\text{bias} [\mathcal{T}_A ( 1 - \mathcal{T} ) + \mathcal{T}_B \mathcal{T}] [1 - \mathcal{T}_A ( 1 - \mathcal{T} ) - \mathcal{T}_B \mathcal{T}]\\
& + \frac{e^3}{h} V_\text{bias} \mathcal{T} (1 - \mathcal{T}) \mathcal{T}_A \mathcal{T}_B ( P_\text{QPC} + P_\text{frac} )\\
& + \frac{e^3}{h} V_\text{bias} \left[ \mathcal{T}^2 \mathcal{T}_A ( 1 - \mathcal{T}_A ) + (1 - \mathcal{T})^2 \mathcal{T}_B ( 1 - \mathcal{T}_B ) \right] \left( 2 - \frac{1}{1-\mathcal{T}} \right) \frac{v^2}{v_F^2}\\
& - 2 \frac{e^3}{h} V_\text{bias}  \mathcal{T} (1 - \mathcal{T}) \left( \mathcal{T}_A + \mathcal{T}_B   \right) \frac{v^2}{v_F^2}\frac{l^2}{\lambda^2}.
\end{aligned}
\label{eq:auto_correlation_corrections}
\end{equation}
Similarly to its role in cross-correlations, $P_\text{int} = P_\text{QPC} + P_\text{frac}$ shows up in Eq.~\eqref{eq:auto_correlation_corrections} of the autocorrelation only when both sources are on. One can then, in principle, evaluate the values of $P_\text{int}$ from autocorrelations following steps similar to those described above for cross-correlations.
\begin{figure}[ht!]
\includegraphics[width= 1\columnwidth]{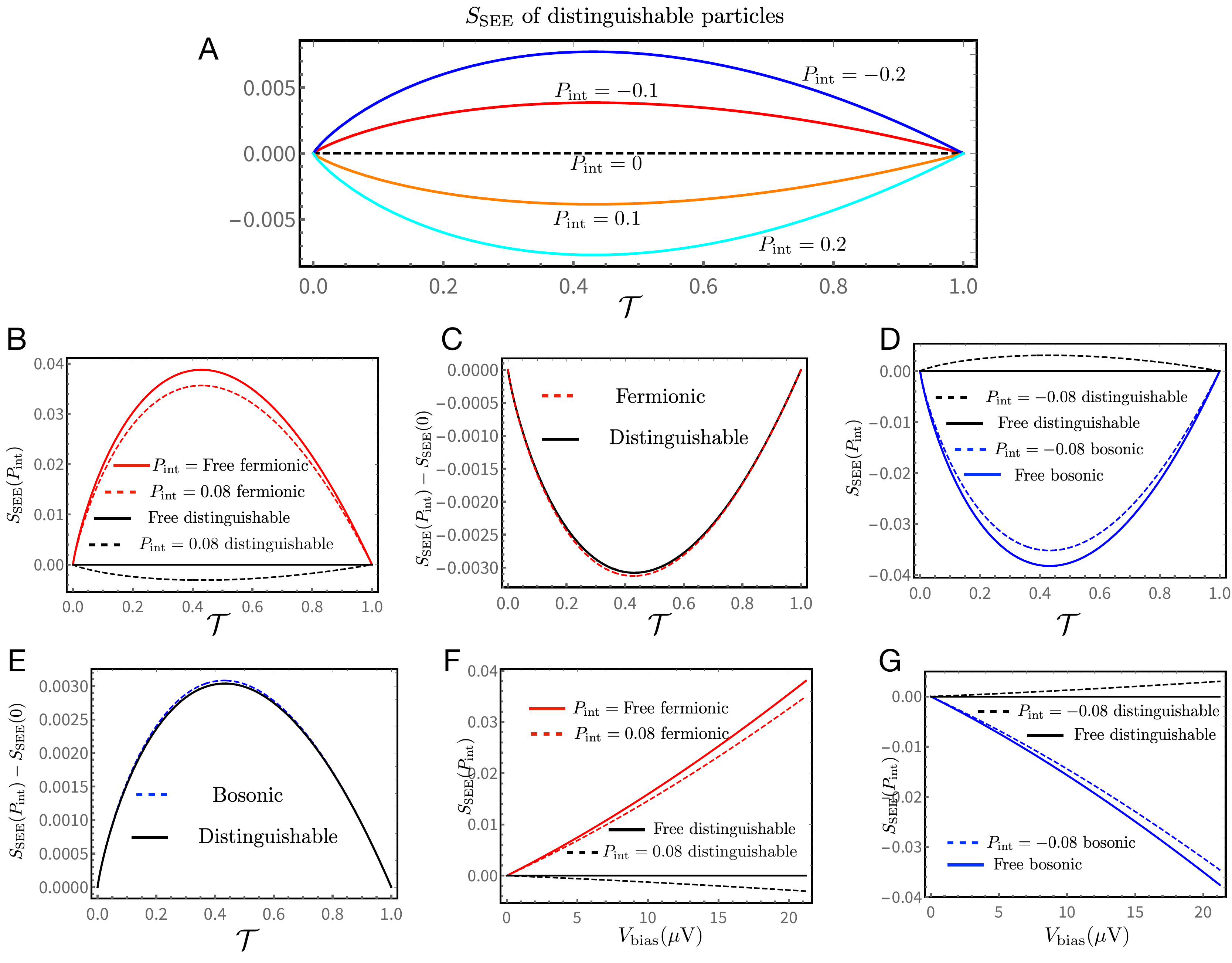}
\caption{Statistics-induced entanglement entropy in the presence of interaction for $\mathcal{T}_A = \mathcal{T}_B = 0.1$.
Interaction is incorporated by the parameter $P_\text{int} = P_\text{QPC} + P_\text{frac}$ 
describing the interaction-induced bunching/anti-bunching probability.
(A) The SEE of distinguishable particles with different values of $P_\text{int}$. The dashed black line indicates the vanishing SEE for free distinguishable particles. (B) The SEE of fermions (red lines) and distinguishable particles (black lines). (C) The interaction-induced SEE of fermionic and distinguishable situations with $P_\text{int} = 0.08$. 
The interaction-induced SEE is defined as the difference between dashed and solid lines with the same color, i.e., $S_\text{SEE} (P_\text{int}) - S_\text{SEE} (0) $, for the dwell time $\tau = 0.01 $ns, and the bias $V_\text{bias} = 20.7\mu$V.
For the given value of $P_\text{int}$, the interaction-induced SEE is almost statistics-insensitive. (D) The SEE of bosons (blue lines) and distinguishable particles (black lines). (E) The interaction-induced SEE of bosons and distinguishable situations, with $P_\text{int}= -0.08$. (F) and (G) show the corresponding data in (B) and (D), respectively, as a function of $V_\text{bias}$. Here we have taken $\mathcal{T} = 0.5$ for the transmission through the central QPC. }
\label{fig:see_interaction}
\end{figure}

As $P_\text{int}$ modifies both the correlation functions and the related FCS generating function, it also modifies the SEE.
Indeed, with interaction involved, three contributions of the non-interacting FCS generating function Eq.~\eqref{eq:sgenerating_function_expression} change to
\begin{equation}
\begin{aligned}
& 1 - \mathcal{T} \mathcal{T}_A (1 - \mathcal{T}_B) - \mathcal{T} \mathcal{T}_B (1 - \mathcal{T}_A) \to 1 - \mathcal{T} f_A (1 - f_B) - \mathcal{T} f_B (1 - f_A),\\
& - f_A f_B \mathcal{T} (1 - \mathcal{T} ) (P_\text{QPC} + P_\text{frac})\\
& \mathcal{T} \mathcal{T}_A (1 - \mathcal{T}_B) \to \mathcal{T} f_A (1 - f_B) + \frac{1}{2} f_A f_B \mathcal{T} (1 - \mathcal{T} ) (P_\text{QPC} + P_\text{frac}),\\
& \mathcal{T} \mathcal{T}_B (1 - \mathcal{T}_A) \to \mathcal{T} f_B(1 - f_A) + \frac{1}{2} f_A f_B \mathcal{T} (1 - \mathcal{T} ) (P_\text{QPC} + P_\text{frac}).
\end{aligned}
\label{eq:interaction_effective_tunnelings}
\end{equation}
This includes two modifications: (i) the modification of the distribution function due to fractionalization, and (ii) the interaction-induced extra bunching.
The first of them can be obtained from the correction to the vertex correlation Eq.~\eqref{eq:interact_vertex}.
The second can be incorporated via introducing effective distribution functions.
Indeed, under the influence of $P_\text{QPC}$ and $P_\text{frac}$, the real distribution functions $f_A$ and $f_B$ in front of the central QPC behave effectively as $\tilde{f}_A$ and $\tilde{f}_B$.
More specifically, one can obtain $\tilde{f}_A$ and $\tilde{f}_B$ by solving
\begin{equation}
    \begin{aligned}
    \mathcal{T} f_A (1 - f_B) + \frac{1}{2} f_A f_B \mathcal{T} (1 - \mathcal{T} ) (P_\text{QPC} + P_\text{frac}) & = \mathcal{T} \tilde{f}_A (1 - \tilde{f}_B), \\
    \mathcal{T} f_B (1 - f_A) + \frac{1}{2} f_A f_B \mathcal{T} (1 - \mathcal{T} ) (P_\text{QPC} + P_\text{frac}) & = \mathcal{T} \tilde{f}_B (1 - \tilde{f}_A).
    \end{aligned}
    \label{eq:effective_distributions}
\end{equation}
For weak interactions, when $P_\text{QPC} + P_\text{frac}$ is a small parameter, the modification of the distribution functions approximately becomes
\begin{equation}
\tilde{f}_A - f_A = \tilde{f}_B - f_B \approx -\frac{f_A f_B}{2 (1 - f_A - f_B)} ( P_\text{QPC} + P_\text{frac} ) (1 - \mathcal{T}).
\label{eq:effective_dis_from_interaction}
\end{equation}

With the effective distribution functions, we can write the SEE of Eq.~\eqref{eq:see_expression} as a sum of two terms: $S_\text{SEE} = S_\text{SEE}^0 + \delta S_\text{SEE}$, where $S_\text{SEE}^0$ follows the structure of Eq.~\eqref{eq:see_expression}, after replacing $\mathcal{T}_A$ and $\mathcal{T}_B$ by $f_A$ and $f_B$.
The other part $\delta S_\text{SEE}$, which is generated by extra bunching at the central QPC, equals
\begin{equation}
\begin{aligned}
    \delta S_\text{SEE} & = - \frac{e \tau}{h}\int d\epsilon  \sum_{i=1,2} \left\{ \ln [ 2 \cosh u_i (f_A,f_B) ] - u_i (f_A,f_B) \tanh [ u_i (f_A,f_B) ] \right. \\
    & \left.  - \ln [ 2 \cosh u_i (\tilde{f}_A,\tilde{f}_B) ] + u_i (\tilde{f}_A,\tilde{f}_B) \tanh [ u_i (\tilde{f}_A,\tilde{f}_B) ] \right\}  .
\end{aligned}
\label{eq:delta_see}
\end{equation}
Following Eq.~\eqref{eq:effective_dis_from_interaction}, $\delta S_\text{SEE}$ becomes negligible for a small value of $P_\text{int} = P_\text{QPC} + P_\text{frac}$, where $f_A - \tilde{f}_A$ and $f_B - \tilde{f}_B$ are negligible quantities.

The SEE influenced by interaction at the central QPC is plotted in Fig.\,\ref{fig:see_interaction}.
In agreement with the analysis above, the correction to SEE from interactions is much smaller than the SEE of free particles, as long as $P_\text{int} = P_\text{QPC} + P_\text{frac}$ is a small quantity.
This is, fortunately, the case of our work, where $P_\text{int} \approx 0.08$ as has been discussed above.

\section*{S3. Statistics-induced entanglement in terms of Bell pairs}
\label{SM:Bell}

In this section, we show that the correlation function measurement is related to the notions of Bell pairs and Bell's inequality.
We consider only the non-interacting fermionic situation in this section.
We begin with the derivation of the two-fermion density matrix and the corresponding EE.
We start with a two-fermion event with the initial wavefunction 
\begin{equation}
    \Psi = | \mathcal{S}_A\rangle | \mathcal{S}_B\rangle
    \label{eq:initial_wavefunction}
\end{equation}
representing a pure state in the second quantization. This function means that one particle appears at the up source and one particle simultaneously appears at the down source. After scattering at two diluters and the central QPC, the initial wavefunction Eq.~\eqref{eq:initial_wavefunction} becomes
\begin{equation}
\begin{aligned}
     |\Psi \rangle & =  r_A r_B |\tilde{\mathcal{D}}_A\rangle |\tilde{\mathcal{D}}_B\rangle + r_A t_B r |\tilde{\mathcal{D}}_A\rangle |\mathcal{D}_B\rangle - r_A t_B t |\tilde{\mathcal{D}}_A,\mathcal{D}_A\rangle |\Omega\rangle \\
  &+ t_A r_B r |\mathcal{D}_A\rangle |\tilde{\mathcal{D}}_B\rangle + t_A r_B t |\Omega\rangle | \mathcal{D}_B,\tilde{\mathcal{D}}_B \rangle + t_A t_B | \mathcal{D}_A\rangle |\mathcal{D}_B\rangle,
\end{aligned}
\label{eq:wavefunction_after_scattering}
\end{equation}
where each term of Eq.~\eqref{eq:wavefunction_after_scattering} is a product of corresponding states in subsystems $A$ and $B$, respectively.
For instance $|\tilde{\mathcal{D}}_A \mathcal{D}_A\rangle |\Omega\rangle$ means that both fermions end up in subsystem $A$ ($|\tilde{\mathcal{D}}_A \mathcal{D}_A\rangle$), leaving subsystem $B$ empty ($|\Omega \rangle$). $t_A$, $t_B$, $r_A$ and $r_B$ refer to the scattering matrix elements at two diluters, while $t$ and $r$ indicate these of the central QPC.
For simplicity all these elements are assumed to be real.

After partial trace over states of subsystem $B$, the reduced density matrix becomes
\begin{equation}
\begin{aligned}
 \rho_A &= \mathcal{R}_A \mathcal{T}_B \mathcal{T} | \tilde{\mathcal{D}}_A,\mathcal{D}_A\rangle \langle \tilde{\mathcal{D}}_A,\mathcal{D}_A | + \mathcal{T}_A \mathcal{R}_B \mathcal{T} |\Omega\rangle \langle \Omega |  \\ 
 & + \mathcal{R}_A( \mathcal{R}_B + \mathcal{T}_B \mathcal{R} ) |\tilde{\mathcal{D}}_A\rangle \langle \tilde{\mathcal{D}}_A| + \mathcal{T}_A (\mathcal{T}_B + \mathcal{R}_B \mathcal{R}) |\mathcal{D}_A\rangle \langle \mathcal{D}_A |\\
 &+  t_A^* r_A r | \tilde{\mathcal{D}}_A\rangle \langle \mathcal{D}_A | + t_A r_A^* r^* |\mathcal{D}_A \rangle \langle \tilde{\mathcal{D}}_A |,
\end{aligned}
\label{eq:reduced_rho_a}
\end{equation}
where all states in Eq.~\eqref{eq:reduced_rho_a} belong to subsystem $A$. Following the definition of the von Neumann entropy, the entanglement entropy is given by 
$$S_\text{ent} = -\sum_{i} \rho_i \ln \rho_i,$$ 
where
\begin{equation}
\begin{aligned}
\rho_1 & = (1 - \mathcal{T}_A) \mathcal{T}_B \mathcal{T}, \quad \rho_2 = \mathcal{T}_A (1 - \mathcal{T}_B ) \mathcal{T}, \\
\rho_3 & = \frac{1}{2} \left[ 1 - (1 - \mathcal{T}_A) \mathcal{T}_B \mathcal{T}- \mathcal{T}_A (1 - \mathcal{T}_B) \mathcal{T} + \zeta_1 \right],\\
\rho_4 & = \frac{1}{2} \left[ 1 - (1 - \mathcal{T}_A) \mathcal{T}_B \mathcal{T}- \mathcal{T}_A ( 1 - \mathcal{T}_B) \mathcal{T} - \zeta_1 \right]
\label{eq:fermion_2p_eigenvalues}
\end{aligned}
\end{equation}
are four eigenvalues of the reduced density matrix~\eqref{eq:reduced_rho_a} and
\begin{equation}
    \zeta_1  = \sqrt{[ 1- \mathcal{T}(\mathcal{T}_A + \mathcal{T}_B)]^2 + 4 \mathcal{T} (1 - \mathcal{T}) \mathcal{T}_A \mathcal{T}_B}.
\end{equation}
The first two eigenvalues $\rho_1$ and $\rho_2$ in Eq.~\eqref{eq:fermion_2p_eigenvalues} correspond, respectively, to the states $|\Omega\rangle$, where both particles end in subsystem $B$, and $| \tilde{\mathcal{D}}_A,\mathcal{D}_A\rangle$, where they instead both stay in the subsystem $A$.
The other two eigenvalues, $\rho_3$ and $\rho_4$, refer to two distinct configurations where each subsystem hosts a single particle. The fact that there are two non-zero eigenvalues indicates an entanglement between subsystems. Indeed, even if we know there is one state in subsystem $B$, the status of that in subsystem $A$ can not be fully determined.
To see the entanglement more clearly, we can rewrite the system state Eq.~\eqref{eq:wavefunction_after_scattering} as
\begin{equation}
\begin{aligned}
|\Psi \rangle = & -r_A t_B t |\tilde{\mathcal{D}}_A,\mathcal{D}_A\rangle |\Omega\rangle + t_A r_B t | \Omega\rangle | \mathcal{D}_B,\tilde{\mathcal{D}}_B\rangle \\
& + \alpha | \uparrow_A \rangle | \uparrow_B \rangle + \beta | \downarrow_A \rangle | \downarrow_B \rangle,
\end{aligned}
\label{eq:two-fermion_state}
\end{equation}
where the first and second lines above refer to wavefunctions $|\tilde{\psi}\rangle $ [corresponding to configurations $(2,0)$ and $(0,2)$] and $|\psi\rangle $ [corresponding to configurations $(1,1)$], respectively.

\textit{Comparison with NOON states.}
In state $|\tilde{\psi} \rangle =|\psi_{2,0}\rangle + |\psi_{0,2}\rangle$, the two particles are in either one of two subsystems, resembling a NOON state (i.e., an entangled state $\sim |N\rangle |0\rangle + |0\rangle |N\rangle$, with $N$ the particle number of the corresponding subsystem).
Akin to the NOON state, the state $|\tilde{\psi}\rangle$ \textit{alone} cannot directly prove the violation of Bell inequality: with only a NOON state, rotation of quasi-spin (with the $z$-direction of spin defined by the particle number) is impossible. Indeed, for a NOON state, the test of Bell inequality requires the fusion of the system with other external states~\cite{STanWallsCollettPRL91,SHessmoPRL04}.
Similarly, for our Pauli-blocked state $|\tilde{\psi}\rangle$, quantum manipulations within the subsystem are not effective; hence, in order to probe Bell's inequality with $|\tilde{\psi}\rangle$, one needs to modify the device by coupling its arms to some external channels.

\textit{Comparison with Bell states.}
As discussed in the main text, the wave function $|\psi\rangle = \alpha | \uparrow_A \rangle | \uparrow_B \rangle + \beta | \downarrow_A \rangle | \downarrow_B \rangle$ is the wave function of an effective Bell pair, with orthonormal pseudospin states $| \uparrow_A \rangle, | \downarrow_A \rangle, | \uparrow_B \rangle$ and $| \downarrow_B \rangle$ [see Eq.~\eqref{eq:four_11_states} below for their definitions]. However, $|\psi\rangle$ is not maximally entangled, since the coefficients  $|\alpha|^2 = \rho_3$ and $|\beta|^2= \rho_4$ are not, in general, equal to each other.
The pseudospin-up and pseudospin-down states are given by
\begin{equation}
\begin{aligned}
| \uparrow_A \rangle & = \frac{2}{\zeta_A} |\tilde{\mathcal{D}}_A\rangle + \frac{-1 + 2 \mathcal{T}_A + \mathcal{T} (\mathcal{T}_B - \mathcal{T}_A) + \zeta_1}{t_A r_A r \zeta_A} |\mathcal{D}_A\rangle\\
| \downarrow_A \rangle & = \frac{2}{\zeta_A'} |\tilde{\mathcal{D}}_A\rangle + \frac{-1 + 2 \mathcal{T}_A + \mathcal{T} (\mathcal{T}_B - \mathcal{T}_A) - \zeta_1}{t_A r_A r \zeta_A'} |\mathcal{D}_A\rangle\\
| \uparrow_B \rangle & = \frac{2}{\zeta_B} |\tilde{\mathcal{D}}_B\rangle + \frac{-1 + 2 \mathcal{T}_B + \mathcal{T} (\mathcal{T}_A - \mathcal{T}_B) + \zeta_1}{t_B r_B r \zeta_B} |\mathcal{D}_B\rangle\\
| \downarrow_B \rangle & = \frac{2}{\zeta_B'} |\tilde{\mathcal{D}}_B\rangle + \frac{-1 + 2 \mathcal{T}_B + \mathcal{T} (\mathcal{T}_A - \mathcal{T}_B) - \zeta_1}{t_B r_B r \zeta_B'} |\mathcal{D}_B\rangle,
\label{eq:four_11_states}
\end{aligned}
\end{equation}
where
\begin{equation}
\begin{aligned}
    \zeta_A & \!=\! \sqrt{4 \!+\! \frac{[-1 + 2 \mathcal{T}_A + \mathcal{T} (\mathcal{T}_B - \mathcal{T}_A) + \zeta_1]^2}{\mathcal{T}_B (1 - \mathcal{T}_B) (1 - \mathcal{T})}}, \ \
    \zeta_A' \! =\! \sqrt{4 \!+\! \frac{[-1 + 2 \mathcal{T}_A + \mathcal{T} (\mathcal{T}_B - \mathcal{T}_A) - \zeta_1]^2}{\mathcal{T}_B (1 - \mathcal{T}_B) (1 - \mathcal{T})}},\\
    \zeta_B & \!=\! \sqrt{4\! +\! \frac{[-1 + 2 \mathcal{T}_B - \mathcal{T} (\mathcal{T}_B - \mathcal{T}_A) + \zeta_1]^2}{\mathcal{T}_A (1 - \mathcal{T}_A) (1 - \mathcal{T})}}, \ \
    \zeta_B' \! =\! \sqrt{4 \!+\! \frac{[-1 + 2 \mathcal{T}_B - \mathcal{T} (\mathcal{T}_B - \mathcal{T}_A) - \zeta_1]^2}{\mathcal{T}_A (1 - \mathcal{T}_A) (1 - \mathcal{T})}}
\end{aligned}
\end{equation}
are normalization factors.
When $\mathcal{T}_A= \mathcal{T}_B$, these functions can be greatly simplified, leading to
\begin{equation}
    \begin{aligned}
    | \uparrow_A \rangle & = \frac{2}{\zeta_A} |\tilde{\mathcal{D}}_A\rangle + \frac{-1 + 2 \mathcal{T}_A + \zeta_1}{t_A r_A r \zeta_A} |\mathcal{D}_A\rangle\\
| \downarrow_A \rangle & = \frac{2}{\zeta_A'} |\tilde{\mathcal{D}}_A\rangle + \frac{-1 + 2 \mathcal{T}_A  - \zeta_1}{t_A r_A r \zeta_A'} |\mathcal{D}_A\rangle\\
| \uparrow_B \rangle & = \frac{2}{\zeta_B} |\tilde{\mathcal{D}}_B\rangle + \frac{-1 + 2 \mathcal{T}_A +  \zeta_1}{t_A r_A r \zeta_B} |\mathcal{D}_B\rangle\\
| \downarrow_B \rangle & = \frac{2}{\zeta_B'} |\tilde{\mathcal{D}}_B\rangle + \frac{-1 + 2 \mathcal{T}_A  - \zeta_1}{t_A r_A r \zeta_B'} |\mathcal{D}_B\rangle,
    \end{aligned}
\end{equation}
where $\zeta_1 = \sqrt{1 - 4\mathcal{T} (1 - \mathcal{T}_A) \mathcal{T}_A}$, and
\begin{equation}
    \begin{aligned}
    \zeta_A & = \zeta_B = \sqrt{4 + \frac{[-1 + 2 \mathcal{T}_A + \sqrt{1 - 4\mathcal{T} (1 - \mathcal{T}_A) \mathcal{T}_A}]^2}{(1-\mathcal{T})(\mathcal{T}_A - 1) \mathcal{T}_A}}\\
    \zeta_A' & = \zeta_B'=\sqrt{4 - \frac{[-1 + 2 \mathcal{T}_A - \sqrt{1 - 4\mathcal{T} (1 - \mathcal{T}_A) \mathcal{T}_A}]^2}{(1-\mathcal{T})(\mathcal{T}_A - 1) \mathcal{T}_A}}
    \end{aligned}
\end{equation}

Now we consider the connection of Eq.~\eqref{eq:four_11_states} to the Bell pairs. In experiments, one measures the correlation $\langle I_A I_B\rangle$, which records the simultaneous appearance of particles in $\mathcal{D}_A$ and $\mathcal{D}_B$.
If we limit ourselves to configurations where each subsystem hosts a single particle of a two-particle state, we can introduce ``spin'' in each subsystem. Indeed, in these situations, we can consider subsystem $A$ as ``spin-up'' along ``$z$''-axis if the particle in $A$ ($B$) is in state $|\!\uparrow_A\rangle$ ($|\!\uparrow_B\rangle$), and ``spin-down'' in state $|\!\downarrow_A\rangle$ ($|\!\downarrow_B\rangle$).
In this sense, the current measurement $\langle I_A I_B\rangle$ can be considered as a ``spin'' measurement along some directions that deviate from the ``$z$''-axis.
It is then related to the measurement of the Bell inequality, similar to that proposed by Ref.\,\cite{SChtchelkatchevPRB02}.

\section*{S4. Entanglement entropy from single-particle and two-particle scattering pictures}

In the previous section, we have shown that for a two-particle scattering event, the outcome can be categorized into two types of configurations: (i) states $|\tilde{\psi}\rangle$, with particles in the same subsystem [i.e., $(2,0)$ and $(0,2)$], and (ii) states $|\psi\rangle$, where each subsystem hosts one particle [i.e., configurations of (1,1)].
In this section, we look into both EE and corresponding SEE contributions from these two types of configurations.

\begin{figure}[ht!]
\includegraphics[width= 0.75\columnwidth]{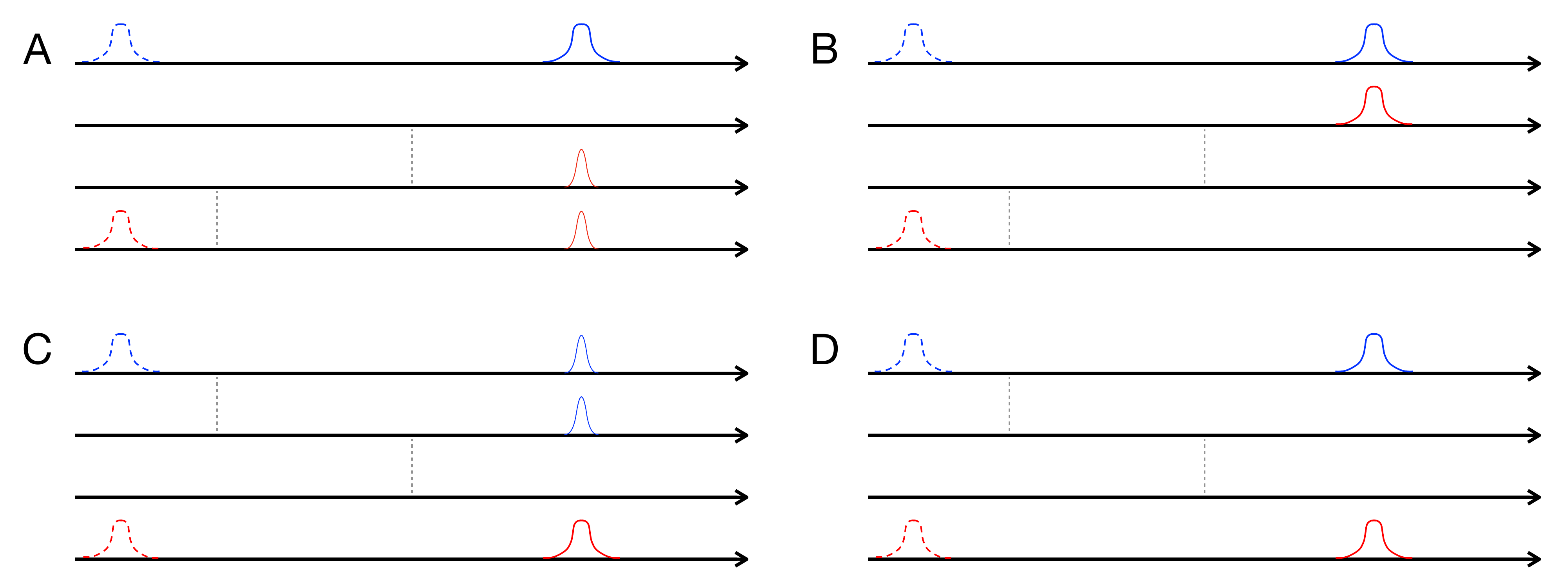}
\caption{Configurations whose EE contributions are removed from SEE. Two initial states of the particles are indicated by dashed pulses. Panels A and B show the final states, for configurations $(1,1)$ and $(2,0)$, respectively, when $\mathcal{T}_A = 0$. Panel C and D present the corresponding configurations when $\mathcal{T}_B = 0$.}
\label{fig:removed_configs}
\end{figure}

Briefly, following Sec.~S3, after scattering events at three QPCs, the entanglement equals $S_\text{ent} = -\sum_{i} \rho_i \ln \rho_i$, with $\rho_i$ [given by Eq.~\eqref{eq:fermion_2p_eigenvalues}] eigenvalues of the reduced density matrix $\rho_A$. Here $\rho_1$, $\rho_2$ correspond to configurations $(2,0)$ and $(0,2)$, respectively. The other two eigenvalues $\rho_3$, $\rho_4$ correspond to configurations of the $(1,1)$ type.
It is thus straightforward to define EE from $|\tilde{\psi}\rangle$ as $S^{(2,0)} + S^{(0,2)} \equiv -\rho_1 \ln \rho_1 - \rho_2 \ln \rho_2$, and that from $|\psi \rangle$ as $S^{(1,1)} \equiv -\rho_3 \ln \rho_3 - \rho_4 \ln \rho_4$.
We can further define SEE of $|\tilde{\psi}\rangle$ and $|\psi\rangle$, following Eq.~(4) of the main text, as
\begin{equation}
    S_\text{SEE}^{(N_A,N_B)} \equiv S^{(N_A,N_B)} (0,\mathcal{T}_B) + S^{(N_A,N_B)} (\mathcal{T}_A,0) - S^{(N_A,N_B)} (\mathcal{T}_A,\mathcal{T}_B),
\end{equation}
for the configuration $(N_A,N_B)$ of the pure two-particle  ($N_A+N_B=2$) state with $N_A$ particle in subsystem $A$ and $N_B$ particles in subsystem $B$.
By doing so, we are removing statistics-irrelevant configurations in Fig.~\ref{fig:removed_configs}, where two involved particles scatter independently.
The obtained EE and SEE are provided in Table~\ref{tab:quantum}.

\begin{center}
\begin{table}[h]
\resizebox{\textwidth}{!}{\begin{tabular}{|c|c|c|}
 \hline
     & Configurations $(2,0)$ and $(0,2)$ & Configuration $(1,1)$  \\ 
 \hline
EE & $ S^{(2,0)} + S^{(0,2)} = -\rho_1 \ln \rho_1 - \rho_2 \ln \rho_2 $ & $ S^{(1,1)} = -\rho_3 \ln \rho_3 - \rho_4 \ln \rho_4 $  \\
\hline
Leading- & $S_\text{SEE}^{(2,0)} + S_\text{SEE}^{(0,2)} $ & $S_\text{SEE}^{(1,1)}$ \\
 order SEE & $\approx - \mathcal{T} \mathcal{T}_A \mathcal{T}_B  [2 + \ln (\mathcal{T}^2 \mathcal{T}_A \mathcal{T}_B)]$ & $ \approx \mathcal{T} \mathcal{T}_A \mathcal{T}_B [2 + \mathcal{T}  \ln (\mathcal{T}_A \mathcal{T}_B \mathcal{T}^2)] $ \\
 \hline
\end{tabular}}
\caption{Contributions of configurations $(N_A=2,N_B=0), (N_A=0,N_B=2)$ and $(N_A=1,N_B=1)$ to the EE and SEE of fermions. SEE is defined following Eq.~(4) of the main text, i.e., $S_\text{SEE}^{(N_A,N_B)} = S^{(N_A,N_B)} (0,\mathcal{T}_B) + S^{(N_A,N_B)} (\mathcal{T}_A,0) - S^{(N_A,N_B)} (\mathcal{T}_A,\mathcal{T}_B)$, for the configuration $(N_A,N_B)$.}
\label{tab:quantum}
\end{table}
\end{center}

As shown in Table~\ref{tab:quantum}, for particles with fermionic statistics, both $|\tilde{\psi}\rangle$ [corresponding to $(2,0)$ and $(0,2)$ configurations] and $|\psi \rangle$ [corresponding to $(1,1)$ configurations] have finite contributions to SEE.
In the strongly diluted limit, their sum equals to $S_\text{SEE} = S_\text{SEE}^{(2,0)} + S_\text{SEE}^{(0,2)} + S_\text{SEE}^{(1,1)} \approx - \mathcal{T} (1 - \mathcal{T}) \mathcal{T}_A \mathcal{T}_B  \ln (\mathcal{T}^2 \mathcal{T}_A \mathcal{T}_B)$, in agreement with Eq.~(11) of the main text.
In addition, in the strongly diluted limit, the ratio of the two contributions to the SEE reads as
\begin{equation}
    \frac{S_\text{SEE}^{(1,1)}}{S_\text{SEE}^{(2,0)} + S_\text{SEE}^{(0,2)}} = -\frac{2 + \mathcal{T} \ln (\mathcal{T}^2 \mathcal{T}_A \mathcal{T}_B) }{2 + \ln( \mathcal{T}^2 \mathcal{T}_A \mathcal{T}_B)}.
    \label{eq:see_ratio}
\end{equation}
Equation~\eqref{eq:see_ratio} demonstrates that, generically, both $S_\text{SEE}^{(1,1)}$ and $S_\text{SEE}^{(2,0)} + S_\text{SEE}^{(0,2)}$ have a finite contribution to SEE.

Before closing this section, we emphasize that, although SEE contains contributions from both $|\tilde{\psi}\rangle$ and $|\psi\rangle$, it does not influence the correlation between Bell nonlocality and a finite SEE. Indeed, if SEE equals zero, two contributing particles become distinguishable from each other. In this case, the two-particle scattered wave function does not contain $|\psi\rangle$, and is thus unable to violate Bell inequality.

\section*{S5. The dwell time and the measuring time}
\label{sec:measuring_time}

In general, we can assume that a certain type of scattering event $i$ produces the time-independent EE $s_{i,A}$ between two subsystems. Here $i$ can be either a single-particle or multi-particle scattering event.
Assuming such an event occurs $N_i$ times during the entire measurement, a total EE $S_\text{ent} = \sum_i N_i s_{i,A}$ is then produced.

In systems with constant sources, $N_i \propto \tau_\text{measure}$ for each type of scattering, where $\tau_\text{measure}$ is the measuring time (of FCS) during which two subsystems become entangled. Theoretically $S_\text{ent} \propto \tau_\text{measure}$ thus also grows linearly in the measuring time.
This time-dependence is indeed naturally expected, as we are evaluating the EE from the FCS generating function that is known to be proportional to $\tau_\text{measure}$.
However, in experiments, EE cannot grow indefinitely in time, as particles lose coherence after entering the drains.
As a consequence, experimentally $S_\text{ent}$ (and also $S_\text{SEE}$) becomes instead proportional to the dwell time $\tau$, which is the time needed for a particle to travel from the central QPC to one of the drains.

In this work, $t_\text{dwell}$ can be considered as the traveling time from the central QPC to either drain. More specifically, it is given by the ratio of the distance between the central QPC and the closest drain (around $5\mu$m) and the traveling velocity (around $4.6 \times 10^5$m/s), which yields $ t_\text{dwell} \approx 0.01 $ns.

\section*{S6. Additional experimental data}

\subsection*{S6A. Experimental data of the $\nu = 3$ integer quantum Hall sample}
\label{sec:exp_correlation}

\begin{figure}[h!]
\includegraphics[width= 0.8  \columnwidth]{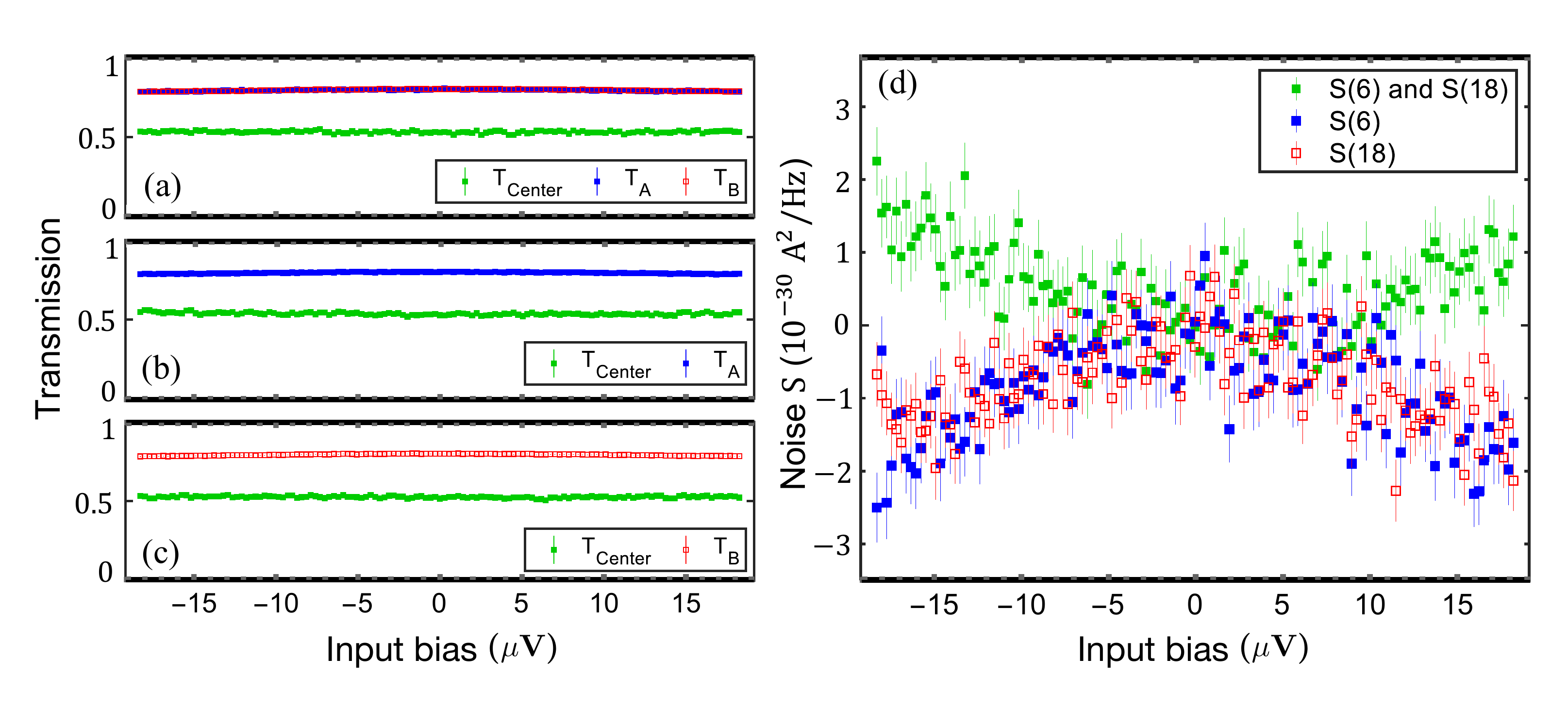}
\caption{Experimental data for the $\nu = 3$ sample. Here, the blue, red, and green data points (with error bars) represent the transmission probabilities at the upper diluter, lower diluter, and central QPC, respectively. (A) Transmission probabilities with both sources on; (B) Transmission probabilities with only the upper source ($S_A$) on, and (C) that with only the lower source ($S_B$) on. (D) Corresponding measured cross-correlations.}
\label{fig_NoiseMeasurementRaw}
\end{figure}

In Sec.~S2, we have analyzed correlation functions in the presence of interaction, and have compared the theoretical results with the measured data. In this section, we present more data in Fig.\,\ref{fig_NoiseMeasurementRaw}.

\subsection*{S6B. Correlations for filling factor $\nu = 1$: emergence of non-linearity}
\label{sec:nu_1_correlations}

\begin{figure}[h!]
\includegraphics[width= 0.8 \columnwidth]{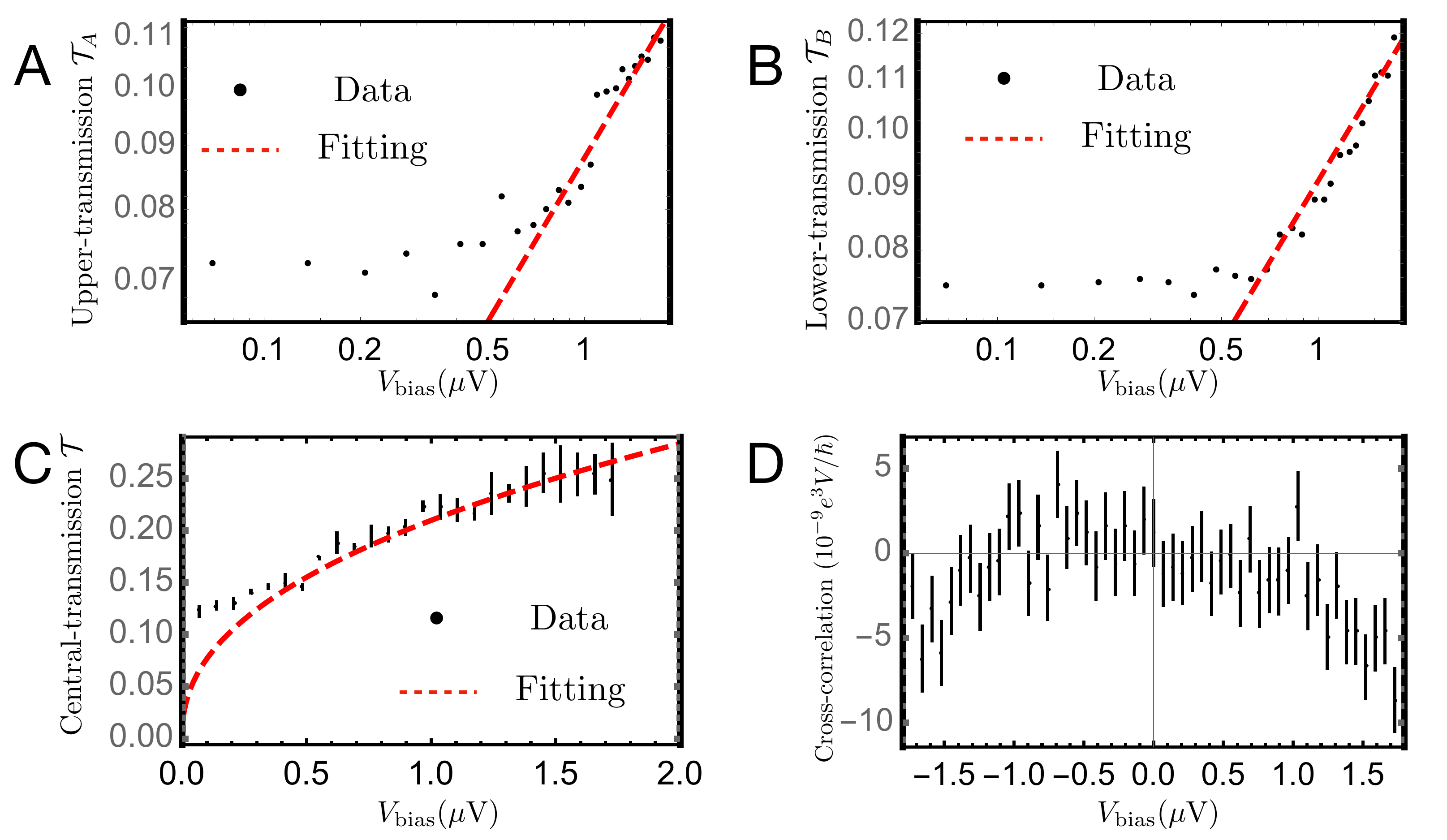}
\caption{Experimental data for the $\nu = 1$ sample. Panels (A), (B), and (C) show the transmission probabilities at three QPCs as functions of the bias voltage. At high bias, the transmissions of all three QPCs can be approximately fitted by a power-law function of the non-equilibrium current $\propto I^{0.435}$ (red dashed) line. This behavior can originate from either Luttinger-liquid renormalization (see, e.g., Ref.\,\cite{SFendleyLudwigSaleurPRL95}) resulting from the interactions between particles in counter-propagating channels, or from excitation emerging from edge reconstruction. Panel (D) shows the irreducible cross-correlation $\langle I_A I_B\rangle - \langle I_A \rangle \langle I_B\rangle$ as a function of the current bias, when two sources are equally diluted. In contrast to the $\nu = 3$ results of Fig.\,5C, the double-source cross-correlation with $\nu = 1$ is negative.}
\label{fig:single_filling}
\end{figure}

In this section, we present the measured correlation functions in an IQH sample with filling factor 1.
The results are shown in Fig.\,\ref{fig:single_filling}.
Surprisingly, the results display strong non-linearity. Indeed, in Figs.\,\ref{fig:single_filling}B, C, and D, transmission probabilities at all three QPCs become current-bias dependent, indicating the possible presence of interaction.
Accompanying the bias-dependence of transmissions, we also observe a negative double-source cross-correlation $\langle I_A I_B\rangle - \langle I_A \rangle \langle I_B\rangle$, as displayed in Fig.\,\ref{fig:single_filling}D.
We anticipate that this negative cross-correlation reflects a modified bunching probability, i.e., a finite $P_\text{QPC}$ of Eq.~\eqref{eq:iu_id_with_interactions}, which becomes manifest owing to the absence of charge fractionalization in a $\nu = 1$ system.
Indeed, in a $\nu = 3$ system (see Fig.~5B), the equal-source cross-correlation is positive. Following Eq.~\eqref{eq:iu_id_with_interactions}, this positive cross-correlation is expected to reflect the sign of the leading interaction-induced contributions, i.e., terms proportional to $\mathcal{T}_A P_A$ and $\mathcal{T}_B P_B$ of Eq.~\eqref{eq:iu_id_with_interactions}.
Importantly, these leading interaction-induced corrections require charge fractionalization between modes of multiple channels and are thus impossible for the $\nu = 1$ system. The next-order correction involving $P_\text{QPC}$, which reflects the interaction-induced modification of the bunching preference due to interactions at the central QPC, becomes then the leading one and thus is easily observable. This negative noise shown in Fig.\,\ref{fig:single_filling}D then perfectly agrees with our analysis of Eq.~\eqref{eq:iu_id_with_interactions}.

\subsection*{S7. Possible SEE application: quantifying entanglement of a mixed state}
\label{SM:SEE_fantasy}

\begin{figure}[h!]
\includegraphics[width= 0.7\columnwidth]{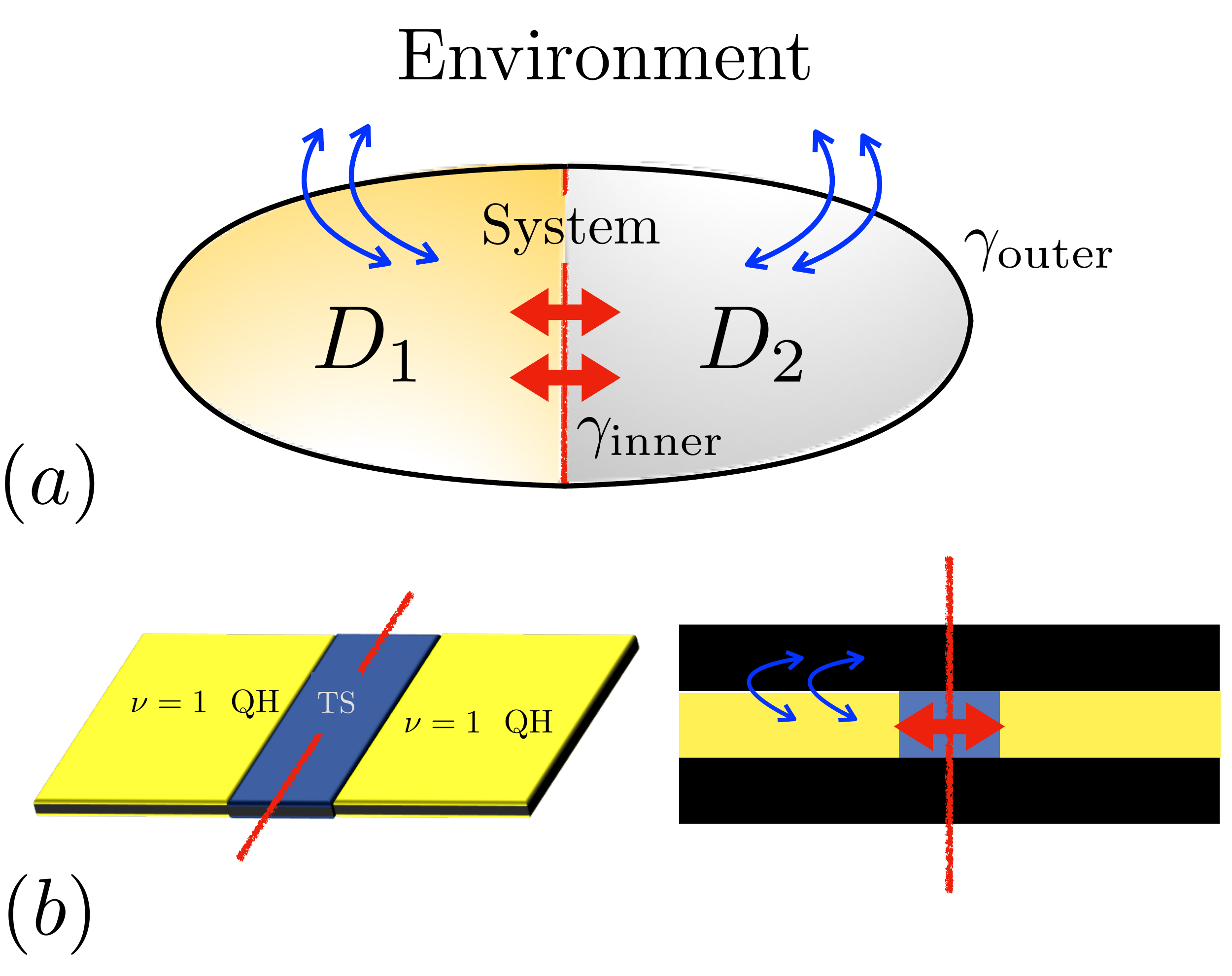}
\caption{Another possible application of our work: quantifying entanglement of mixed states. (A) A mixed state where the system comprising subsystems $D_1$ and $D_2$ interacts with its environment. After the partial trace over states in $D_2$, the calculated EE involves contributions from both interfaces $\gamma_\text{inner}$ and $\gamma_\text{outer}$. The communication between the system and the environment and the $D_1$--$D_2$ communications involve different types of quasiparticles, indicated by different types of arrows. (B) An example to illustrate our point. Left panel: the system that contains Majorana chiral modes (marked by the black arrow) at the interface of a $\nu=1$ quantum Hall (QH) sample and a topological superconductor (TS)~\cite{SFisherPRX14}. Right panel: experimentally, the EE involves contributions from both chiral Majorana modes  (the thick red double arrow) and the system-environment communication with fermions (the blue double arrows). The latter entanglement contribution can be excluded in case one focuses on only SEE from chiral Majorana modes.}
\label{fig:mixed_state}
\end{figure}

Our proposed framework can potentially help the quantification of the entanglement of mixed states, such as those in the setup displayed in Fig.\,\ref{fig:mixed_state}A, where the system $D_1 + D_2$ couples to the environment.
In this situation, the formally evaluated entanglement entropy
of the system comprising $D_1$ and $D_2$ does not quantify entanglement between the subsystems, but is rather dominated by the interaction with the environment. Indeed, after a naive partial trace over states in $D_2$, the EE contains entanglement from both interfaces $\gamma_\text{inner}$ and $\gamma_\text{outer}$.

To quantify entanglement in mixed states, various potential candidates, such as logarithmic entanglement negativity, mutual information, and
topological EE (see, e.g., Refs.~\cite{SPlenioPRL05,SHorodeckiRevModPhys09}) have been discussed in the literature.
In addition, the hierarchy of quantum correlations, including the EPR steering and Bell nonlocality were introduced, to classify entanglement of mixed states: entanglement $<$ EPR steering $<$ Bell nonlocality (see, e.g., Ref.~\cite{SWisemanJonesDohertyPRL07}, and related discussions of Ref.~\cite{SCollinsPRL02,STehPRA16}).
Denoted as hierarchy, it means that Bell nonlocality of a mixed state is sufficient to perform EPR steering (and to prove entanglement), but not the other way around (i.e., not all entangled mixed state can realize EPR steering; not every EPR steerable mixed state is subject to Bell's nonlocality inequality).
The entanglement indicators proposed in the present work give an alternative idea for obtaining the entanglement of mixed states. Indeed, the redundant entropy from system-environment communication is naturally excluded following our proposed functions (EP and SEE), similarly to the cancellation of such contributions in mutual information. This is particularly true if the two involved subsystems become entangled through exchanging  anyons or chiral Majorana modes (Fig.\,\ref{fig:mixed_state}B), as they are normally not involved in the system-environment entanglement.

\subsection*{S8. EE and SEE with a finite scattering area}
\label{SM:scattering_area}

\begin{figure}[h!]
\includegraphics[width=0.8  \columnwidth]{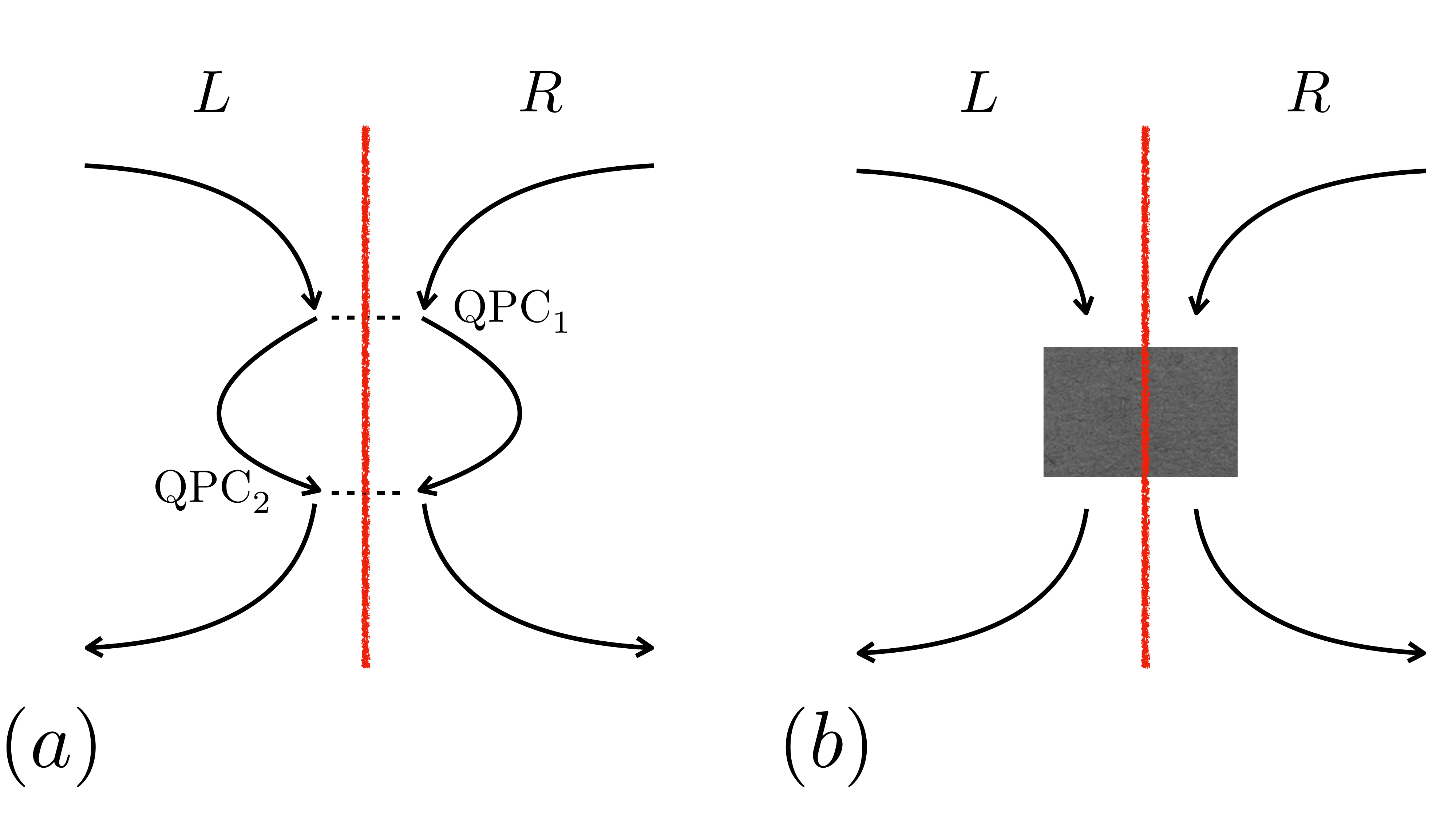}
\caption{Situations with (A) two scattering QPCs and (B) a scattering area (the grey box). Other required elements, e.g., diluters, are not shown in the figure.}
\label{fig:scattering_area}
\end{figure}

In our work, we extract the information on entanglement, i.e., the EE and SEE between two subsystems of an HOM interferometer, from current correlations.
In our method, the obtained SEE reflects entanglement induced by particles after the central QPC.
In general, when scattering between subsystems occurs at multiple positions (Fig.~\ref{fig:scattering_area}), one also has to consider the EE of particles in the scattering regime.

To see this point more explicitly, we start with the simple situation of Fig.~\ref{fig:scattering_area}A. We consider the EE produced by a single particle from the right part, with the system described by the initial wavefunction $|R\rangle$.
This setup is equivalent to a Mach-Zehnder interferometer.
After scattering at both QPCs, the final wavefunction can be written as
\begin{equation}
    (t_1 r_2 + r_1 t_2) |L\rangle + ( - t_1 t_2 + r_1 r_2) |R\rangle,
\end{equation}
where $t_1$, $r_1$, $t_2$, and $r_2$ are the scattering matrix parameters for the first and second QPCs, respectively.
Consider, for instance, the case of destructive interference $t_1 r_2 = - r_1 t_2$. In this case, when the particle has traveled after the second QPC, it always stays in the previous (right) subsystem, leading to a vanishing EE. However, one cannot conclude that these two subsystems never become entangled. Indeed, when the particle is between the two QPCs, it introduces finite EE $\mathcal{T}_1 \ln \mathcal{T}_1 + (1 - \mathcal{T}_1 ) \ln (1-\mathcal{T}_1)$, where $\mathcal{T}_1 \equiv t_1^2$ is the transmission probability of the first QPC.

Now, if we go to the more general case presented in Fig.~\ref{fig:scattering_area}B, the EE and SEE contain two contributions: from particles after the scattering area and from those within the scattering area.
The former contribution can be figured out by obtaining the effective transmission probability $\mathcal{T}_\text{sa}$ of the entire scattering area. Indeed, one can obtain EE of the former contribution by substituting transmission $\mathcal{T}$ of the HOM geometry by $\mathcal{T}_\text{sa}$.
This contribution grows linearly with the measurement time, as long as this time is smaller than other possible cutoffs in time, e.g., the dephasing time.

In contrast, entanglement from particles within the scattering area is constant if the measurement time $\tau$ is larger than the traveling time within the QPC. Specifically, in the two-QPC structure shown in Fig.~\ref{fig:scattering_area}, the EE and SEE from particles between two QPCs can be obtained by (i) replacing $\mathcal{T}$ by $\mathcal{T}_1$ of the first QPC and (ii) replacing the measuring time $\tau$ by the traveling time between two QPCs.
As a consequence, the EE and SEE from inter-QPC particles are negligible once the inter-QPC traveling time is much smaller than the dephasing time of the system.

\end{document}